\documentclass[11pt]{article}
\usepackage{jheppub}
\usepackage{avant}
\usepackage[dvipsnames]{xcolor}
\usepackage{mathtools,setspace,braket,parskip}
\usepackage[none]{hyphenat}
\usepackage[bb=boondox]{mathalpha}
\usepackage{float}
\usepackage{tikz-feynman}
\tikzfeynmanset{compat=1.1.0}

\tikzset{
    graviton/.style={
        double,
        decoration={snake,aspect=0.75,mirror,segment length=1.5mm},
        decorate
    }
}

\newcommand{\sd}{\mathrm{d}}
\newcommand{\pd}{\partial}
\newcommand{\bb}[1]{\mathbb{#1}}
\newcommand{\cl}[1]{\mathcal{#1}}
\renewcommand{\to}{\longrightarrow}
\newcommand{\dd}{\hat{\mathrm{d}}}
\newcommand{\del}{\hat{\delta}}
\newcommand{\Del}{\hat{\delta}_{\Phi}}
\newcommand{\normord}[1]{:\mathrel{\mspace{2mu}#1\mspace{2mu}}:}
\renewcommand{\Re}{\operatorname{Re}}
\renewcommand{\Im}{\operatorname{Im}}
\newcommand{\Tr}[1]{\text{Tr}\left(#1\right)}
\newcommand{\no}[1]{:\mathrel{\mspace{2mu}#1\mspace{2mu}}:}
\renewcommand{\[}{\begin{equation}\begin{aligned}}
\renewcommand{\]}{\end{aligned}\end{equation}}
\newcommand{\DC}[1]{{C_{#1}}}

\hypersetup{
    colorlinks=true,
    linkcolor=Mahogany,
    anchorcolor=Mahogany,
    citecolor=Mahogany,
    filecolor=Mahogany,
    urlcolor=Mahogany
}

\title{Inclusive Radiation and Backreaction from the Phase-Space S-Matrix}
\author{Nathan Moynihan}
\affiliation{School of Physical and Chemical Sciences, Queen Mary University of London, Mile End Road, London E1 4NS, United Kingdom}
\emailAdd{n.moynihan@qmul.ac.uk}

\begin{document}
\abstract{
We develop a phase-space description of classical scattering in which the matter sector of the Dyson S-matrix is partially Weyl transformed while the radiation sector remains operator valued. The resulting S-matrix symbol provides a common origin for several classical observables: we show that the inclusive waveform, the classical displacement in phase-space (i.e. the impulse and position shift), and the angular momentum arise as different projections of the same object. We show that the partially Weyl-transformed $S$-matrix admits an exponential organisation in terms of an elastic phase and connected radiation kernels, while its inclusive one-point projection admits a coherent representative with waveshape $\alpha_I$ describing classical radiation, extending our recent proposal in ref. \cite{Cristofoli:2021jas}. Furthermore, we show explicitly how nonlinear gravitational memory can be derived from an inclusive coherent waveshape. The dependence of the waveshape on the hard scattering data naturally leads to a quantum geometry, including an induced Berry connection on the space of waveforms, whose contribution to the phase-space displacement captures both radiation-reaction and static-field effects. We illustrate this structure by deriving static contributions to the position shift and field angular momentum in both scalar QED and gravity.
}

\maketitle
\section{Introduction} 
\label{sec:introduction} 
Classical two-body scattering can be organised in several closely related ways, and many modern scattering amplitude and worldline \cite{Brandhuber:2023hhy,Brunello:2026rdk,Haiashi:2026eyn,DeAngelis:2026ymn,Doran:2026bng,Aoude:2026bek,Damgaard:2026kqg,Bjerrum-Bohr:2026fhx,Bjerrum-Bohr:2026fhs,Cipriani:2026myb,Scheopner:2023rzp,Cho:2023kux,Bautista:2026qse,Alessio:2026bdi,Aoki:2026eos,Heissenberg:2025fcr,DeAngelis:2025vlf,Akpinar:2025byi,Fernandes:2025cog,Brunello:2025eso,Lopez-Arcos:2025cly,Aoude:2025xxq,Georgoudis:2025vkk,Bjerrum-Bohr:2025bqg,Alessio:2025flu,Bern:2026oqp,Alaverdian:2025jtw,Akpinar:2025huz,Aoude:2024sve,Falkowski:2024yuy,Bohnenblust:2024hkw,Brandhuber:2024qdn,Falkowski:2024bgb,Cheung:2024byb,Menezes:2026fmj,Cangemi:2026cjy,Gatica:2025uhx,Moynihan:2019bor,Akhtar:2025fil,Aoki:2025ihc,Ilderton:2025gug,Gatica:2024mur,Herrmann:2024yai,Copinger:2024pai,Gonzo:2024zxo,Aoki:2024bpj,Adamo:2024oxy,Luna:2023uwd,Brandhuber:2023hhl,DeAngelis:2023lvf,Jakobsen:2026ayw,Bohnenblust:2026ujk,Driesse:2026qiz,Wilson-Gerow:2023syq,Ben-Shahar:2025tiz,He:2025how,Haddad:2025cmw,Mogull:2025cfn,Hoogeveen:2025tew,Bohnenblust:2025gir,Biswas:2024ept,Haddad:2024ebn,Wilson-Gerow:2025xhr,Gonzo:2026yha,Bern:2023ity,Saavedra:2026ttk,Biendarra:2026nig} approaches are now very mature, making high-order contributions to e.g. gravitational wave physics. Amongst the most efficient are the \textit{eikonal} formulations, which represent the classical limit of the $S$-matrix as a phase whose derivatives generate a range of classical observables. The precise definition of the eikonal has become somewhat polluted over the years, and several contending exponential representations of the $S$-matrix exist: the `standard' eikonal, which we will take to mean the Fourier transform of the elastic $2\to2$ amplitude \cite{Fernbach:1949zz,Wallace:1971zz,Wallace:1977ae,Amati:1987wq,Amati:1990xe,Amati:1992zb,Kabat:1992tb,DiVecchia:2019myk,DiVecchia:2020ymx,Haddad:2021znf,Adamo:2021rfq,DiVecchia:2021bdo,DiVecchia:2022owy} (see \cite{DiVecchia:2023frv} for a review and many more references), the HEFT phase $\delta_{HEFT}$ \cite{Brandhuber:2021eyq} and the $N$-operator or Magnusian \cite{Magnus:1954zz,Damgaard:2021ipf,Damgaard:2023ttc,Kim:2024grz,Kim:2024svw,Kim:2025olv,Kim:2025magnusian,Brandhuber:2025igz,Guo:2026xaw}, to name a few. Within these formulations, it is common to perform a reparametrisation of the kinematic data, expressing them in terms of the `half-way' or `barred' momenta, which are the midpoints of the incoming and outgoing momenta usually defined as \cite{Parra-Martinez:2020dzs}
\[
p_i = \tilde p_i + \frac{q_i}{2},~~~~~p_i' = \tilde p_i' - \frac{q_i}{2}.
\]
This strongly motivates working in phase-space, where such variables are usually referred to as `midpoint' or `center' variables, and where they naturally show up in the Weyl transform of operators \cite{Moyal1949,Hillery:1983ms,OzorioDeAlmeida1998,Curtright:2011vw}, and has recently received an $S$-matrix treatment \cite{Kim:2025ebl}. Recently, there has been a lot of interest in extending the eikonal to include radiative observables and radiation-reaction effects \cite{Ciafaloni:2018uwe,Addazi:2019mjh,DiVecchia:2021bdo,Cristofoli:2021jas,DiVecchia:2022piu,Kim:2025hpn}. In \cite{Cristofoli:2021jas}, we proposed a radiative extension of the eikonal, in which a coherent radiation factor with waveshape $\alpha_\lambda(k)$ is convolved with the conservative eikonal phase. The proposal was motivated by the requirement that classical field observables have negligible variance, which produced an infinite set of relations between amplitudes with different loop orders and legs. The construction was deliberately minimal: the precise final state was proposed rather than derived in general, and its exponentiation was established at leading order. One purpose of the present paper is to explore this proposed final state in phase-space, obtaining it directly from the Dyson $S$-matrix, and to determine precisely what the coherent state does, and does not, represent.

In the KMOC formulation, observables are computed directly from in-in matrix elements of $\bb S^\dagger\bb O\bb S$ \cite{Kosower:2018adc}. The relation between this and the eikonal descriptions is transparent enough for conservative observables, but it becomes less obvious once radiation is included, because scattering amplitudes naturally organise outgoing radiation into exclusive fixed-multiplicity sectors, whereas a deterministic classical waveform is an inclusive one-point function, 
\[ 
\alpha_{I,\lambda}(k) = \langle \bb S^\dagger a_\lambda(k)\bb S\rangle. 
\] 
The coefficient of a one-radiation final state and the expectation value $\langle a_\lambda(k)\rangle$ coincide at leading order, but they need not coincide once connected higher-radiation processes contribute. A natural exponential formulation of the $S$-matrix is organised in terms of exclusive amplitudes, reproducing the inclusive waveform order by order in perturbation theory \cite{Caron-Huot:2023vxl,Georgoudis:2023lgf,Kim:2025hpn,Alessio:2025flu,Gonzo:2024zxo,Gonzo:2026yha,Blanco:2026prj}. Nevertheless, it should be possible to derive the classical, deterministic waveform from a purely coherent state. This issue is particularly timely following the recent amplitude calculation of nonlinear gravitational memory \cite{Christodoulou:1991cr,Thorne:1992sdb,Georgoudis:2025vkk}. Ref.~\cite{Georgoudis:2025vkk} showed that the leading nonlinear memory contains a classical contribution from a six-point amplitude inside a cut, composed with the five-point amplitude. Earlier analyses found that the directly resolved tree-level two-radiation contribution is $\hbar$-suppressed in the field-strength variance and graviton-number projections \cite{Cristofoli:2021jas,Britto:2021pud}. In particular, ref.~\cite{Britto:2021pud} showed that, by evaluating the $6{\rm pt}^*\times6{\rm pt}$ cut term, the leading departure of the graviton-number distribution from Poissonianity at $\cl{O}(G^4)$ is purely quantum. Importantly, however, these conclusions do not exclude the six-point amplitude from ever contributing to a different inclusive observable, such as the waveform via a $5{\rm pt}^*\times6{\rm pt}$ cut. Ref.~\cite{Georgoudis:2025vkk} argued that classical gravitational radiation is not completely described by an ordinary non-squeezed coherent state, and that nonlinear memory requires going beyond the coherent-eikonal proposals of refs.~\cite{Cristofoli:2021jas,DiVecchia:2022piu}. We agree with the first statement when it refers to the complete exclusive outgoing state. Nonlinear interactions generate connected two- and higher-radiation components, so the exact state is generally not coherent and a coherent state cannot reproduce its number statistics, noise, or higher connected correlators. This does not, however, mean there is not a coherent representative of the classical field: any finite-frequency, classical asymptotic waveform can be reproduced as the one-point function of a coherent state. The relevant coherent displacement is not the exclusive one-radiation coefficient, but the full inclusive waveshape 
\[ 
\alpha_{I,\lambda}(k) = \langle a_\lambda(k)\rangle, 
\] 
which itself receives classical contributions from higher-multiplicity amplitudes through inclusive cuts. The nonlinear-memory term is the first such nontrivial example of this contribution. Importantly, the coherent radiative state survives as a representative of the inclusive one-point field, provided it is not identified with the exact exclusive state. We will show that this distinction places the proposal of ref.~\cite{Cristofoli:2021jas,DiVecchia:2022piu} on firmer ground while also expanding its domain of validity.

In ref.~\cite{Luna:2023uwd}, we introduced a translation operator that shifts the on-shell conditions via 
\[ 
e^{q\cdot Y}, \qquad Y_\mu = \frac12\frac{\partial}{\partial p_2'{}^\mu} - \frac12\frac{\partial}{\partial p_1'{}^\mu}, 
\] 
reparameterising the eikonal in the half-way momenta and generating the iterations required for the classical impulse and spin kick. Although this translation proved efficient for calculations, it was proposed by observation and its underlying origin was not manifest. We will show here that this translation is a manifestation of a two-particle Bopp shift \cite{Bopp1956}, and a natural object on phase-space. Under a Weyl transform, operator composition becomes a Moyal star product, and the star product of Fourier-resolved symbols translates their momentum arguments by half of the transfer carried by the neighbouring factor \cite{Bayen:1977ha,Curtright:2011vw}. The half-way momenta are therefore not just a neat reparametrisation, but also the natural midpoint variables of the Weyl transform. With this in mind, we will develop a phase-space formalism for obtaining classical observables directly from the Dyson $S$-matrix, and show that the Bopp shifts are a natural consequence of the star product structure.

We will take the effective field theory approach that matter and radiation ought to be treated very differently in the classical limit: massive matter particles, say charged spheres or black holes, are treated as point particles with fixed particle number, whereas radiation is allowed in any amount, with the classical limit corresponding to a large occupation number. In the quantum theory, this is reflected in the fact that the matter sector is restricted to a fixed particle-number Hilbert space, while the radiation sector is a Fock space, following the philosophy of \cite{Cristofoli:2021jas,DiVecchia:2022piu}. With this in mind, we will perform a partial Weyl transform of the $S$-matrix over the matter sector \cite{Wigner:1932eb,Hillery:1983ms}, mapping it to an on-shell phase-space symbol, while leaving the radiation Fock space operator valued.

At fixed phase-space data 
\[ 
\tilde z=(x_1,\tilde p_1;x_2,\tilde p_2), 
\] 
this gives the state-valued matter symbol 
\[ 
\ket{\Psi_{\rm out}(\tilde z)} \equiv S_W(\tilde z)\ket{0_{\rm rad}}. 
\] 
This object is a ket in radiation Fock space, but as we will see, it is not by itself a physical outgoing state at each fixed $\tilde z$. The main object of study will be the symbol $S_W(\tilde z)$, which, as we will show, has a convenient exponential form.

The central result of this paper is that the inclusive waveform and the displacement in phase-space are two different projections of this same symbol. Defining 
\[ 
\cl N(\tilde z) = \langle0| S_W^\dagger(\tilde z) \star_m S_W(\tilde z) |0\rangle, 
\] 
we find the exact relations 
\[ 
\alpha_{I,\lambda}(k;\tilde z) = \frac{1}{\cl N} \langle0| S_W^\dagger(\tilde z) \star_m a_\lambda(k) \star_m S_W(\tilde z) |0\rangle, 
\] 
and 
\[ 
\Delta z_A = \frac{i\hbar}{2\cl N} \langle0| S_W^\dagger(\tilde z) \star_m \overleftrightarrow{\partial}_A S_W(\tilde z) |0\rangle, 
\] 
where $\star_m$ is the matter star product.

Upon expanding the outgoing state in fixed radiation multiplicity, the first formula pairs the $n$-radiation coefficient with the $(n+1)$-radiation coefficient: the annihilation operator removes the measured particle, while the remaining radiation is summed inclusively. The second formula pairs equal multiplicities because differentiation with respect to the phase-space variables preserves radiation number, and contains the usual impulse as well as the asymptotic position shift. The waveform and the displacement are therefore not independent structures appended to an eikonal, but projections of the same partially Weyl-transformed $S$-matrix. Neither result assumes a coherent outgoing state, however as we will see it is convenient to assume a coherent representative of the inclusive waveform, which is defined by the displacement $\alpha_{I,\lambda}(k)$.

Once this inclusive waveshape has been determined, it defines a coherent representative which reproduces the classical deterministic field but not the higher correlators of the exclusive state. Interestingly, the dependence of this state on the hard scattering data also allows us to construct a quantum geometry on the space of waveforms \cite{Berry:1984jv,Simon:1983mh}. In particular, the space is endowed with a Berry connection $\cl A_A$, which contributes directly to the phase-space displacement, 
\[ 
\Delta z_A^{\rm coh} = -\partial_A\operatorname{Re}\chi + \cl A_A, \qquad \cl A_A = -\frac{\hbar}{2i} \sum_\lambda\int d\Phi(k)\, \alpha_{I,\lambda}^*(k) \overleftrightarrow{\partial}_A \alpha_{I,\lambda}(k). 
\] 
The curvature of this connection measures the obstruction to describing the outgoing radiation state by the derivative of a single scalar phase (i.e. the eikonal), matching our expectation from \cite{Cristofoli:2021jas}, where the derivative of the waveshape gave rise to the radiation reaction part of the impulse. However, we will also see that this Berry connection corrects the position shift, and matching classical results there requires us to extend our proposal in \cite{Cristofoli:2021jas}, specifically requiring that the waveshape also depends explicitly on the transverse momentum transfer $q$. Using this, we will show that the Berry connection is responsible for the leading static-field contributions to the position shift and field angular momentum in scalar QED and gravity. We will also show that the usual symplectic area of phase space that is typically preserved under conservative Hamiltonian flow is still conserved even when radiation is present, given a suitible area defined on the space of waveforms. We show that this gives rise to a symplectic balance law.

The paper is organised as follows. We first develop the projected on-shell phase-space formalism and derive the Bopp translations from the partial Weyl transform. We then obtain the phase-space eikonal from the Dyson $S$-matrix and explain why local recoil must be resolved before approximations related to the local history can be made. Next, we derive the exact multiplicity projections for the inclusive waveform and hard displacement, and construct connected radiation kernels from the full on-shell amplitudes. We apply the resulting inclusive cut formula to nonlinear gravitational memory, before studying the quantum geometry of the coherent representative, momentum and angular-momentum balance, and the static position shifts in scalar QED and gravity. Technical details of the positive-energy star product, Dyson ordering, endpoint reduction, and amplitude reconstruction are collected in the appendices. 

\section{Phase-Space Symbols, Bopp Shifts and KMOC Observables}
\label{sec:phase-space-symbols}
\label{sec:phase-space-weyl-kmoc}

We use the following conventions throughout. Hats absorb factors of $2\pi$ but \emph{not} factors of $\hbar$, which we will track explicitly in this section,
\[
\hat{\sd}^n q\equiv \frac{\sd^n q}{(2\pi)^n},\qquad
\del^{(n)}(q)\equiv (2\pi)^n\delta^{(n)}(q),\qquad
\del^+(p^2-m^2)\equiv 2\pi\,\theta(p^0)\delta(p^2-m^2),
\]
and the on-shell measure is $\sd\Phi(p)=\hat{\sd}^4p\,\del^+(p^2-m^2)$, corresponding to the relativistic normalization $\braket{p|p'}=2E_p(2\pi)^3\delta^{(3)}(\mathbf p-\mathbf p')$. Note that this corresponds to normalized creation and annihilation operators so that
\[
[a_\eta(k), a^\dagger_{\eta'}(k')] = \delta_{\eta \eta'} \Del(k-k') \,,
\]
where $\Del(k-k')$ is
\[
\Del(k-k') \equiv 2k^0\,\hat\delta^{(3)}(\vec k-\vec k').
\]
We define the Fourier transform and its inverse by
\[
\widetilde f(q)
=
\int\sd^n x\,e^{iq\cdot x/\hbar}f(x),
\qquad
f(x)
=
\frac{1}{\hbar^n}
\int\hat{\sd}^n q\,e^{-iq\cdot x/\hbar}\widetilde f(q),
\]
so that
\[
\int\sd^n x\,e^{iq\cdot x/\hbar}
=
\hbar^n\del^{(n)}(q).
\]

\subsection{Weyl Symbols and the On-Shell Phase-Space Pairing}
Classical observables are naturally described by phase-space distributions, where their expectation values are computed as phase-space integrals weighted by probability densities \cite{Hillery:1983ms}
\[
\braket{A}_{cl} = \int d^4x d^4 p \theta(p^0) ~\delta(p^2-m^2)A(x,p)P_{cl}(x,p)
\]
where $P_{cl}(x,p)$ is a probability distribution function. Given the uncertainty principle, it is not possible to define a joint probability distribution for quantum systems: we cannot assign a probability that a given quantum object will simultaneously have definite position $x$ and momentum $p$. However, we can define \textit{quasi}-probability distributions, which behave like classical probability distributions, but can take negative values in some regions, signalling a breakdown of classicality. The most well-known of these is the \textit{Wigner function} \cite{Wigner:1932eb}, which is the phase-space representation of the density matrix. The goal of this section is to establish a formalism for describing classical, relativistic observables on phase space in terms of the $S$-matrix. We begin by introducing the \textit{Wigner--Weyl transform}, which maps operators on Hilbert space to functions on phase space called \textit{Weyl symbols}.

For an operator $\bb A$ acting on a Hilbert space, its Weyl symbol is
\[
\cl{W}[\bb A](x,p) = A_W(x,p) =
\int d^4y\,e^{-ip\cdot y/\hbar}
\left\langle x+\frac{y}{2}\right|\bb A\left|x-\frac{y}{2}\right\rangle .
\label{eq:canonical-weyl-symbol}
\]
Using $\langle x|p\rangle \sim e^{+ip\cdot x/\hbar}$ and inserting two momentum resolutions of the identity gives
\[
\left\langle x+\frac y2\right|\bb A\left|x-\frac y2\right\rangle
\propto
\int d^4p_+\,d^4p_-\,
e^{+i(p_+-p_-)\cdot x/\hbar}
e^{+i(p_++p_-)\cdot y/(2\hbar)}
\langle p_+|\bb A|p_-\rangle .
\]
The $y$ integral therefore fixes $p=(p_++p_-)/2$; defining $q=p_+-p_-$ yields the equivalent momentum-space form
\[
\cl{W}[\bb A](x,p) = A_W(x,p)
=
\int \hat{\sd}^4q\,e^{+iq\cdot x/\hbar}
\left\langle p+\frac{q}{2}\right|\bb A\left|p-\frac{q}{2}\right\rangle .
\label{eq:canonical-weyl-symbol-momentum}
\]

The Wigner function, denoted $\rho_W(x,p) = \cl{W}[\rho](x,p)$, is the Weyl symbol of the density matrix, and the expectation value of an operator can then be computed as an integral over phase space of the product of the Wigner function and the Weyl symbol of that operator, which is a direct analogue of the classical expectation value formula above. This is the basic mechanism used throughout the paper: the state supplies a phase-space density and the operator supplies a symbol.

With these conventions the Hilbert-space trace becomes the phase-space trace pairing
\[
\Tr{\rho\bb A(x)}
=
\frac{1}{\hbar^4}\int \hat \sd^4p~\sd^4y
\rho_W(y,p)A_W(y+x,p).
\label{eq:canonical-trace-pairing}
\]

The unconstrained Weyl transform maps products of operators to \textit{star products} of their symbols, which are non-commutative deformations of the usual product of functions on phase space. For two operators $\bb A$ and $\bb B$, we have
\[
\cl W[\bb A\bb B]=A_W\star B_W,
\label{eq:weyl-product-star}
\]
where
\[
A\star B
=
A\exp\left[
\frac{i\hbar}{2}
\left(
\overleftarrow\partial_{x^\mu}\overrightarrow\partial_{p_\mu}
-
\overleftarrow\partial_{p_\mu}\overrightarrow\partial_{x^\mu}
\right)
\right]B .
\label{eq:moyal-star-product}
\]
The corresponding Moyal bracket is given by
\[
\{A,B\}_{\rm M}
\equiv
\frac{[A,B]_\star}{i\hbar}
=
\frac{1}{i\hbar}(A\star B-B\star A)
=
\cl W\!\left[\frac{[\bb A,\bb B]}{i\hbar}\right]
=
\{A,B\}_{\rm PB}+O(\hbar^2).
\label{eq:moyal-bracket}
\]
so the Poisson bracket is the leading classical term.

In particular, these definitions give the standard result
\[
\cl W[\hat x^\mu\hat p_\nu]
=x^\mu p_\nu+\frac{i\hbar}{2}\delta^\mu{}_\nu,
\qquad
x^\mu\star p_\nu-p_\nu\star x^\mu
=i\hbar\delta^\mu{}_\nu,
\qquad
\{x^\mu,p_\nu\}_{\rm M}=\delta^\mu{}_\nu.
\]

We will often encounter symbols containing Fourier phases. For such objects, for example if we have
\[
F_q(x,p)=e^{+iq\cdot x/\hbar}f(p),
\qquad
G_\ell(x,p)=e^{+i\ell\cdot x/\hbar}g(p),
\]
then the star product gives the \textit{Bopp-shift identity} \cite{Bopp1956,Curtright:2011vw,deGosson:2011book}
\[
F_q\star G_\ell
=
e^{+i(q+\ell)\cdot x/\hbar}
f\left(p+\frac{\ell}{2}\right)
g\left(p-\frac{q}{2}\right).
\label{eq:bopp-shift-identity}
\]
Note that the explicit factor of $\hbar$ in \eqref{eq:moyal-star-product} can be cancelled by derivatives acting on the phases. This is important, since it tells us that the expansion parameter is then not simply $\hbar$, but the ratio $q/p$. As we will see, this is the natural expansion parameter in the eikonal regime, where the momentum exchange $q$ is small compared to the hard momentum $p$.

Since we will largely be interested in the two-particle sector, it is useful to consider the Bopp shift acting purely in momentum space on two-particle symbols, for example by defining
\[
F(x_1,p_1;x_2,p_2) &= \int \hat{\sd}^4q_1\hat{\sd}^4q_2\, e^{+i(q_1\cdot x_1+q_2\cdot x_2)/\hbar}\del^{(4)}(q_1+q_2)F(q_1,p_1;q_2,p_2)\\
&= \int \hat{\sd}^4q\, e^{+iq\cdot x_1/\hbar-iq\cdot x_2/\hbar}F(p_1,p_2;q),
\]
and similarly for $G$, where we have used the fact that the system has total translational invariance. The star product of two such symbols is then
\[
F\star G
&=
\int \hat{\sd}^4q\,\hat{\sd}^4\ell\,
e^{+i(q+\ell)\cdot (x_1-x_2)/\hbar}\,
F\!\left(p_1+\frac{\ell}{2},p_2-\frac{\ell}{2};q\right)
G\!\left(p_1-\frac{q}{2},p_2+\frac{q}{2};\ell\right).
\]
In other words, the star product acts as a translation operator in momentum space:
\begin{equation}
\begin{aligned}
F(p_1,p_2;q) \rightarrow F\left(p_1+\frac12\ell,p_2-\frac12\ell;q\right)=e^{-\ell\cdot Y}F(p_1,p_2;q),
\end{aligned}
\end{equation}
where 
\[
Y^\mu = \left(
\frac12\frac{\partial}{\partial p_{2\mu}}
-
\frac12\frac{\partial}{\partial p_{1\mu}}
\right).
\]
This is the observed Bopp translation used in \cite{Luna:2023uwd} to define the eikonal phase beyond leading order, and here we identify its phase-space origin: it is the two-particle Bopp shift expressed in momentum space.

For a product of many Fourier-resolved insertions on the same worldline the same identity gives the useful formula
\[
\prod_{r=1}^{N}{}_{\star}
\left[ e^{iq_r\cdot x/\hbar}j_r(\tilde p,q_r)
\right]
=e^{iQ\cdot x/\hbar}
\prod_{r=1}^{N}
j_r\!\left(
\tilde p
-\frac12\sum_{a<r}q_a
+\frac12\sum_{a>r}q_a,
q_r
\right),
\qquad Q=\sum_{r=1}^Nq_r .
\label{eq:N-current-bopp-shift}
\]
This is the local recoil content of the star product, and below it will give the local differential representation of the positive-energy matter product.  The insertion labelled by $r$ is evaluated at the midpoint momentum appropriate to its position in the ordered sequence: it sees half of the kicks inserted before it and half of the kicks inserted after it.  Thus the star product is not a quantum correction that can be discarded blindly: it defines a sequence of small momentum kicks, whose sum can be large and classical.

\subsection{On-Shell Relativistic Phase-Space}
Consider the expectation value of a local Hermitian operator $\bb{O}(x)$, evaluated at spacetime point $x$, for an on-shell state $\rho$ of a relativistic particle of mass $m$. We have then
\[
\braket{\bb{O}(x)}
=\Tr{\rho\bb{O}(x)}
=\int \sd\Phi(p)\sd\Phi(p')\,
\braket{p|\rho|p'}\braket{p'|\bb{O}(x)|p},
\]
where we have inserted a complete set of single-particle momentum eigenstates.

Translation invariance gives $\braket{p'|\bb{O}(x)|p}=e^{i(p'-p)\cdot x/\hbar}\braket{p'|\bb{O}(0)|p}$, so the $x$-dependence factors out as a plane-wave phase, motivating the definition $q=p-p'$. Hermiticity of $\rho$ and $\bb{O}$ forces the integrand $F(p,p')\equiv\braket{p|\rho|p'}\braket{p'|\bb{O}(0)|p}$ to satisfy $F(p,p')^*=F(p',p)$. Locality and hermiticity then suggest passing to midpoint variables
\[
\tilde p=\tfrac12(p+p'),\qquad q=p-p',\qquad
p=\tilde p+\tfrac q2,\quad p'=\tilde p-\tfrac q2,
\]
with unit Jacobian. The reality of $\braket{\bb{O}(x)}$ is manifest after the $q$ integration. Writing $p^2-m^2=A+B$ and $p'^2-m^2=A-B$ with $A=\tilde p^2+q^2/4-m^2$ and $B=\tilde p\cdot q$, the identity $\delta(A+B)\delta(A-B)=\tfrac12\delta(A)\delta(B)$ gives
\[
\sd\Phi(p)\sd\Phi(p')
=\hat{\sd}^4\tilde p\,\hat{\sd}^4q\,\hat\delta_\Gamma(\tilde p,q),
\qquad
\hat\delta_\Gamma\equiv
\Theta(\tilde p,q)\,
\del\!\left(\tilde p^2+\tfrac{q^2}{4}-m^2\right)\del(2\tilde p\cdot q),
\]
where
\[
\Theta(\tilde p,q)\equiv
\theta\!\left(\tilde p^0-\tfrac{q^0}{2}\right)
\theta\!\left(\tilde p^0+\tfrac{q^0}{2}\right) = \theta(p^0)\theta(p'^0).
\]
The idea is then that these conditions project us to the positive energy, transverse and on-shell part of phase space. 

The two on-shell constraints are equivalently
\[
\tilde p\cdot q=0,\qquad
\tilde p^2=m^2-\frac{q^2}{4}.
\label{eq:midpoint-onshell-constraints}
\]
The first is the transverse support of the off-diagonal momentum and the second fixes the norm of the midpoint of two exactly on-shell momenta, with the usual on-shell condition recovered at $q=0$.

We can now define the midpoint momentum space \textit{symbols} for the state and the observable,
\[
\rho_W(\tilde p,q)\equiv\braket{\tilde p+\tfrac q2|\rho|\tilde p-\tfrac q2},
\qquad
O_W(\tilde p,q)\equiv\braket{\tilde p+\tfrac q2|\bb{O}|\tilde p-\tfrac q2},
\]
and, using $\braket{\tilde p-\tfrac q2|\bb{O}(x)|\tilde p+\tfrac q2}=e^{-iq\cdot x/\hbar}O_W(\tilde p,-q)$, the trace formula becomes
\[
\braket{\bb{O}(x)}=\int\hat{\sd}^4\tilde p\,\hat{\sd}^4 q\,
\hat\delta_\Gamma(\tilde p,q)\,
\rho_W(\tilde p,q)\,
e^{-iq\cdot x/\hbar}\,O_W(\tilde p,-q),
\label{eq:trace-momentum}
\]
To carry this construction over to the on-shell setting, we choose to distribute the two constraints in $\hat\delta_\Gamma$ between the state and the observable. Although both constraints come from the requirement that the two endpoint momenta are on shell, in our chosen representation we assign them different roles: $\del(\tilde p^2+\tfrac{q^2}{4}-m^2)$ enforces the on-shell projection with the state, while $\del(2\tilde p\cdot q)$ carries the transverse projection with the observable. The positive-energy step functions on the external momenta are part of the final projection and assigned to $\rho_W$. We define the matter star product $\star_m$ to be the star product induced by operator composition on the positive-energy one-particle matter Hilbert space. Essentially, $\star_m$ is the unconstrained Moyal star product \eqref{eq:moyal-star-product} restricted to the positive-energy on-shell domain, and we take $m = 1,2$ to label the specific matter line. This is so it matches the usual positive-energy operator product, which we usually obtain by inserting a resolution of the identity which carries $\sd\Phi(r)$ and hence $\theta(r^0)$. 

This asymmetric assignment is not forced, but it is convenient for mmaking contact with the literal half-way momenta commonly used in amplitude calculations. It differs from the usual self-dual Stratonovich--Weyl construction \cite{Stratonovich:1956,Brif:1998}: the state and observable are represented here by `mutually dual' rather than identical maps, with the on-shell and transverse constraints assigned asymmetrically. Our construction therefore does not satisfy the strict single-kernel, self-dual Stratonovich--Weyl axioms. That said, self-duality may nevertheless be relaxed in more general phase-space constructions by introducing dual quantisation and dequantisation kernels. In particular, the projected trace pairing used here realises a generalised-traciality relation of the type discussed in \cite{Brif:1998}, and belongs to the broader class of phase-space representations reviewed in \cite{Balazs:1983hk,Brif:1998}. In a paper released concurrently with this one \cite{Canxin_Amps26}, a strict self-dual Stratonovich--Weyl correspondence \textit{is} employed to construct classical observables from amplitudes. The construction outlined there has some differences in how constraints are applied and in how the phase-space symbols are constructed, but nontheless both constructions yield well-defined classical limits in the applications considered. It would be interesting to explore the relationship between these two constructions in more detail, and we leave this for future work.

Assigning one delta to each factor produces two useful phase-space objects:
\[
\rho_W(x,\tilde p)
\equiv
\frac{1}{\hbar^s}\int\hat{\sd}^4 q\,
e^{i q\cdot x/\hbar}\,
\Theta(\tilde p,q)\,
\del\!\left(\tilde p^2+\tfrac{q^2}{4}-m^2\right)
\rho_W(\tilde p,q),
\label{eq:covariant-wigner}
\]
\[
O_W(x,\tilde p)
\equiv
\frac{1}{\hbar^{4-s}}\int\hat{\sd}^4 q\,
e^{i q\cdot x/\hbar}\,
\del(2\tilde p\cdot q)\,
O_W(\tilde p,q),
\label{eq:covariant-weyl-symbol}
\]
where $x$ is the phase-space position variable and $O_W(\tilde p,q)$ is the kernel for $\bb{O}(0)$. The distinction between these objects, defined on phase-space, and the momentum space symbols defined above, is understood by their differing arguments. In our chosen representation, the Wigner function carries the mass-shell projection while the operator symbol carries the transverse projection, ensuring that no particular position on the worldline is preferred: the delta function enforces the equivalence $x \sim x + \lambda \tilde p$. Indeed, shifting $x\to x+\lambda\tilde p$ multiplies each Fourier mode by $e^{i\lambda\tilde p\cdot q/\hbar}=1$ due to the transverse delta function. In the rest frame of $\tilde p$, the same delta sets $q^0=0$, so the observable is independent of the coordinate along the line. 

The split of $\hbar$ factors is general, and in principle choosing any $s$ leads to a well-defined observable. However, it is most convenient to choose $s=4$, such that star products of operators obey the standard convolution rules and that the expectation value of the identity operator is normalized to unity, i.e.
\[
\braket{\bb{1}}
&=\int\hat{\sd}^4\tilde p\,\sd^4 x\,
\rho_W(x,\tilde p)1_W(x,\tilde p)
\\
&=\frac{1}{\hbar^4}
\int\hat{\sd}^4\tilde p\,\sd^4x\,
\hat{\sd}^4q\,\hat{\sd}^4q'\,
e^{i(q+q')\cdot x/\hbar}\,
\Theta(\tilde p,q)\,
\del\!\left(\tilde p^2+\tfrac{q^2}{4}-m^2\right)
\rho_W(\tilde p,q)\del(2\tilde p\cdot q')\,
2\left(\tilde p^0+\tfrac{q'^0}{2}\right)
\del^{(3)}(\mathbf q')
\\
&=\int\hat{\sd}^4\tilde p\,
\del^+(\tilde p^2-m^2)\,
\braket{\tilde p|\rho|\tilde p}
=\Tr{\rho}=1
\] 
where we have used
\[
\int d^4x\,e^{i(q+q')\cdot x/\hbar} = \hbar^4\del^{(4)}(q+q').
\]  
The identity then gives $1_W(x,\tilde p)=1$ for the projected positive-energy symbols. The usual star product does not naively restrict itself to the positive energy patch, and so it is understood that the $\star_m$ product is the positive-energy restriction of the usual Moyal star product. For completeness, this is explicitly discussed in appendix~\ref{app:bopp-matter-poles}.

For $s=4$, the on-shell Weyl transform obeys
\[
\cl{W}_m[\bb A\bb B]
=
\cl{W}_m[\bb A]\star_m\cl{W}_m[\bb B],
\label{eq:rescaled-onshell-weyl-product}
\]
up to terms which are always projected away in the on-shell pairing. The corresponding differential representation therefore comes with no $\hbar$ prefactor, and is given by
\[
C_W
&=
A_W\exp\left[
\frac{i\hbar}{2}
\left(
\overleftarrow\partial_{x^\mu}\overrightarrow\partial_{\tilde p_\mu}
-
\overleftarrow\partial_{\tilde p_\mu}\overrightarrow\partial_{x^\mu}
\right)
\right]B_W
\\
&=
A_WB_W
+\frac{i\hbar}{2}\{A_W,B_W\}_{\rm PB}
+O(\hbar^2).
\]

The positive-energy composition rule is $C_W=A_W\star_m B_W$, as opposed to the general case where $C_W=\hbar^{4-s}(A_W\star_m B_W)$.  We should remember, however, that derivatives acting on the Fourier phases can cancel the explicit $\hbar$ in the differential representation, so this formal expansion is not by itself a classical expansion when the momentum transfers are held fixed.

The expectation value of $\bb{O}(x)$ is then the projected on-shell Weyl pairing,
\[
\braket{\bb{O}(x)}=\int\hat{\sd}^4\tilde p\,\sd^4 x'\,\rho_W(x',\tilde p)\,O_W(x'+x,\tilde p),
\label{eq:trace-phase-space}
\]
where $O_W(x'+x,\tilde p)$ is the Weyl symbol of $\bb{O}(0)$ evaluated at the shifted argument $x'+x$, encoding the translation from origin to insertion point $x$.

It should be noted, however, that neither of these objects is unique on its own: they are auxiliary continuations, and only their projected pairing \eqref{eq:trace-phase-space} is physical. We will usually leave this final matter pairing implicit in what follows. In particular, when we refer to $\Delta O$ as an observable, we mean that the corresponding Weyl expression has been evaluated using the pairing \eqref{eq:trace-phase-space}, and in this paper we will always take the Wigner function to be in the sharp-packet limit, as detailed in section~\ref{sec:incoming-matter-state}. Before the projected pairing, we note that any symbol is simply a local phase-space function, which need not depend on the incoming Wigner function. You might worry that this definition could lead to ambiguities, for example when derivatives of the symbol are taken, but the transverse delta function in \eqref{eq:covariant-weyl-symbol} ensures that the symbol is only ever evaluated on the transverse subspace, making the projected pairing well-defined and unambiguous, as we show explicitly in appendix \ref{app:off-shell-extensions}. For a general on-shell function, we take the derivatives to be evaluated while continuing to satisfy the on-shell and transverse conditions. When derivatives occur inside a star product, its final pairing is independent of the chosen continuation, as we show explicitly in appendix~\ref{app:off-shell-extensions}.

\subsection{Operator-Valued Matter Symbols}
The S-matrix is an operator on the full Fock space of all the fields that make up the system. We are specifically interested in a system of classical bodies interacting via a massless messenger, and so we do not need the full Fock space of the classical bodies: the classical sector is restricted to fixed hard-particle number. Our strategy to simplify this problem, then, will be to utilise the phase-space formalism developed in section~\ref{sec:phase-space-weyl-kmoc} to define a partial symbol: we perform a Weyl transform only over the classical matter sector, leaving the massless particle Fock space untouched. As discussed above, the Weyl transform is a linear map from operators on Fock space to functions on phase space, and so the partial Weyl transform of the S-matrix will be an operator on the massless particle Fock space, but a function on the classical matter phase space. In other words, we will compute
\[
\cl W_m[\mathbb S] = \cl W_m[\mathbb S](x_1,\tilde p_1;x_2,\tilde p_2).
\]
For fixed phase-space coordinates $\tilde z=(x_1,\tilde p_1;x_2,\tilde p_2)$, we define the outgoing radiation ket by
\[
\ket{\Psi_{\rm out}(\tilde z)}
\equiv
S_W(\tilde z)\ket{0_{\rm rad}},
\qquad
S_W(\tilde z)
\equiv
\cl W_m[\mathbb S](\tilde z).
\]
At fixed $\tilde z$ this is a ket in the radiation Fock space. The physical observables are obtained by placing this symbol inside the projected matter pairing developed in section~\ref{sec:phase-space-weyl-kmoc}. It should be noted that this state is not necessarily a genuinely physical state on its own, but its paired projection onto the physical matter sector is.

In what follows, we will work in the classical limit when necessary, being particularly careful not to take the limit too early. For example, the local Bopp shifts scale as $q/p$, however these are perfectly classical when summed to sufficiently high number, and should not be arbitrarily taken to vanish too early. 
\subsection{The Incoming Matter State}
\label{sec:incoming-matter-state}
We are interested in an initial state containing exactly two matter particles and zero photons (or gravitons) is given by 

\[
\ket{\psi} = \int \sd\Phi(p_1) \sd\Phi(p_2) \, \varphi_1 (p_1) \varphi_2(p_2) e^{-i b \cdot p_1 / \hbar}\, \ket{p_1, p_2;0} \,.
\label{eq:psi}
\]
The impact parameter phase translates particle one relative to particle two, and these states can be used to define the Wigner densities for each matter particle
\[
\rho_1(x,\tilde p) = \rho_{\varphi_1}(x-b,\tilde p),~~~~~\rho_2(x,\tilde p) = \rho_{\varphi_2}(x,\tilde p).
\]
For a pure one-particle state, these are defined as
\[
\rho_{\varphi}(x,\tilde p)
&=
\frac{1}{\hbar^4}\int\hat{\sd}^4q\,
e^{iq\cdot x/\hbar}\,
\Theta(\tilde p,q)\,
\del\!\left(\tilde p^2+\frac{q^2}{4}-m^2\right)
\varphi\!\left(\tilde p+\frac q2\right)
\varphi^*\!\left(\tilde p-\frac q2\right).
\]

Inside the wavefunctions, we will choose to drop the $q$ corrections, however we recognise that this is a choice: it is not necessarily the case that no classical observable is sensitive to it, and indeed we expect that next-to-eikonal effects could indeed be sensitive to the $q$-dependence of the wavefunction. We leave this interesting problem to the future and for now we will take
\[
\varphi_i\!\left(\tilde p_i+\frac q2\right)
\varphi_i^*\!\left(\tilde p_i-\frac q2\right)
\longrightarrow
|\varphi_i(\tilde p_i)|^2.
\]
For the same reason, we can also use the leading on-shell support of the Wigner function,
\[
\del\!\left(\tilde p_i^2+\frac{q^2}{4}-m_i^2\right)
\longrightarrow
\del\!\left(\tilde p_i^2-m_i^2\right).
\]
This second step only neglects the quantum smearing of the mass shell and is not a license to set $q=0$ elsewhere in the integrand.

In the sharp-packet limit the Wigner functions localise on the classical initial data. Writing $b = b_1-b_2$, and denoting the classical momenta by $p_i^{\rm cl}$,
\[
\rho_i(x_i,\tilde p_i)
\longrightarrow
\delta^{(4)}(x_i-b_i)\,
\hat{\delta}_{\Phi_i}(\tilde p_i-p_i^{\rm cl}),
\qquad
\int\hat{\sd}^4\tilde p_i\,
\hat{\delta}_{\Phi_i}(\tilde p_i-p_i^{\rm cl})F(\tilde p_i)
=F(p_i^{\rm cl}).
\]
Note, however, that this classical momentum and impact parameter are evaluated at the midpoint, after projecting $b$ onto the plane transverse to the two hard momenta. In particular, one should impose $b_\perp\cdot\tilde p_1=b_\perp\cdot\tilde p_2=0$, rather than treating the unprojected separation as physical.
\subsection{$S$-matrix Observables and the Emergence of KMOC}
Now consider the operator 
\[
\bb A(x) = S^\dagger\bb O(x)S.
\]
This is \textit{not} a strictly local operator due to the presence of the S-matrix, however it still retains global translation covariance, since $[S,\bb P^\mu] = 0$, and hence
\[
\bb A(x+a) = e^{i\bb P\cdot a/\hbar}\bb A(x)e^{-i\bb P\cdot a/\hbar} = S^\dagger\bb O(x+a)S.
\]
Thus the same midpoint and Weyl-symbol reasoning applies. We are interested in $\Delta\bb O(x) = S^\dagger\bb O(x)S - \bb O(x)$, which is also translation covariant and Hermitian. So far, we have assumed that $\bb P^\mu = \sum_i \bb P_i^\mu$ and that $\bb O(x)$ acts on a single-particle Hilbert space, but this need not be the case, and the generalisation is straightforward.
In the two-particle case, the expectation value of $\Delta\bb O(x)$ in a state $\rho = \rho_1\otimes\rho_2$ is given by 
\[
\braket{\Delta\bb O(x)} &= \Tr{\rho_1\otimes\rho_2\Delta\bb O(x)} \\
&= \int\hat{\sd}^4\tilde p_1\,\hat{\sd}^4 q_1\hat{\sd}^4\tilde p_2\,\hat{\sd}^4 q_2\,\hat\delta_\Gamma(\tilde p_1,q_1)\hat\delta_\Gamma(\tilde p_2,q_2)e^{+i(q_1+q_2)\cdot x/\hbar}\\
&~~~~~~\times \rho_W(\tilde p_1,-q_1)\rho_W(\tilde p_2,-q_2)\,\Delta O_W(\tilde p_1,\tilde p_2,q_1,q_2),
\]
where we used 
\[
\langle p_1',p_2'|S^\dagger\bb O_1(x)S|p_1,p_2\rangle
=
e^{+i(q_1+q_2)\cdot x/\hbar}
\langle p_1',p_2'|S^\dagger\bb O_1(0)S|p_1,p_2\rangle,
\]
defining the independent transfer momentum by $q_i=p_i'-p_i$, and where
\[
\Delta O_W(\tilde p_1,\tilde p_2,q_1,q_2) = \braket{\tilde p_1+\frac12q_1,\tilde p_2+\frac12q_2|S^\dagger [\bb O,S]|\tilde p_1-\frac12q_1,\tilde p_2-\frac12q_2},
\]
satisfying $\Delta O_W(\tilde p_1,\tilde p_2,q_1,q_2)^* = \Delta O_W(\tilde p_1,\tilde p_2,-q_1,-q_2)$ by hermiticity. Global translation is generated by $\bb P_1+\bb P_2$, which fixes the operator phase to $e^{+i(q_1+q_2)\cdot x/\hbar}$.

We now consider the density matrix for a pure state $\ket{\psi}$, where $\rho=|\psi\rangle\langle\psi|$. The momentum-space kernel is given by 
\[\label{wigner-pure}
\rho_W(\tilde p,q)=\braket{\tilde p+\tfrac q2|\psi}\braket{\psi|\tilde p-\tfrac q2} = \varphi(\tilde p+\tfrac q2)\varphi^*(\tilde p-\tfrac q2).
\] 
The corresponding on-shell Wigner density is then
\[
\rho_W(x,\tilde{p}) 
&= \frac{1}{\hbar^4}\int\hat{\sd}^4 q\,
e^{i q\cdot x/\hbar}\,
\Theta(\tilde p,q)\,
\del\!\left(\tilde p^2+\tfrac{q^2}{4}-m^2\right)\varphi(\tilde p+\tfrac q2)\varphi^*(\tilde p-\tfrac q2).
\]
For the formal manipulations below we assume an admissible family of KMOC wavepackets whose leading Wigner densities admit a classical probabilistic interpretation and are sharply supported around the chosen classical momenta and impact parameter. 

Plugging the pure-state kernel into the phase-space trace formula, we find
\[
\braket{\Delta\bb O(x)} &= \Tr{\rho_1\otimes\rho_2\Delta\bb O(x)} \\
&= \int\hat{\sd}^4\tilde p_1\,\hat{\sd}^4 q_1\hat{\sd}^4\tilde p_2\,\hat{\sd}^4 q_2\,\hat\delta_\Gamma(\tilde p_1,q_1)\hat\delta_\Gamma(\tilde p_2,q_2)e^{+i(q_1+q_2)\cdot x/\hbar}\\
&~~~~~~\times \varphi(\tilde p_1-\tfrac{q_1}{2})\varphi^*(\tilde p_1+\tfrac{q_1}{2})\varphi(\tilde p_2-\tfrac{q_2}{2})\varphi^*(\tilde p_2+\tfrac{q_2}{2})\,\Delta O_W(\tilde p_1,\tilde p_2,q_1,q_2).
\]
Changing variables back to $p_i = \tilde{p}_i - \tfrac{q_i}{2}$ and $p_i' = \tilde{p}_i + \tfrac{q_i}{2}$, the positive energy $\Theta$ reassembles the $\delta^+$'s and the expectation value becomes
\[
\braket{\Delta\bb O(x)} &= \int\hat{\sd}^4 p_1\,\hat{\sd}^4 p_1'\hat{\sd}^4 p_2\,\hat{\sd}^4 p_2'\,\hat\delta^{(+)}(p_1^2-m_1^2)\hat\delta^{(+)}(p_1'^2-m_1^2)\hat\delta^{(+)}(p_2^2-m_2^2)\hat\delta^{(+)}(p_2'^2-m_2^2)\\
&~~~~~~\times e^{+i[(p_1'-p_1)+(p_2'-p_2)]\cdot x/\hbar}\varphi(p_1)\varphi^*(p_1')\varphi(p_2)\varphi^*(p_2')\,\braket{p_1',p_2'|S^\dagger[\bb O,S]|p_1,p_2}.
\]
Defining the two-particle KMOC state with impact parameter $b$ as
\[
\ket{\psi} = \int\hat{\sd}^4 p_1\,\hat{\sd}^4 p_2\,\hat\delta^+(p_1^2-m_1^2)\hat\delta^+(p_2^2-m_2^2)\varphi(p_1)\varphi(p_2)e^{-ip_1\cdot b/\hbar}\ket{p_1,p_2},
\]
the translated density matrix $\rho_b=|\psi\rangle\langle\psi|$ has the off-diagonal phase
\[
\begin{aligned}
\langle p_1,p_2|\rho_b|p_1',p_2'\rangle
&=
e^{-ip_1\cdot b/\hbar}e^{+ip_1'\cdot b/\hbar}
\varphi(p_1)\varphi^*(p_1')\varphi(p_2)\varphi^*(p_2')
\\
&=
e^{+iq_1\cdot b/\hbar}
\varphi(p_1)\varphi^*(p_1')\varphi(p_2)\varphi^*(p_2').
\end{aligned}
\]
The operator insertion point remains independent of $b$. Setting it to $x=0$ gives
\[
\begin{aligned}
\braket{\psi|S^\dagger[\bb O,S]|\psi}
&= \int\hat{\sd}^4 p_1\,\hat{\sd}^4 p_1'\hat{\sd}^4 p_2\,\hat{\sd}^4 p_2'\,\hat\delta^{(+)}(p_1^2-m_1^2)\hat\delta^{(+)}(p_1'^2-m_1^2)\hat\delta^{(+)}(p_2^2-m_2^2)\hat\delta^{(+)}(p_2'^2-m_2^2)\\
&\quad\times e^{+iq_1\cdot b/\hbar}\varphi(p_1)\varphi^*(p_1')\varphi(p_2)\varphi^*(p_2')\,\braket{p_1',p_2'|S^\dagger[\bb O,S]|p_1,p_2}.
\end{aligned}
\]
This is exactly the usual KMOC formula for the change in an observable. 
\section{From the Dyson $S$-Matrix to the Phase-Space Eikonal}
\label{sec:dyson-eikonal-main}
\label{sec:partial-weyl-dyson}

In ref. \cite{Cristofoli:2021jas}, we assumed a particular form for the eikonal, and motivated its minimal extension to include radiation. In this section, we will show that our proposed state can be derived directly from the Dyson $S$-matrix when partially Weyl transformed over the matter sector. In particular, we will show that the ``half-way'' variables $\tilde p_i$ that appeared there are the natural variables of the Weyl transform, and that radiation is a natural consequence of the operator $S$-matrix.

In quantum field theory, one definition of the $S$-matrix is as a time-ordered exponential operator
\[
\mathbb{S} = \cl T\exp\!\Bigl[-\frac{i}{\hbar}\!
   \int \sd^4x \hat{\cl{H}}_{int}(x)\Bigr].
\]
We will for now focus on scalar QED, however we will keep the discussion general enough that it can be applied to other theories, such as gravity, with minimal modification. We will write $\hat{\cl{H}}_{int} = -\bb{J}^\mu\bb{A}_\mu - V(\hat\phi,\bb{A})$, so that the Dyson formula becomes
\[
\mathbb{S} = \cl T\exp\!\Bigl[\frac{i}{\hbar}\!
   \int \sd^4x \left(\bb{J}^\mu(x)\bb{A}_\mu(x)+V(\hat{\phi}(x),\bb{A}(x))\right)\Bigr].
\]
We will take the current operator to be
\[
\bb J_i^\mu(x)= i e_i\hbar\bigl[\phi_i^\dagger \partial^\mu\phi_i-(\partial^\mu\phi_i^\dagger)\phi_i\bigr],
\]
where $i=1,2$ labels two scalar particles. The potential $V$ contains the seagull interaction
\[
V(x) = e^2|\hat \phi(x)|^2 \bb A_\mu(x) \bb A^\mu(x)
\]
as well as any other interactions that may be present in the theory, such as a non-minimal coupling to gravity $\xi R |\phi|^2$. We will keep $V$ general for now, but we will return to the specific case of scalar QED later on.
Ignoring the potential term for the time being, let's consider the linear part of the S-matrix
\[
\bb S_0 &= \cl T\exp\!\Bigl[\frac{i}{\hbar}\int d^4x  \bb{A}_\mu(x)\bb J^\mu(x)\Bigr]\\ &= 
\sum_{n=0}^\infty \frac{i^n}{n!\hbar^n}\prod_i\int d^4x_i \cl{T}\Big[\bb J^{\mu_1}\cdots \bb J^{\mu_n}\bb{A}_{\mu_1}\cdots \bb{A}_{\mu_n}\Big].
\]
\subsection{The Scalar Current Symbol}
To see that these definitions make some sense, let us take $\bb{O}(x)=\bb{J}^\mu(x)$, the scalar QED current at leading order, with
\[
\bb{J}^\mu(x)=ie\hbar\bigl[\phi^\dagger\partial^\mu\phi-(\partial^\mu\phi^\dagger)\phi\bigr].
\]
With the relativistically normalised one-particle states specified above, and fixing the current such that $\bb Q\ket p=e\ket p$, the matrix element is $\braket{p'|\bb{J}^\mu(x)|p}=\frac{e}{\hbar^3}(p'+p)^\mu e^{i(p'-p)\cdot x/\hbar}$. The explicit $\hbar$ in $\bb{J}^\mu$ cancels the $1/\hbar$ produced by differentiating the plane waves, while the remaining $\hbar^{-3}$ is

\[
\braket{p'|\bb Q|p}
\equiv
\int\sd^3x\,\braket{p'|\bb J^0(x)|p}
=
e\,\braket{p'|p},
\qquad
\int\sd^3x\,e^{i(\mathbf p'-\mathbf p)\cdot\mathbf x/\hbar}
=
\hbar^3(2\pi)^3\delta^{(3)}(\mathbf p'-\mathbf p).
\]

It is tempting to interpret this as a current of a particle with momentum $p$, but in reality it should be thought of as the Weyl symbol of an operator that connects to nearby momentum eigenstates, hence why the midpoint momentum $p'+p$ appears in the symbol.

We use $J_W^\mu(\tilde p,q)=2e\tilde p^\mu\del(2\tilde p\cdot q)$ for the momentum-space current coefficient. The same symbol with a spacetime argument denotes the finite symbol. Since the transverse delta function leaves three independent Fourier directions, we have
\[\label{eq:current-symbol}
J_W^\mu(x,\tilde p)
\equiv
\cl W_m[\bb J^\mu](x,\tilde p)
&=
\frac{1}{\hbar^3}
\int\hat{\sd}^4 q\,
e^{i q\cdot x/\hbar}\,
J_W^\mu(\tilde p,q)
\\
&=
\frac{e}{\hbar^3}\tilde u^\mu
\int\hat{\sd}^4 q\,
e^{i q\cdot x/\hbar}\,
\del(\tilde u\cdot q),
\]
using $\tilde p^\mu=m\tilde u^\mu$ together with $\del(2m\tilde u\cdot q)=\del(\tilde u\cdot q)/2m$. We can rewrite the delta function as $\del(\tilde u\cdot q)=\int d\tau\,e^{i\tau\tilde u\cdot q}$, so the coefficient integral becomes
\[
J_W^\mu(x,\tilde p)
&=
\frac{e}{\hbar^3}\tilde u^\mu
\int \sd\tau\hat{\sd}^4 q\,
e^{i q\cdot x/\hbar-i\tilde u\cdot q\tau}
\\
&=
e\hbar\tilde u^\mu
\int d\tau\,
\delta^{(4)}(x-\hbar\tau\tilde u)
\\
&=
e\tilde u^\mu
\int\sd\tau\,
\delta^{(4)}(x-\tilde u\tau).
\]
This is the finite classical point-particle current of a worldline with affine velocity $\tilde u^\mu$ passing through the origin. It is important to note that this worldline representation encodes the recoil that has been injected by operating with $\bb J^\mu$, since the velocity $\tilde u$ appearing in the symbol is the average of the incoming and outgoing four-velocities.

For the current inserted at $x'$, the observable symbol appearing in \eqref{eq:trace-phase-space} is the shifted expression
\[
J_W^\mu(x'+x,\tilde p)
=
e\tilde u^\mu
\int\sd\tau\,
\delta^{(4)}(x+x'-\tilde u\tau),
\]
which is a worldline passing through $x'=-x$ rather than the origin, confirming that insertion at $x$ is equivalent to translating the worldline to $-x$ in the phase-space argument.
The Weyl transform maps ordered products of operators to ordered star-products of symbols. For symbols associated with matter line $i$, the positive-energy product has the following local differential representation

\[
F\star_i G
=
F\exp\!\left[
\frac{i\hbar}{2}
\left(
\overleftarrow\partial_{x_i^\mu}\overrightarrow\partial_{\tilde p_{i\mu}}
-
\overleftarrow\partial_{\tilde p_{i\mu}}\overrightarrow\partial_{x_i^\mu}
\right)
\right]G.
\]
where 
\[
\star_m = \star_1\star_2.
\]
Different matter lines have different phase-space variables, so $J_{1,W}^\mu\star_m J_{2,W}^\nu = J_{1,W}^\mu J_{2,W}^\nu$. The crucial point here, as discussed at length in appendix \ref{app:time-order-weyl-brackets}, is that ordinary time ordering becomes star-product time ordering of the matter symbols together with ordinary photon time ordering.

Performing a partial Weyl transform on the matter sector, we can express the linear part of the Dyson series as
\[
\cl W_m[\mathbb S_0]
=
\sum_{n=0}^\infty
\frac{i^n}{n!\hbar^n}
\prod_{a=1}^n\int d^4y_a\,
\cl T
\left[
\left(
J_W^{\mu_1}(y_1)\star_m\cdots\star_m J_W^{\mu_n}(y_n)
\right)
\mathbb A_{\mu_1}(y_1)\cdots\mathbb A_{\mu_n}(y_n)
\right]
\]
where
\[
J_W^\mu(y)
\equiv
J_{1,W}^\mu(y;x_1,\tilde p_1)
+
J_{2,W}^\mu(y;x_2,\tilde p_2)
\]
and
\[
\cl W_m[\mathbb S] = \cl W_m[\mathbb S](x_1,\tilde p_1;x_2,\tilde p_2) = \cl W_m[\mathbb S](\tilde z)
\]
is a function of the phase spaces of particles one and two, and is operator valued in the photon Fock space. 

Factorising the time-ordering, the Dyson series can therefore be written as
\[
\cl W_m[\mathbb S_0]
&=
\sum_{n=0}^\infty
\frac{i^n}{n!\hbar^n}
\prod_{a=1}^n\int d^4y_a\,
\sum_{i_1,\ldots,i_n=1}^2
\cl T_{\star}[J_{i_1,W}^{\mu_1}(y_1)
\star_m\cdots\star_m
J_{i_n,W}^{\mu_n}(y_n)]
\,
\cl T[\mathbb A_{\mu_1}(y_1)\cdots\mathbb A_{\mu_n}(y_n)].
\\
&= \cl T \cl{T}_{\star}\exp_{\star_m}\!\Bigl[\frac{i}{\hbar}\int d^4y\,\sum_{i=1}^2 J_{i,W}^\mu(y)\mathbb A_\mu(y)\Bigr],
\]
where the $\exp_\star$ is defined as a power series in the matter star product \cite{Bayen:1977ha}. Since this is now linear in the photon field and time-ordered, we can apply the star-product version of Wick's theorem to exchange time ordering for normal ordering and contractions. Writing
\[
\Delta^F_{\mu\nu}(x-y)
&=
\hbar D_{\mu\nu}^F(x-y)
=
-\frac{i}{\hbar}
\int\hat{\sd}^4q\,
\frac{\eta_{\mu\nu}}{q^2+i\epsilon}
e^{-iq\cdot(x-y)/\hbar},
\label{eq:photon-propagator-kmoc-coefficient}
\]
we obtain
\[
\cl W_m[\mathbb S_0]
=
\cl{T}_{\star}\left\{\exp_{\star_m}\!\left[
-\frac{1}{2\hbar^2}
\int_{x,y}\,
\Delta^F_{\mu\nu}(x-y)\,
J_W^\mu(x)\star_m J_W^\nu(y)
\right] \star_m \normord{\exp_{\star_m}\!\Bigl[\frac{i}{\hbar}\int_w\,J_W^\mu(w)\mathbb A_\mu(w)\Bigr]}\right\},
\]
where we have used $\int_x = \int d^4x$ for brevity.

We note that this resummation is possible because the matter symbol does not act on the state, so the star product of the symbols is just a function that can be resummed into this exponential form, using Wick's theorem to express the time-ordering of the photon operator in terms of its contractions with the matter sector and a normal ordered star-product, as shown explicitly in appendix \ref{app:wicks-theorem}.

We write the current in momentum space as
\[
J_W^\mu(x)
&=
\frac{1}{\hbar^3}
\sum_i
\int\hat{\sd}^4q\,
e^{iq\cdot x/\hbar}\,
J_{(i),W}^\mu(q).
\label{eq:dyson-current-coefficient-representation}
\]
where $i$ labels the line of the particle on which the current is inserted, and $J_{(i),W}^\mu(q)$ is its momentum-space symbol. We now consider the scenario where each line has an impact parameter $b_i$, and we are interested in the limit where the Wigner functions localise on the classical initial data, at large impact parameter and small momentum transfer, so that the symbol of the current is well approximated by the classical point-particle current. In this limit, the star product of the symbol reduces to the standard product of opposite worldline currents, since the same-line star products are classically subleading. 

\[
\int_{x,y}\,
\Delta^F_{\mu\nu}(x-y)\,
J_W^\mu(x)\star_m J_W^\nu(y) = -2i\hbar\int\hat{\sd}^4q \,\frac{J_{(1),W}^\mu(q)J_{\mu,(2),W}(-q)}{q^2+i\epsilon}\, .
\]
As for the normal-ordered term, we can write it by expanding the gauge field in momentum space
\[
A_\mu(x)
=
\sum_\lambda
\int d\Phi(k)
\left[
u_{\lambda\mu}(x;k)a_\lambda(k)
+
u_{\lambda\mu}^*(x;k)a^\dagger_\lambda(k)
\right],~~~~~~u_{\lambda\mu}(x;k)
=
\frac{1}{\sqrt\hbar}\epsilon_\mu^{(\lambda)}(k)e^{-ik\cdot x/\hbar},
\] 
and use the current symbol in eq. \eqref{eq:dyson-current-coefficient-representation} to obtain 
\[
\frac{i}{\hbar}\int J_W\cdot\mathbb A
=
\Sigma_W[a^\dagger]
+
\overline{\Sigma}_W[a],
\]
where
\[
\Sigma_W[a^\dagger]
=
\sum_\lambda\int d\Phi(k)\,
\alpha_\lambda(k;\tilde z)\,
a_\lambda^\dagger(k),
\]
with the leading waveshape defined as
\[
\alpha_\lambda(k;\tilde z)
=
\frac{i}{\hbar}
\int d^4x\,
J_W^\mu(x;\tilde z)
u_{\lambda\mu}^*(x;k).
\]
After making these identifications, the linear part of the Dyson series can be written as
\[
\cl W_m[\mathbb S_0]
=
\cl T_{\star_m}\left\{
\exp_{\star_m}\!\left[\frac{i}{\hbar}\int\hat{\sd}^4q \,\frac{J_{(1),W}^\mu(q)J_{\mu,(2),W}(-q)}{q^2+i\epsilon}\right]
\star_m
\normord{\exp_{\star_m}\!\Bigl[\Sigma_W[a^\dagger]+\overline{\Sigma}_W[a]\Bigr]}
\right\}.
\label{eq:dyson-wick-resummation-leading}
\]
The final linear state, then, after normal ordering, is 
\[
\cl W_m[\mathbb S_0]\ket{0} = \cl T_{\star_m}\left\{
\exp_{\star_m}\!\left[\frac{i}{\hbar}\chi_0(b)\right]
\star_m
\exp_{\star_m}\!\Bigl[\sum_\lambda\int d\Phi(k)\,
\alpha_\lambda(k;\tilde z)\,
a_\lambda^\dagger(k)\Bigr]
\right\}\ket{0},
\label{eq:dyson-wick-resummation-leading-final-state}
\]
where we define the leading order eikonal
\[
\chi_0(b)
=
\int\hat{\sd}^4q\,
\frac{
J_{(1),W}^\mu(q)
J_{\mu,(2),W}(-q)
}{q^2+i\epsilon}.
\label{eq:dyson-eikonal-phase}
\]
The nonlinear interactions can be restored to all orders by replacing the messenger field in $V$ with a functional derivative with respect to the unresolved current symbol, as reviewed in more detail in appendix \ref{app:dyson-star-ordering}.
However, the functional derivatives require us to keep the linear source symbol unresolved (i.e. unintegrated) until they have acted. We therefore write
\[
\cl W_m[\mathbb S]
=
\cl T_{\star_m}\left\{
\exp_{\star_m}\!\left[\frac{i}{\hbar}
V\!\left(
-i\hbar\frac{\overrightarrow{\delta}}{\delta J_W};
\tilde z
\right)
\right]
\star_m
\cl W_m[\mathbb S_0[J_W]]
\right\}.
\]
Importantly for us, the potential $V$ acts on the eikonal and the waveshape, but also generates higher-order $(a^\dagger)^n$ terms, meaning that the final state is not in general a coherent state.

To see that eq. \eqref{eq:dyson-eikonal-phase} matches the usual eikonal after integration, we can express the current symbols in momentum space using eq. \eqref{eq:current-symbol}
\[
J_{i,W}^\mu(q;b_i)
=
2e_i\tilde p_i^\mu\,
e^{iq\cdot b_i/\hbar}\,
\del(2\tilde p_i\cdot q),
\]
such that 
\[
J_{1,W}^\mu(q;b_1)J_{2,W,\mu}(-q;b_2)
&=
4e_1e_2\tilde p_1\cdot\tilde p_2\,
e^{iq\cdot (b_1-b_2)/\hbar}\,
\del(2\tilde p_1\cdot q)\del(2\tilde p_2\cdot q).
\]
Therefore, the tree-level eikonal phase is given by
\[
\chi_0
&=
\int \hat{\sd}^4q\,
\del(2\tilde{p}_1\cdot q)\del(2\tilde{p}_2\cdot q)\,
e^{+iq\cdot b/\hbar}\,
\cl{A}^{(0)}_{2\rightarrow 2}(q),
\qquad b=b_1-b_2,
\]
where
\[
\cl{A}^{(0)}_{2\rightarrow 2}(q)
&=
\frac{4e_1e_2\tilde p_1\cdot\tilde p_2}{q^2+i\epsilon} = \cl{A}_4[\tilde p_1,\tilde p_2;q]
\]
is the scalar QED 4pt amplitude at tree level. This is the standard formula for the eikonal phase in the recoil shifted plane \cite{DiVecchia:2023frv,Kabat:1992tb,Cristofoli:2021vyo}, and here we see it arise naturally.

After the nonlinear derivatives have acted and the lower elastic and radiative iterations have been removed, it is convenient to collect the zero-radiation terms into the elastic symbol $E_W$ and the connected creation-only terms into the normal-ordered radiation operator $F_\star$, which is dimensionless, writing the final state as
\[
\cl W_m[\mathbb S]\ket{0}
=
E_W(\tilde z)
\star_m
\quad
\exp_{\star_m}\!F_\star[J_W,a^\dagger;\tilde z]\ket{0}.
\]
Here we have discarded the normal-ordering symbol with the understanding that we are acting only on the vacuum. For a generic potential $V$, the radiation functional contains arbitrarily many creation operators, and the final state is not in general a coherent state.
\subsubsection{The Endpoint Approximation}
The astute reader will have noticed that the exponential form of the eikonal in eq.~\eqref{eq:dyson-wick-resummation-leading} is not an ordinary exponential, but rather a star-product exponential, unlike the eikonal proposed in ref.~\cite{Cristofoli:2021jas}. The star product is inherited from operator composition, and its practical effect here is to route all of the local momentum through the ordered current insertions. It therefore records a sequence of kicks along the worldline,
\[
z_{\rm in}
\longrightarrow z_1
\longrightarrow z_2
\longrightarrow\cdots
\longrightarrow z_{\rm out}.
\]
At the two endpoints of the trajectory, the Weyl transform means we naturally describe things in terms of the midpoint and displacement variables, rather than the initial and final positions, where
\[
z_- \equiv z_{\rm in}
=
\tilde z-\frac12\Delta z,
\qquad
z_+ \equiv z_{\rm out}
=
\tilde z+\frac12\Delta z,
\]
where
\[
\tilde z
=
\frac12\left(z_{\rm in}+z_{\rm out}\right),
\qquad
\Delta z
=
z_{\rm out}-z_{\rm in}.
\]
The individual insertions are then evaluated at the local midpoints $z_r=\tilde z+\delta z_r$, whose shifts $\delta z_r$ record how the total recoil is distributed throughout the interaction.

Most conservative observables are insensitive to the precise local history of the scattering process, and so it is often convenient to simply ignore the local momentum routing and replace each midpoint momentum with the global midpoint momentum, i.e. we replace $z_r\rightarrow \tilde{z}$, effectively dropping the $\delta z_r$ terms. We call this the endpoint approximation (or endpoint reduction). After all internal star products required to construct a complete amplitude or cut kernel have been evaluated, endpoint reduction replaces the remaining matter star products by ordinary products. Functions that are defined perturbatively, such as star-exponentials and star-logarithms, will be replaced by their ordinary functions. Where the eikonal is concerned, we first evaluate the star-products of the symbol, then endpoint-reduce to obtain an ordinary exponential of the reduced \textit{centre-generating function} or \textit{phase action} \cite{Marinov1979PhaseAction,deAlmeida:2013wal,deAlmeida:2020hgy}. It should not be imposed before these internal Bopp shifts have performed their momentum routing. The precise construction is given in appendix~\ref{app:no-local-recoil-endpoint}.

In the purely classical, conservative case, this endpoint description is not new. Marinov introduced a phase-space analogue of Hamilton's principal function, related to it by a Legendre transformation \cite{Marinov1979PhaseAction}. This analogue, which ought to be called the \textit{Marinovian} following recent naming trends \cite{Kim:2025magnusian}, is naturally a function of the midpoint $\tilde z$, rather than the initial and final positions, on which Hamilton's principal function, or equivalently the on-shell action, depends \cite{Kim:2025magnusian}. Derivatives of the Marinovian implicitly determine the finite canonical transformation between the two endpoints, while in the semiclassical limit the Marinovian gives the leading phase of the Weyl symbol of the evolution operator. It should be distinguished from the Magnusian: the Magnusian generates a canonical transformation through an exponentiated Poisson bracket, whereas the Marinovian generates it implicitly via the midpoint variables. In essence, the Marinovian encodes the local Hamiltonian trajectory in a single midpoint generating function. This is precisely the endpoint description we wish to recover after the local Bopp routing has been resolved. It is also natural to ask what we are throwing away when we make the endpoint approximation, and the answer is local terms that correspond to a $q/p$ expansion, that is, a ratio of the transfer momentum and the hard momentum. Therefore, the endpoint approximation discards local terms beginning at `next-to-eikonal' order \cite{Laenen:2008gt,Laenen:2010uz,Fernandes:2024xqr}. Such terms are certainly worthy of study in their own right, but are not the focus of this paper and so we leave this to future work.

The endpoint reduction will be made only after the full elastic and radiative symbols have been constructed, i.e. after the Bopp shifts have performed all of the relevant momentum routing. The elastic factor can then be written as an ordinary midpoint exponential, as in \cite{Cristofoli:2021vyo,Luna:2023uwd}, while its outer matter star product with radiation is retained, as in eq.~\eqref{eq:radiation-exclusive-state-new}. For a generic potential $V$, the radiation contains arbitrarily many creation operators, and so the final state is not in general a coherent state. 

The Dyson construction tells us what has to be computed, while the scattering amplitudes provide the simplest way of computing it. Before turning to that prescription, we first isolate the reason that endpoint reduction cannot be imposed immediately, and why evaluating the Bopp shifts is necessary to resolve the local recoil history of the scattering process.

\subsection{Straight Lines Do Not Radiate}
\label{sec:endpoint-radiation}
The endpoint approximation discards all local recoil information in favour of the global momentum transfer, and is well suited to observables which don't depend on the precise local scattering history. The radiation, however, is sensitive to the local history: radiation emitted at $t = -\infty$ will have a vastly different profile to radiation emitted at $t = 0$, and so the endpoint approximation is not expected to be valid for radiation, at least at finite frequencies. The first place to see this is in the linearised S-matrix, where we have a single photon emitted from one of the two worldlines.

At first order, the partially transformed Dyson series is
\[
\cl W_m^{(1)}[\bb S_0]
=
\frac{i}{\hbar}\int_x
J_W^\mu(x)\bb A_\mu(x).
\]
Projecting onto a single outgoing photon, and using $J_W^\mu=J_{1W}^\mu+J_{2W}^\mu$, gives
\[
\begin{aligned}
\bra{k,\lambda}
\cl W_m^{(1)}[\bb S_0]
\ket0
&=
\frac{i}{\hbar}
\sum_i\int_x
u^*_{\lambda\mu}(x;k)J_{iW}^\mu(x)
\\
&=
\frac{i}{\sqrt\hbar}
\sum_i
2e_i(\widetilde p_i\cdot\epsilon_\lambda^*(k))
e^{-\frac{i}{\hbar}k\cdot b_i}
\del(2\widetilde p_i\cdot k)
\\
&\equiv
\frac{i}{\sqrt\hbar}
\sum_i\cl A_{3,i}^{\rm ep}(k).
\end{aligned}
\label{eq:first-order-endpoint-three-point}
\]
There is only one current insertion on each term, so there is no matter ordering or local recoil history to resolve, and the endpoint reduction is exact. Writing $k=\omega n$, with $n^2=0$, the three-point block becomes
\[
\cl A_{3,i}^{\rm ep}(k)
=
e_i\frac{\widetilde p_i\cdot\epsilon_\lambda^*(k)}
{\widetilde p_i\cdot n}
\del(\omega),
\]
where the impact-parameter phase is unity on the support of $\del(\omega)$. In other words: straight lines do not radiate at finite frequency. We shall return to the zero-frequency case in a later section.

Let's now consider the next order in radiation, where we have one exchanged photon and one emitted, with a massive pole. We treat the contact vertex separately below. The third order term is
\[
\cl{W}_m[\bb S_0] = \frac{1}{3!}\left(\frac{i}{\hbar}\right)^3\int_{x_1,x_2,x_3}\cl{T}_\star\left[J_W^{\mu_1}(x_1)\star_m J_W^{\mu_2}(x_2)\star_m J_W^{\mu_3}(x_3)\right]\cl{T}_\gamma\left[\bb A_{\mu_1}(x_1)\bb A_{\mu_2}(x_2)\bb A_{\mu_3}(x_3)\right].
\]
Consider an outgoing photon with momentum $k$ and polarisation $\epsilon$, obtained by sandwiching the photon part between \(\bra{k,\lambda}\) and \(\ket0\). Using Wick's theorem, we find
\[
 \bra{k,\lambda}
 \cl{T}_\gamma\!\left[A_{\mu_1}(x_1)A_{\mu_2}(x_2)A_{\mu_3}(x_3)\right]
 \ket0 &= u^*_{\lambda\mu_1}(x_1;k)\Delta^F_{\mu_2\mu_3}(x_2-x_3)
+u^*_{\lambda\mu_2}(x_2;k)\Delta^F_{\mu_1\mu_3}(x_3-x_1)\\
&+u^*_{\lambda\mu_3}(x_3;k)\Delta^F_{\mu_1\mu_2}(x_1-x_2),
 \label{eq:three-photon-Wick}
\]
such that 
\[
\bra{k,\lambda}
\cl{W}^{(3)}_m[\bb S_0]
\ket0
=
\frac12\left(\frac{i}{\hbar}\right)^3
\int_{x_1,x_2,x_3}&
u^*_{\lambda\mu_3}(x_3;k)\Delta^F_{\mu_1\mu_2}(x_1-x_2)\\
&\times \cl{T}_\star\left[J_W^{\mu_1}(x_1)\star_m J_W^{\mu_2}(x_2)\star_m J_W^{\mu_3}(x_3)\right].
\]
We now recall that $J_W = J_{1W} + J_{2W}$, and note that terms in which all three currents lie on one worldline do not survive the connected two-body classical limit we consider. Assuming one photon is emitted to infinity, we are left with two distinct contributions: the remaining photon is emitted from particle 1 and absorbed by particle 2, or it is emitted from particle 2 and absorbed by particle 1. This ensures that no matter what, the star product of the currents is non-trivial, and hence the radiation explicitly depends on the local history of the scattering, unlike the tree four-point exchange, where each worldline carries only one current insertion and the star product is trivial. Assuming the radiation is emitted from $x_3$, the result is 
\[
\bra{k,\lambda}
\cl{W}^{(3)}_m[\bb S_0]
\ket0 = \left(\frac{i}{\hbar}\right)^3
\int_{x_1,x_2,x_3}&
u^*_{\lambda\mu_3}(x_3;k)\Delta^F_{\mu_1\mu_2}(x_1-x_2)\\
&\times \cl{T}_\star\left[J_{1W}^{\mu_1}(x_1)\star_m J_{2W}^{\mu_2}(x_2)\star_m J_{1W}^{\mu_3}(x_3)\right] + (1\leftrightarrow 2).
\label{eq:three-current-product}
\]
The first term corresponds to the photon being emitted from particle 1, while the second term corresponds to the photon being emitted from particle 2. There is also a contact term, where the photon is emitted from the seagull vertex, which we will come back to.

Let's now impose the endpoint reduction to get a feel for the consequences. At the level of the resolved current product, the endpoint reduction trivialises matter time ordering and local momentum routing, replacing the resolved current by the endpoint current
\[
J_{iW}^{\mu}
\bigl(x;\widetilde p_{i,\mathrm{local}}\bigr)
\longrightarrow
J_{i,\mathrm{ep}}^{\mu}
\bigl(x;\widetilde p_i\bigr).
\label{eq:endpoint-current-replacement}
\]

which means
\[
&\left.
\cl T_\star\left[
J_{1W}^{\mu_1}(x_1)
\star_m
J_{2W}^{\mu_2}(x_2)
\star_m
J_{1W}^{\mu_3}(x_3)
\right]
\right|_{\mathrm{ep}} =
J_{1,\mathrm{ep}}^{\mu_1}(x_1)
J_{2,\mathrm{ep}}^{\mu_2}(x_2)
J_{1,\mathrm{ep}}^{\mu_3}(x_3).
\label{eq:endpoint-three-current-product}
\]

Substituting eq.~\eqref{eq:endpoint-three-current-product} into the radiation matrix element gives
\begin{align}
\left.
\bra{k,\lambda}
\cl W_m^{(3)}[\bb S_0]
\ket0
\right|_{\mathrm{ep}}
={}&
\left(\frac{i}{\hbar}\right)^3
\int_{x_1,x_2,x_3}
u^*_{\lambda\mu_3}(x_3;k)
\Delta^F_{\mu_1\mu_2}(x_1-x_2)
\notag\\
&\qquad\times
J_{1,\mathrm{ep}}^{\mu_1}(x_1)
J_{2,\mathrm{ep}}^{\mu_2}(x_2)
J_{1,\mathrm{ep}}^{\mu_3}(x_3).
\label{eq:endpoint-third-order-before-factor}
\end{align}
The \(x_3\) dependence is now completely independent of the \(x_1\) and \(x_2\) dependence. The integral therefore factorises:
\[
\bra{k,\lambda}
\cl W_m^{(3)}[\bb S_0]
\ket0
\bigg|_{\mathrm{ep}}
={}&
\bigg[
\left(\frac{i}{\hbar}\right)^2
\int_{x_1,x_2}
J_{1,\mathrm{ep}}^{\mu_1}(x_1)
\Delta^F_{\mu_1\mu_2}(x_1-x_2)
J_{2,\mathrm{ep}}^{\mu_2}(x_2)
\bigg]
\notag\\
&\times
\bigg[
\frac{i}{\hbar}
\int_{x_3}
u^*_{\lambda\mu_3}(x_3;k)
\left(J_{1,\mathrm{ep}}^{\mu_3}(x_3) + J_{2,\mathrm{ep}}^{\mu_3}(x_3)\right)
\bigg].
\label{eq:endpoint-factorised-integrals}
\]
We recognise the first integral as the endpoint eikonal, while the second integral is the sum of two endpoint-integrated 3pt blocks.
In momentum space this is the product of the endpoint eikonal and two endpoint-integrated 3pt blocks,

\[
\bra{k,\lambda}
\cl W_m^{(3)}[\bb S_0]
\ket0
\bigg|_{\mathrm{ep}}
={}&-\frac{1}{\hbar^{3/2}}
\int\dd^4\ell\del(2\tilde p_1\cdot\ell)
\del(2\tilde p_2\cdot\ell)\,
e^{+\frac{i}{\hbar}\ell\cdot(b_1-b_2)}
\frac{4e_1e_2(\tilde p_1\cdot\tilde p_2)}{\ell^2+i\epsilon}
\\
&~~~~~\times
\left(
2e_1(\tilde p_1\cdot\epsilon_\lambda^*(k))
e^{-\frac{i}{\hbar}k\cdot b_1}
\del(2\tilde p_1\cdot k)
+(1\leftrightarrow2)\right)\\
&=\frac{i}{\hbar}\chi_0(b_1-b_2)
\left(
\frac{i}{\sqrt\hbar}\cl{A}_{3,1}^{\rm ep}(k)
+
\frac{i}{\sqrt\hbar}\cl{A}_{3,2}^{\rm ep}(k)
\right).\label{eq:three-current-product2}
\]
where the 3pt block is the same as the one given in eq. \eqref{eq:first-order-endpoint-three-point}.

We recognise this as an iteration term, essentially the leading order eikonal and leading order $A\cdot J$ 3pt term, and hence it is removed by the connected classical projection. In other words, the \textit{premature} endpoint approximation produces no finite-frequency radiation either: we need to evaluate all Bopp shifts and momentum routing before the approximation is applied. 

Given how much of the conservative sector is captured by the endpoint approximation, it is natural to ask whether we can perform a \textit{partial} endpoint approximation, where we only apply the endpoint approximation to the conservative eikonal, but not radiation. The idea is to keep the resolved matter ordering until the full radiative block has been constructed, only then taking the endpoint limit. This ordering transfers recoil between the exchange and emission vertices and is evaluated through the Bopp shift. After the complete radiative block has been integrated, stripped, and matched to an endpoint kernel at the order under consideration, its ordering data are already contained in that kernel. At this point, the ordered star-products for the kernel have been resolved, while its outer composition with the completed elastic symbol remains a matter star product unless the final global endpoint reduction is also taken.

We have introduced a conceptual derivation of the eikonal structure and its radiative extension, but as we will see, this is not practically efficient. Evaluating the star products retains the local recoil information needed by radiation, while the endpoint approximation gives the ordinary eikonal phase. Evaluating this for every Dyson ordering for every fixed-multiplicity coefficient would be a tedious process, but thankfully we don't need to: as we will see, the full on-shell scattering amplitudes have already performed the ordered momentum routing.
\subsection{Zero-Frequency Dressings}
\label{subsec:zero-frequency-dressings}

We saw earlier that straight worldlines don't radiate, but that zero-frequency dressings are automatically present (see eq.~\eqref{eq:first-order-endpoint-three-point}), and encode a time-independent long-range field. This is perfectly sensible: the Fourier transform of a time-independent field has support at $\omega=0$, rather than at a finite radiative frequency. A charged asymptotic particle should therefore be thought of as the hard particle together with its accompanying static field. The static field dressing is already part of the linear Dyson resummation we performed in eq.~\eqref{eq:dyson-wick-resummation-leading}, and is not something that needs to be imposed by hand.

We begin by defining the zero-frequency part of the one-radiation coefficient, using eq.~\eqref{eq:first-order-endpoint-three-point}, as
\[
\alpha_{{\rm ZF},\lambda}(k;\tilde z)
\equiv
S_{1,W,\lambda}^{\rm ZF}(k;\tilde z)
=
\frac{i}{\sqrt{\hbar}}
\sum_i
\cl A_{3,i}^{\rm ep}(k;\tilde z).
\]
The corresponding part of the creation generator is then part of eq.~\eqref{eq:dyson-wick-resummation-leading-final-state}, defined via
\[
\Sigma_{{\rm ZF},W}[a^\dagger;\tilde z]
\equiv
\sum_\lambda\int d\Phi(k)\,
\alpha_{{\rm ZF},\lambda}(k;\tilde z)
a_\lambda^\dagger(k).
\label{eq:zero-frequency-creation-generator}
\]
The formal zero-frequency part of the final state is then
\[
\left.
\cl W_m[\bb S_0]\ket0
\right|_{\rm ZF}
=
\cl T_{\star_m}
\left\{
\exp_{\star_m}\!\left[
\frac{i}{\hbar}\chi_0
\right]
\star_m
\exp_{\star_m}\!\left[
\Sigma_{{\rm ZF},W}[a^\dagger]
\right]
\right\}
\ket0.
\label{eq:linear-zero-frequency-dyson-state}
\]

We can see the exponentiation directly from our example in the previous section, at least at leading-soft level. On the support of $\del(\omega)$, the impact-parameter phase in eq.~\eqref{eq:first-order-endpoint-three-point} is one, so $\Sigma_{{\rm ZF},W}$ depends on the hard momenta but not on the hard positions. Its matter star-products with itself therefore reduce to ordinary products,
\[
\Sigma_{{\rm ZF},W}
\star_m
\Sigma_{{\rm ZF},W}
=
\Sigma_{{\rm ZF},W}^{\,2},
\qquad
\exp_{\star_m}\!\left[
\Sigma_{{\rm ZF},W}
\right]
=
\exp\!\left[
\Sigma_{{\rm ZF},W}
\right].
\]
The corresponding formal displacement obeys the usual coherent-state relation
\[
a_\lambda(k)
\exp\!\left[
\Sigma_{{\rm ZF},W}
\right]
\ket0
=
\alpha_{{\rm ZF},\lambda}(k;\tilde z)
\exp\!\left[
\Sigma_{{\rm ZF},W}
\right]
\ket0.
\]
This immediately raises two problems, however. First, the strict zero-frequency displacement does not define a coherent state in the usual sense: its normalisation involves the product of two delta functions, which is ill-defined. Second, it isn't clear that this defines a nonzero field after we naively perform the $d\Phi(k)$ integral. To give these expressions meaning, we follow the logic of ref.~\cite{Elkhidir:2024izo}: we regulate the zero-frequency modes, take the far-zone limit and perform the momentum-space integrals at fixed regulator, and only then remove it. This order of limits defines the asymptotic zero-frequency sector and reproduces the corresponding Coulombic field. Importantly, the zero-frequency ket below is not literally a Fock coherent state \cite{Prabhu:2022zcr}, but rather a shorthand for the regulated asymptotic dressing defined by this prescription. 

The creation-only exponential is only the normal-ordered radiation part of the state. At fixed regulator, the corresponding normalised coherent state is
\[
\ket{\alpha_{{\rm ZF},\varepsilon}}
=
\exp\!\left[
-\frac12N_{{\rm ZF},\varepsilon}
\right]
\exp\!\left[
\Sigma_{{\rm ZF},W,\varepsilon}
\right]
\ket0,
\qquad
N_{{\rm ZF},\varepsilon}
=
\sum_\lambda\int d\Phi(k)\,
\left|
\alpha_{{\rm ZF},\lambda,\varepsilon}(k)
\right|^2.
\]
In the complete linear Dyson evolution the zero-radiation coefficient generated by the Wick contractions supplies the corresponding normalisation and phase, and the $S$-matrix is unitary.  That being said, the product of two zero-frequency terms must be regulated, since $N_{\rm ZF}$ otherwise contains products of distributions. The same coherent-state mechanism appears in the traditional treatment of long-wavelength radiation and IR divergences \cite{Bloch:1937pw,Yennie:1961ad,Weinberg:1965nx,Kulish:1970ut}. There is, however, an important distinction: the term we discussed fixes the time-independent asymptotic field, but does not by itself regulate the IR divergences in the way that e.g. Faddeev--Kulish does \cite{Kulish:1970ut}.

It is useful to make contact here with the notation of ref.~\cite{Elkhidir:2024izo} at this point. There, they define a soft-dressed asymptotic state
\[
\ket{\psi} = \int \sd\Phi(p)\varphi(p)e^{iQ_s/\hbar}\ket{p},
\]
where $Q_s$ is the soft charge operator.

In our formalism, the equivalent state is given in terms of the symbol of this soft charge, which is given in terms of $\Sigma_{{\rm ZF},W}$ and its complex conjugate
\[
Q_{s,W}
\equiv
-i\hbar
\left(
\Sigma_{{\rm ZF},W}
-
\Sigma_{{\rm ZF},W}^\dagger
\right),
\qquad
\exp\!\left[
\frac{i}{\hbar}Q_{s,W}
\right]
=
\exp\!\left[
\Sigma_{{\rm ZF},W}
-
\Sigma_{{\rm ZF},W}^\dagger
\right],
\label{eq:zero-frequency-soft-charge}
\]
where this $Q_{s,W}$ is Hermitian. At fixed regulator, its exponential is unitary and gives the coherent dressing on the vacuum, while it can also act on an arbitrary radiation state. In particular, the Baker--Campbell--Hausdorff series terminates after the first commutator and gives
\[
e^{-\frac{i}{\hbar}Q_{s,W}}
a_\lambda(k)
e^{\frac{i}{\hbar}Q_{s,W}}
=
a_\lambda(k)
+
\alpha_{{\rm ZF},\lambda}(k;\tilde z).
\label{eq:zero-frequency-soft-charge-displacement}
\]

This makes the physical interpretation clear: the dressing translates the radiation field by precisely the zero-frequency mode obtained from the lowest order of the Dyson expansion. In gravity, ref.~\cite{Elkhidir:2024izo} shows explicitly that the analogous displacement produces the linearised Schwarzschild field of the asymptotic massive particle. The static field is therefore not an extra ingredient attached to the $S$-matrix in our formalism: the regulated zero-mode sector of the Dyson symbol encodes the asymptotic static field, assuming the far-zone prescription of ref. \cite{Elkhidir:2024izo}.

The reason for retaining the zero-frequency coefficient will become clear when we study the geometry of the outgoing state, which includes both radiative and static modes. At the order considered below, the inclusive waveshape can be expanded as
\[
\alpha_{I,\lambda}
=
\alpha_{I,\lambda}^{\rm ZF}
+
\alpha_{I,\lambda,0}
+\cdots,
\qquad
\alpha_{I,\lambda}^{\rm ZF}
=
S_{1,W,\lambda}^{\rm ZF},
\label{eq:inclusive-waveshape-zero-frequency-expansion}
\]
where $\alpha_{I,\lambda,0}$ is the leading regular soft waveform. We can equivalently write the zero-frequency dressing by

\[
\ket{\alpha_{\rm ZF}}
=
\exp\!\left[
\frac{i}{\hbar}Q_{s,W}
\right]
\ket0.
\]
We use this ket as shorthand for the regulated dressing inside observables.

Finally, we note that the soft displacement by itself is not the complete asymptotic symmetry. Ref.~\cite{Elkhidir:2024izo} shows that commuting the soft charge through the $S$-matrix generates a hard contribution $Q_h$, so that the complete soft transformation is organised by
\[
\left[
S_W,
Q_{s,W}
+
Q_{h,W}
\right]_{\star_m}
=
0.
\]
We will not need the explicit hard charge in what follows, but it is interesting to note that in gravity its action on finite-frequency radiation is related to an angle-dependent translation of retarded time, and hence to a BMS supertranslation. This concept will become relevant later when we discuss the frame dependence of the static gravitational field.
\section{The Partial-Weyl Outgoing State}
\label{sec:parent-state-projections}

\subsection{Radiation and the Inclusive Waveform}
The deterministic waveform is an in-in observable, and for an incoming radiation vacuum, using $\bb S=1+i\bb T$, its waveshape is
\[
\alpha_{I,\lambda}(k)
=
\langle\bb S^\dagger a_\lambda(k)\bb S\rangle
=
i\langle[a_\lambda(k),\bb T]\rangle
+
\langle\bb T^\dagger[a_\lambda(k),\bb T]\rangle.
\label{eq:inclusive-in-in-waveshape}
\]
After acting on the final states with $\bb T$, we obtain amplitudes, with the first term above the uncut one-radiation term, built from the full five-point amplitude, while the second is the cut completion, a product of an amplitude and its conjugate. Crucially, the second term is an adjacent multiplicity cut, where the left and right amplitudes differ by one radiative particle.

We will refer to $\braket{a}$ as the \textit{inclusive} waveshape, and in more generality it is defined in terms of the symbol $S_W$ as
\[
\alpha_{I,\lambda}(k,\tilde z)
&\equiv
\frac{
\langle0|
S_W^\dagger(\tilde z)
\star_m
a_\lambda(k)
\star_m
S_W(\tilde z)
|0\rangle
}{
\langle0|
S_W^\dagger(\tilde z)
\star_m
S_W(\tilde z)
|0\rangle
}.
\label{eq:inclusive-waveshape-amplitude-response}
\]
Exact star unitarity sets the denominator to one, but it is useful to retain it when the amplitude series is truncated: we will need to keep the terms that ensure unitarity to the required order, or perform the star-product and possibly endpoint reduce. The elastic factor will be stripped only after the complete cut waveshape has been formed. Importantly, this object entirely characterises the classical field, for example the classical photon, satisfying Maxwell's equations, is given by
\[
A_\mu(x)
=
\sum_\lambda\int d\Phi(k)
\left[
 u_{\lambda\mu}(x;k)\alpha_{I,\lambda}(k,\tilde{z})
+u^*_{\lambda\mu}(x;k)\alpha^*_{I,\lambda}(k,\tilde{z})
\right],
\]
where we define
\[
u_{\lambda\mu}(x;k)
=
\frac{1}{\sqrt\hbar}\epsilon_\mu^{(\lambda)}(k)e^{-ik\cdot x/\hbar},
\]
so the positive-frequency part of the classical radiation field is obtained by replacing $a_\lambda(k)$ by $\alpha_{I,\lambda}(k)$ in the photon operator.  The distinction between the exclusive kernel and the inclusive waveshape is invisible at leading order, but is crucial to include once nonlinear radiation interactions are present. 

The exact waveshape is most conveniently expanded as a sum over the $n$-point S-matrix symbols obtained by projecting the outgoing state onto fixed radiative multiplicity, i.e.
\[
S_W(\tilde z)|0\rangle
=
\sum_{n=0}^{\infty}
\frac1{n!}
\int_{1\cdots n}
S_{n,W}(1,\ldots,n;\tilde z)
a_1^\dagger\cdots a_n^\dagger
|0\rangle,
\qquad
S_{0,W}\equiv E_W.
\]
Here, $S_{n,W}(1,\ldots,n;\tilde z)=\langle1,\ldots,n|S_W(\tilde z)|0\rangle$ is the direct projection of the partially Weyl-transformed $S$-matrix onto the complete $n$-radiation sector. It is a coefficient of the photon Fock state and contains no remaining photon operators, while the ordinary on-shell amplitude $\cl A_{2\to2+n}$ enters via a partial Weyl transform.

We can act with the annihilation operator, using $[a_\lambda(k),a_{\lambda_i}^\dagger(k_i)] = \delta_{\lambda\lambda_i} \delta_\Phi(k,k_i),$ to obtain
\[
a_\lambda(k)S_W(\tilde z)|0\rangle
={}&
\sum_{n=1}^{\infty}
\frac1{n!}
\int_{1\cdots n}
S_{n,W}(1,\ldots,n;\tilde z)\sum_{r=1}^{n}
\delta_{\lambda\lambda_r}
\delta_\Phi(k,k_r)
a_1^\dagger\cdots a_{r-1}^\dagger a_{r+1}^\dagger\cdots a_n^\dagger
|0\rangle.
\]
Since $S_{n,W}$ is symmetric in its labels, each of the $n$ terms gives the same contribution after integration, and after relabelling we find
\[
a_\lambda(k)S_W(\tilde z)|0\rangle
=
\sum_{n=0}^{\infty}
\frac1{n!}
\int_{2\cdots n+1}
S_{n+1,W}
\bigl(
(k,\lambda),2,\ldots,n+1;\tilde z
\bigr)
a_2^\dagger\cdots a_{n+1}^\dagger
|0\rangle.
\]
The annihilation operator removes the measured photon from the $(n+1)$-particle component, leaving an $n$-particle state which can overlap with the $n$-particle component of the conjugate amplitude symbol $S_{n,W}^\dagger$.
The normalisation term in the denominator gives
\[
\langle0|
S_W^\dagger(\tilde z)
\star_m
S_W(\tilde z)
|0\rangle
=
\sum_{n=0}^{\infty}
\frac1{n!}
\int_{1\cdots n}
S_{n,W}^\dagger(1,\ldots,n;\tilde z)
\star_m
S_{n,W}(1,\ldots,n;\tilde z),
\]
while the numerator gives
\[
&\langle0|
S_W^\dagger(\tilde z)
\star_m
a_\lambda(k)
\star_m
S_W(\tilde z)
|0\rangle
\\
&\qquad=
\sum_{n=0}^{\infty}
\frac1{n!}
\int_{2\cdots n+1}
S_{n,W}^\dagger(2,\ldots,n+1;\tilde z)
\star_m
S_{n+1,W}
\bigl(
(k,\lambda),2,\ldots,n+1;\tilde z
\bigr).
\]
The exact inclusive waveshape may therefore be written as
\[
\begin{aligned}
\alpha_{I,\lambda}(k,\tilde z)
=
\frac{
\displaystyle
\sum_{n=0}^{\infty}
\frac1{n!}
\int_{2\cdots n+1}
S_{n,W}^\dagger(2,\ldots,n+1;\tilde z)
\star_m
S_{n+1,W}
\bigl(
(k,\lambda),2,\ldots,n+1;\tilde z
\bigr)
}{
\displaystyle
\sum_{n=0}^{\infty}
\frac1{n!}
\int_{1\cdots n}
S_{n,W}^\dagger(1,\ldots,n;\tilde z)
\star_m
S_{n,W}(1,\ldots,n;\tilde z)
}.
\end{aligned}
\label{eq:inclusive-amplitude-adjacent-multiplicity}
\]
This expression has the structure of an inclusive cut. The coefficient on the right of the cut contains the measured radiation mode $(k,\lambda)$ and $n$ additional photons. The conjugate coefficient contains the same $n$ unobserved photons, hence it has one less radiating particle overall. Integrating over their on-shell phase space performs the inclusive sum over unobserved radiation. The matter star product is the on-shell sewing of two completed amplitude symbols, essentially performing the cut.
\subsection{Phase Space Displacement: The Impulse and Position Shift}
\label{subsec:phase-space-displacement}
We can derive the displacement in phase space directly from the original partially Weyl-transformed state. We use the usual phase-space conventions
\[
z^A=(x_i^\mu,p_{i\mu}),
\qquad
\Omega^{AB}
=
\begin{pmatrix}
0&I\\
-I&0
\end{pmatrix},
\qquad
\Omega_{AB}
=
\begin{pmatrix}
0&-I\\
I&0
\end{pmatrix}.
\label{eq:canonical-phase-space-matrices-radiation}
\]
Importantly, the lowering of $\Delta z_A\equiv\Omega_{AB}\Delta z^B$ is performed with the inverse canonical phase-space matrix $\Omega$ and \textit{not} the spacetime metric. In components,
\[
\Delta z_A
=
\left(-\Delta p_{i\mu},\Delta x_i^\mu\right),
\qquad
\Delta p_{i\mu}=-\Delta z_{x_i^\mu},
\qquad
\Delta x_i^\mu=\Delta z_{p_{i\mu}}.
\label{eq:lowered-displacement-components}
\]
We define the local phase-space displacement function by 
\[
\Delta z^A\equiv \frac{\braket{0|S_W^\dagger\star_m z^A\star_m S_W|0}}{\langle0|S_W^\dagger\star_m S_W|0\rangle}-z^A
\]
This function can be defined without choosing the incoming Wigner function, although it is not yet physical. The \textit{observable} impulse and position shift are obtained by placing the numerator and denominator, separately, in the projected matter pairing before taking their ratio, and then subtracting the incoming coordinate. For a Wigner function sharply centred on a single phase-space point, as defined in section \ref{sec:incoming-matter-state}, this simply reduces to the local expression above, and so we will omit this procedure in what follows. Of course, it would be interesting to study the effect of a more general incoming Wigner function, say a function with some statistical classical uncertainty (a distribution of initial or final states, for example), and we leave this to future work.
The star product of a coordinate $z$ truncates after its first derivative, and so applying this to the outgoing expectation value and using the derivative of $\cl N$ gives an exact result
\[
z^A\star_m S_W
&=
z^A S_W
+
\frac{i\hbar}{2}\Omega^{AB}\partial_B S_W,
\qquad
S_W^\dagger\star_m z^A
=
z^A S_W^\dagger
-
\frac{i\hbar}{2}\Omega^{AB}\partial_B S_W^\dagger,
\\
S_W^\dagger\star_m z^A\star_m S_W
&=
z^A\left(S_W^\dagger\star_m S_W\right)
+
\frac{i\hbar}{2}\Omega^{AB}
\left[
S_W^\dagger\star_m\partial_B S_W
-
(\partial_B S_W^\dagger)\star_m S_W
\right].
\]
The norm derivative is
\[
\partial_A\cl N
=
\langle0|
(\partial_A S_W^\dagger)\star_m S_W
+
S_W^\dagger\star_m(\partial_A S_W)
|0\rangle,
\]
so the symmetric derivative cancels from the normalised expectation. Lowering with $\Omega_{CA}\Omega^{AB}=\delta_C{}^B$ then gives
\[
\Delta z_A
=
\frac{i\hbar}{2\cl N}
\langle0|
S_W^\dagger
\star_m
\overleftrightarrow{\partial}_A
S_W
|0\rangle,
\label{eq:exact-lowered-phase-space-displacement}
\]
where $\cl N(\tilde z) \equiv \langle0|S_W^\dagger(\tilde z)\star_m S_W(\tilde z)|0\rangle$ and 
\[
S_W^\dagger\star_m\overleftrightarrow{\partial}_A S_W
\equiv
S_W^\dagger\star_m(\partial_A S_W)
-(\partial_A S_W^\dagger)\star_m S_W.
\]
Note that we did \textit{not} assume that star-unitarity holds to obtain this result, nor has any coherent state or classical assumption been made.

In the purely elastic endpoint limit, $S_W=e^{i\chi/\hbar}$ and $\cl N=1$, so
\[
\Delta z_A=-\partial_A\chi,
\label{eq:elastic-lowered-displacement}
\]
which verifies the impulse and position-shift signs above.

Inserting a complete set of radiation states, and using the definition of the $n$-radiation kernels, we can write the exact lowered displacement as
\[
\Delta z_A
=
\frac{i\hbar}{2\cl N}
\sum_{n=0}^{\infty}
\frac{1}{n!}
\int_{1\cdots n}
S_{n,W}^\dagger(1,\ldots,n)
\star_m
\overleftrightarrow{\partial}_A
S_{n,W}(1,\ldots,n).
\label{eq:diagonal-lowered-displacement}
\]
Every term has the same number of radiative particles either side of the cut, unlike the inclusive one-point function, which has $n$ and $n+1$ particles on either side. In other words, outgoing radiation changes multiplicity by one, while the derivative preserves it, and both expressions are projections of the same partially Weyl transformed outgoing state.
\section{Connected Radiation Kernels from On-Shell Amplitudes}
\label{sec:connected-radiation-kernels}

\subsection{Radiation from On-Shell Amplitudes}
The functional derivative form of the potential is useful for systematically generating arbitrarily high-order conservative and radiative kernels, but it is time-consuming to evaluate. As we show explicitly in appendix~\ref{app:dyson-star-ordering}, evaluating the leading order radiation coefficients for each of the star products, time orderings and endpoint approximations results in the usual five-point amplitude. This is much simpler to evaluate, especially using modern on-shell amplitude techniques, and so we should obviously use this approach. Given that the exclusive radiation kernels can be expressed in terms of on-shell amplitudes at any order, we will now see how this works in our formalism. Crucially, however, everything \textit{can} be derived explicitly from the resummed Dyson expansion after a partial Weyl transform and the explicit evaluation of the Bopp shifts, with the amplitudes obtained from this directly.

We use hats to denote operators acting only on the matter Hilbert space, with the photon Fock space projected out, defining
\[
\widehat E = \braket{ 0|\bb S| 0},
\qquad
\widehat S_{n,\vec\lambda}(\vec k) = \braket{\vec k,\vec\lambda|\bb S| 0},
\]
where $\widehat E$ is the elastic zero-photon operator and $\widehat S_{n,\vec\lambda}(\vec k)$ is the projection of the $S$-matrix onto the $n$-photon sector, with $\vec k$ and $\vec\lambda$ the $n$-dimensional vectors of the momenta and polarisations of the outgoing photons. Using $\bb S = \bb 1 + i\bb T$, we can write these as 
\[
\widehat E = 1+i\widehat T_4,~~~~~\widehat S_{n,\vec\lambda}(\vec k) = i\widehat T_{4+n,\vec\lambda}(\vec k).
\]
Their Weyl transforms are then
\[
E_W(\tilde z)
&=
1+iT_{4,W}(\tilde z)
\\
&=
1 + \int
\hat{\sd}^4\ell e^{+\frac{i}{\hbar}\ell\cdot(x_1-x_2)}
\del(2\tilde p_1\cdot\ell)
\del(-2\tilde p_2\cdot\ell)\cl A_{2\to2}
\left(
\tilde p_a\pm\frac{\ell}{2},
\to
\tilde p_a\mp\frac{\ell}{2},
\right).
\label{eq:partial-weyl-elastic-amplitude}
\]
and
\[
S_{n,W,\vec\lambda}(\vec k_n;\tilde z)
&=
iT_{4+n,W,\vec\lambda}(\vec k_n;\tilde z)
\\
&=
\int
\prod_{a=1}^{2}\hat{\sd}^4q_a\,
e^{\frac{i}{\hbar}
\sum_{a=1}^{2}q_a\cdot x_a}
\prod_{a=1}^{2}
\del(2\tilde p_a\cdot q_a)\cl A_{2\to2+n,\vec\lambda}
\left(
\tilde p_a-\frac{q_a}{2}
\to
\tilde p_a+\frac{q_a}{2};
\vec k_n
\right),
\label{eq:partial-weyl-amplitude}
\]
where $q_a=p_{a,\rm out}-p_{a,\rm in}$ is the momentum change of worldline $a$, so momentum conservation reads $q_1+q_2+K_n=0$ for $K_n=\sum_{r=1}^nk_r$. The momentum-conserving delta function in the amplitude performs the $q_2$ integral, giving
\[
q_1\equiv q,
\qquad
q_2\equiv-q-K_n.
\]
We can now make a slightly formal construction of a symbol with desirable properties that will be used later\footnote{All star inverses and star logarithms in this construction are understood as formal perturbative series about the identity, and we don't assume any global branch properties.}.
The useful point is that the complete on-shell amplitude has already done the local momentum routing calculation for us, since it contains every ordering of emission relative to exchange, the corresponding matter poles and the nonlinear contact terms. Rather than treating each radiative multiplicity separately, we reassemble the fixed-multiplicity symbols into a generating functional made up of creation operators (valid on the vacuum state)
\[
S_W[a^\dagger;\tilde z]| 0\rangle
=
\left[
E_W(\tilde z)
+
\sum_{n\geq1}\frac1{n!}
\int_{1\cdots n}
S_{n,W}(1,\ldots,n;\tilde z)
a_1^\dagger\cdots a_n^\dagger
\right]
| 0\rangle.
\label{eq:onshell-fixed-multiplicity-symbol}
\]
We then strip the complete elastic factor from the left and take the star logarithm,
\[
\bb R_W[a^\dagger;\tilde z]
&\equiv
E_W^{-\star_m}(\tilde z)
\star_m
S_W[a^\dagger;\tilde z],
\\
F_{\tilde z}[a^\dagger]
&\equiv
\left.
\log_{\star_m}\bb R_W[a^\dagger;\tilde z]
\right|_{a^\dagger},
\qquad
E_W^{-\star_m}\star_m E_W=1.
\label{eq:onshell-radiation-logarithm}
\]
The connected radiation generator $F_{\tilde z}$ is dimensionless, and as formal series we can write
\[
\bb R_W[a^\dagger;\tilde z]
=
\exp_{\star_m}F_{\tilde z}[a^\dagger].
\]
The notation $\left.\cdots\right|_{a^\dagger}$ removes the vacuum coefficient and retains only terms containing at least one creation operator after normal ordering. We similarly define the elastic connected generator by
\[
\Gamma_{\rm el}(\tilde z)
=
\log_{\star_m}E_W(\tilde z).
\]
We can now exponentiate these objects to give us the connected factorisation, which we derive directly from the Dyson expansion in appendix~\ref{app:dyson-star-ordering},
\[
S_W[a^\dagger;\tilde z]| 0\rangle
=
\exp_{\star_m}\left(\Gamma_{\rm el}(\tilde z)\right)
\star_m
\exp_{\star_m}\left(F_{\tilde z}[a^\dagger]\right)
| 0\rangle.
\]
Projecting this identity onto a fixed $n$-particle radiation state gives
\[
&\delta_{n0}\bb 1
+iT_{4+n,W,\vec\lambda}(\vec k_n;\tilde z)=
\braket{\vec k_n,\vec\lambda|
\exp_{\star_m}\left(\Gamma_{\rm el}(\tilde z)\right)
\star_m
\exp_{\star_m}\left(F_{\tilde z}[a^\dagger]\right)
| 0}.
\]
Since $\bb R_W$ has unit vacuum coefficient, its logarithm cannot change the term linear in $a^\dagger$. We immediately find
\[
C_{1,\lambda}(k;\tilde z)
=
\left.
\frac{\delta F_{\tilde z}[a^\dagger]}
{\delta a_\lambda^\dagger(k)}
\right|_{a^\dagger=0}
=
\left[
E_W^{-\star_m}
\star_m
S_{1,W,\lambda}(k)
\right](\tilde z).
\label{eq:onshell-full-C1}
\]
We compute this using Bopp shifts in appendix \ref{sec:radiation-from-star-ordering}, and the result is the full connected five-point one-radiation kernel. The word ``full'' is important here: $S_{1,W,\lambda}$ is built from the complete fixed-multiplicity amplitude, including its contact terms, while the inverse elastic factor removes its lower elastic iterations. There are no radiative iterations to remove at this order, although they will appear at higher orders.  For scalar QED, we write the phase-space positions as the impact parameters $b_i$, and solve momentum conservation by
\[
q_a=\ell-k,
\qquad
q_b=-\ell.
\]
Using the stripped five-point normalisation of eq.~\eqref{eq:partial-endpoint-full-five-point-amplitude}, for which the fixed-multiplicity $S$-matrix coefficient carries the overall factor $i/\hbar^{3/2}$, eq.~\eqref{eq:partial-weyl-amplitude} then gives
\[
C_{1,\lambda}^{(0)}(k;\tilde z)
&=
S_{1,W,\lambda}^{(0)}(k;\tilde z)
\\
&=
\frac{i}{\hbar^{3/2}}
\sum_{a\neq b}
\int\dd^4\ell\,
e^{+\frac{i}{\hbar}[(\ell-k)\cdot b_a-\ell\cdot b_b]}
\del\!\left(2\tilde p_a\cdot(\ell-k)\right)
\del\!\left(2\tilde p_b\cdot\ell\right)
\cl A_{5,a}^{(0),\rm full}(k,\ell)
\\
&=
C_{1,\lambda}^{(0),\rm pole}(k;\tilde z)
+
C_{1,\lambda}^{(0),\rm sg}(k;\tilde z).
\label{eq:onshell-sqed-full-C1}
\]
This is precisely eq.~\eqref{eq:star-ordering-full-C1}. In particular, the radiating line carries the combined transfer $\ell-k$, rather than the separate straight-line support $\del(2\tilde p_a\cdot\ell)\del(2\tilde p_a\cdot k)$ we saw earlier, and the two matter poles and the seagull contact are already packaged into $\cl A_{5,a}^{(0),\rm full}$. We see then that the on-shell construction reaches the same conclusion as the resolved star-ordering calculation with much less work: compute the complete gauge-invariant amplitude, take its partial Weyl transform, and, beyond leading order, remove the elastic iterations with the star inverse. The perturbative form of this last operation is described next.
\subsection{Connected Kernels and Iteration Subtractions}
Equations~\eqref{eq:onshell-radiation-logarithm} and \eqref{eq:onshell-full-C1} define the connected radiation generator and the full one-radiation kernel directly from complete amplitudes, and it's worth expanding these relations perturbatively to see what's going on. We use $r$ for the outgoing-radiation multiplicity, $L$ for the perturbative order and $s$ for the nonlinear-insertion order. At fixed multiplicity we write
\[
E_W
&=
1+\sum_{L\geq0}\cl E_L,
\qquad
S_{r,W}(1,\ldots,r)
=
\sum_{L\geq0}S_{r,W}^{(L)}(1,\ldots,r),
\\
S_W
&=
E_W
+
\sum_{r\geq1}\frac1{r!}
\int_{1\cdots r}
\left[
\sum_{L\geq0}S_{r,W}^{(L)}(1,\ldots,r)
\right]
a_1^\dagger\cdots a_r^\dagger.
\label{eq:fixed-multiplicity-perturbative-expansion}
\]
Here $\cl E_L$ is the fixed-order elastic coefficient at perturbative order $L$, while $S_{r,W}^{(L)}$ is the fixed $r$-point amplitude symbol, including the contact terms. The connected or subtracted elastic coefficients are the coefficients of $\Gamma_{\rm el}=\log_{\star_m}E_W$ below.

There are two distinct notions of connectedness in these expressions, elastic and inelastic. Expanding the elastic logarithm gives us a definition for the first kind
\[
\Gamma_{\rm el}
=
\log_{\star_m}E_W
={}&
\cl E_0
+
\left(
\cl E_1-\frac12\cl E_0\star_m\cl E_0
\right)
\\
&+
\left[
\cl E_2
-\frac12\left(
\cl E_0\star_m\cl E_1
+
\cl E_1\star_m\cl E_0
\right)
+
\frac13
\cl E_0\star_m\cl E_0\star_m\cl E_0
\right]
+\cdots.
\]
Endpoint reduction --- omitting the star products, in this case --- is only allowed once each term has been composed with any radiation kernels that might require it. The same is true of the radiative sector: each $S_{r,W}^{(L)}$ must be sufficiently composed and stripped before endpoint reduction. Writing the coefficient of radiation multiplicity $r$ in the stripped symbol $E_W^{-\star_m}\star_m S_W$ as $\cl R_r=\sum_{L\geq0}\cl R_r^{(L)}$, we have
\[
\bb R_W
=
1
+
\sum_{r\geq1}\frac1{r!}
\int_{1\cdots r}
\left[
\sum_{L\geq0}\cl R_r^{(L)}(1,\ldots,r)
\right]
a_1^\dagger\cdots a_r^\dagger.
\]
If we only care about conservative dynamics, we can take the endpoint limit immediately to obtain the centre-generating eikonal
\[
E_W(\tilde z)\big|_{\rm ep}
=
\exp\!\left[\frac{i}{\hbar}\chi(\tilde z)\right].
\]

For one emitted quantum, the elastic star inverse is the only source of subtractions, and its expansion is
\[
E_W^{-\star_m}
=
1-\cl E_0
+
\left(
\cl E_0\star_m\cl E_0-\cl E_1
\right)
+\cdots,
\]
and hence
\[
\DC{1}^{(0)}(i)
&=
\cl R_1^{(0)}(i)
=
S_{1,W}^{(0)}(i),
\\
\DC{1}^{(1)}(i)
&=
\cl R_1^{(1)}(i)
=
S_{1,W}^{(1)}(i)
-
\cl E_0\star_m S_{1,W}^{(0)}(i),
\\
\DC{1}^{(2)}(i)
&=
\cl R_1^{(2)}(i)
=
S_{1,W}^{(2)}(i)
-
\cl E_0\star_m S_{1,W}^{(1)}(i)
+
\left(
\cl E_0\star_m\cl E_0-\cl E_1
\right)
\star_m S_{1,W}^{(0)}(i).
\label{eq:C1-elastic-subtractions}
\]
This is the perturbative form of eq.~\eqref{eq:onshell-full-C1}. We have $\DC{1}=\cl R_1$ at every perturbative order because the radiation logarithm cannot factorise a coefficient containing only one outgoing quantum.

The second notion is radiative connectedness. Expanding the logarithm in eq.~\eqref{eq:onshell-radiation-logarithm}, we write
\[
F_{\tilde z}[a^\dagger]
=
\sum_{r\geq1}\frac1{r!}
\int_{1\cdots r}
\DC{r}(1,\ldots,r;\tilde z)
a_1^\dagger\cdots a_r^\dagger,
\qquad
\DC{r}
=
\sum_{L\geq0}\DC{r}^{(L)}.
\label{eq:connected-radiation-generator-expansion}
\]
The two-radiation kernel is the first sector in which this logarithm makes a nontrivial subtraction. At the first such order,
\[
\DC{2}^{(1)}(i,j)
=
\cl R_2^{(1)}(i,j)
-
\frac12\left[
\DC{1}^{(0)}(i)\star_m\DC{1}^{(0)}(j)
+
\DC{1}^{(0)}(j)\star_m\DC{1}^{(0)}(i)
\right].
\label{eq:C2-radiative-subtraction}
\]
The two operations therefore remove different factorisations: left multiplication by the inverse elastic factor removes lower elastic iterations from each fixed-multiplicity amplitude, while the radiation logarithm removes factorised radiation production from the stripped coefficients.

A nonlinear insertion of order $s$ changes the connected radiation generator by
\[
\Delta_sF[a^\dagger]
=
\sum_{r\geq1}\frac1{r!}
\int_{1\cdots r}
\Delta_s\DC{r}(1,\ldots,r)
a_1^\dagger\cdots a_r^\dagger.
\label{eq:DeltaF-connected-kernel-expansion}
\]
The terms containing two or more creation operators must be retained: although they are exclusive multi-radiation kernels, their cuts contribute to the inclusive one-point waveshape considered later. The subscripts ``irr'' and ``cl'' used below mean that these two kinds of iteration have been removed before taking the classical limit. This connected projection keeps the genuine scalar-QED triangle and radiative seagull corrections while discarding products of lower exchanges \cite{Cristofoli:2021jas}.
\subsection{Comparison with the Eikonal-Waveshape Convolution}
The star product of the eikonal with the radiation kernel is the way in which the conservative recoil is routed through the matter legs, preserving the momentum of the outgoing radiation. In ref. \cite{Cristofoli:2021jas}, we performed the same routing via a convolution of the eikonal and waveshape in Fourier space. To compare this with the approach here, we can Fourier expand the waveshape of ref.~\cite{Cristofoli:2021jas} as
\[
\alpha_{\lambda}(k;x_1,x_2)
=
\int
\hat{\sd}^4q_1\hat{\sd}^4q_2\,
e^{\frac{i}{\hbar}(q_1\cdot x_1+q_2\cdot x_2)}
\del^{(4)}(q_1+q_2-k)\,
\widetilde\alpha_{{\rm ex},\lambda}(k;q_1,q_2).
\]
The conservative eikonal depends on the same two positions, and its Fourier transform is
\[
E_{\rm el}(x_1,x_2;\tilde p_1,\tilde p_2)
&\equiv
\int\hat{\sd}^4\ell\,\sd^4x\,
e^{\frac{i}{\hbar}\ell\cdot(x+x_2-x_1)}
e^{\frac{i}{\hbar}\chi(x_\perp;s)}
\\
&=
\int\hat{\sd}^4\ell\,
e^{\frac{i}{\hbar}\ell\cdot(x_2-x_1)}
\widetilde E_{\rm el}
(\tilde p_1,\tilde p_2;\ell),
\]
where
\[
\widetilde E_{\rm el}
(\tilde p_1,\tilde p_2;\ell)
\equiv
\int\sd^4x\,
e^{\frac{i}{\hbar}\ell\cdot x}
e^{\frac{i}{\hbar}\chi(x_\perp;s)}.
\]
Importantly, we also assume that both the eikonal and waveshape depend on the external matter momentum $\tilde p_1$ and $\tilde p_2$, which we suppress in the notation.

Now, consider the one-radiation coefficient obtained from the five-point expression in eq.~(4.42) of ref.~\cite{Cristofoli:2021jas}. Equating the coherent-state overlap in eq.~(4.40) with the direct five-point result in eq.~(4.42) gives eq.~(4.43), which may equivalently be written as
\[
S_{1,\lambda}(k;x_1,x_2)
&= E_{\rm el}(x_1,x_2)\alpha_{\lambda}(k;x_1,x_2)\\
&= \int
\hat{\sd}^4q_1\hat{\sd}^4q_2\hat{\sd}^4\ell\,
e^{\frac{i}{\hbar}((q_1-\ell)\cdot x_1+(q_2+\ell)\cdot x_2)}
\del^{(4)}(q_1+q_2-k)\,
\widetilde E_{\rm el}(\ell)\widetilde\alpha_{\lambda}(k;q_1,q_2).
\]
The shared dependence on $x_1$ and $x_2$ therefore routes the elastic transfer $\ell$ through the two matter legs of the radiation kernel. In particular,
\[
(q_1-\ell)+(q_2+\ell)=k,
\]
so this routing redistributes momentum between the two matter lines without changing the total momentum carried by the radiation. The ordinary product of the eikonal and waveshape in position space is consequently a convolution in momentum space. To make this concrete, we can change variables and write the Fourier transform as
\[
S_{1,\lambda}(k;x_1,x_2)
= \int
\hat{\sd}^4r_1\hat{\sd}^4r_2\hat{\sd}^4\ell\,
e^{\frac{i}{\hbar}(r_1\cdot x_1+r_2\cdot x_2)}
\del^{(4)}(r_1+r_2-k)\,
\widetilde E_{\rm el}(\ell)\widetilde\alpha_{\lambda}(k;r_1+\ell,r_2-\ell).
\]
We see then that convolution has shifted the arguments of the waveshape from $(r_1,r_2)$ to $(r_1+\ell,r_2-\ell)$, while the total radiated momentum remains fixed.

In the phase-space formulation we have developed, this convolution is represented by the star product and Bopp shifts, although it is perhaps simpler to see this if we define the waveshape directly as
\[
\alpha_{\lambda}=C_{1,\lambda}=E_{\rm el}^{(-\star_m)}\star_m S_{1,\lambda},
\]
where $E_{\rm el}^{(-\star_m)}\star_m E_{\rm el}=1$.

Consider the momentum-space representation of the star product of the inverse elastic factor and the one-radiation coefficient, evaluated via the Bopp shift identity in eq.~\eqref{eq:bopp-shift-identity}. Suppressing the transfer integrals and the factor $\del^{(4)}(q_1+q_2-k)$, we have
\[
&\left[e^{+\frac{i}{\hbar}q\cdot(x_2-x_1)}
E^{-1}_{\rm el}(\tilde p_1,\tilde p_2,q)\right]
\star_m
\left[e^{+\frac{i}{\hbar}(q_1\cdot x_1+q_2\cdot x_2)}
S_{1,\lambda}(\tilde p_1,\tilde p_2,q_1,q_2,k)\right]
\\
&\quad=
e^{+\frac{i}{\hbar}[(q_1-q)\cdot x_1+(q_2+q)\cdot x_2]}
E^{-1}_{\rm el}(\tilde p_1 +\frac{q_1}{2},\tilde p_2+\frac{q_2}{2},q)
S_{1,\lambda}(\tilde p_1 +\frac{q}{2},\tilde p_2-\frac{q}{2},q_1,q_2;k).
\label{eq:C1-bopp-momentum-routing}
\]
Again changing variables to $r_1=q_1-q$ and $r_2=q_2+q$, we find 
\[
&\alpha_{\lambda}(\tilde p_1,\tilde p_2,r_1,r_2;k)\\
&\qquad=
\int \hat\sd^4q \tilde E^{(-\star_m)}_{\rm el}(\tilde p_1 +\frac{r_1 + q}{2},\tilde p_2+\frac{r_2-q}{2},q)
\tilde S_{1,\lambda}(\tilde p_1 +\frac{q}{2},\tilde p_2-\frac{q}{2},r_1+q,r_2-q;k)
\]
This is the output of the phase-space formalism as we have developed it, but it is not yet the same as the convolution of ref.~\cite{Cristofoli:2021jas}. The difference is that the Bopp shifts have introduced local half-shifts into the arguments of the elastic factor and the one-radiation coefficient: we need to take the endpoint limit. Doing so then gives
\[
\alpha_{\lambda}(\tilde p_1,\tilde p_2,r_1,r_2;k)\bigg|_{\rm ep} =\int \hat\sd^4q \tilde E^{(-1)}_{\rm el}(\tilde p_1,\tilde p_2,q)
\tilde S_{1,\lambda}(\tilde p_1,\tilde p_2,r_1+q,r_2-q;k).
\label{eq:C1-bopp-momentum-routing-endpoint}
\]
This is precisely the transfer-space convolution obtained by Fourier-transforming eq.~(4.44) of ref.~\cite{Cristofoli:2021jas}, with the same routing $(r_1,r_2)\longrightarrow (r_1+q,r_2-q)$. The full star product keeps, in addition, the full history of the half-way shifts generated by the Bopp identity. Endpoint reduction removes these outer shifts and recovers the Fourier transform of eq.~(4.44). It should be stressed at this point that $\alpha_{\lambda}$ is defined as the coefficient of $a^\dagger_\lambda(k)$ in the earlier coherent ansatz, and is matched to the five-point amplitude. It therefore agrees with the connected exclusive kernel $C_1$ in that construction, but only with the leading regular sector of the inclusive waveshape. As we have argued here, at nonlinear order $\alpha_I$ also contains adjacent-multiplicity cuts.
\section{Inclusive Waveforms and the Coherent Classical Representative}
\label{sec:radiation-inclusive-cuts}

\subsection{The $N$-Operator Organisation of Exclusive Radiation}
The exclusive $S$-matrix may be organised perturbatively as a single operator exponential \cite{Magnus:1954zz,Damgaard:2021ipf,Damgaard:2023ttc},
\[
\bb S
=
\exp\!\left(\frac{i}{\hbar}\bb N\right),
\]
with $\bb N = \bb N^\dagger$ Hermitian for a unitary $S$-matrix\footnote{Defined as a log, we take $\bb N=-i\hbar\log\bb S$ to again be the formally perturbative logarithm close to the identity.}.
After the partial Weyl transform, the same statement becomes
\[
S_W(\tilde z)
=
\exp_{\star_m}\!\left[\frac{i}{\hbar}N_W(\tilde z)\right],
\]
where $N_W$ remains an operator on the radiation Fock space. It contains zero-, one-, two- and higher-radiation components, and therefore organises the exact exclusive outgoing state used in the nonlinear-memory calculation of ref. \cite{Georgoudis:2025vkk}. The $N$-operator is not itself the deterministic waveform: that waveform is the inclusive one-point function obtained by sewing adjacent radiation multiplicities below.

Let's explore the first nontrivial cut. We can first evaluate the star products in eq.~\eqref{eq:inclusive-amplitude-adjacent-multiplicity} and then take the endpoint limit, where the common elastic factor cancels between the numerator and denominator. The full-$S$ denominator in eq.~\eqref{eq:inclusive-waveshape-amplitude-response} is one by star unitarity, but the norm of the stripped radiation operator $e^{F[a^\dagger]}$ is nontrivial, and we will see that it is related to $\Im\chi_{\rm cut}$. At fixed perturbative order, the stripped coefficients are related via
\[
\cl R_0=1,
\qquad
\cl R_1(1)=\DC{1}(1),
\qquad
\cl R_2(1,2)
=
\DC{2}(1,2)
+
\DC{1}(1)\DC{1}(2).
\label{eq:R2-late-endpoint-reduction}
\]
The first term is the connected two-radiation contribution, while the second describes factorised production of two independent radiated photons or gravitons.
The corresponding stripped numerator begins as
\[
\begin{aligned}
&\langle0|
\mathbb R_W^\dagger
a_\lambda(k)
\mathbb R_W
|0\rangle
\\
&\quad=
\DC{1}_{\lambda}(k;\tilde z)
+
\sum_{\lambda_2}
\int d\Phi(k_2)\,
\DC{1,\lambda_2}^*(k_2;\tilde z)
\DC{2,\lambda\lambda_2}(k,k_2;\tilde z)
\\
&\qquad
+
\DC{1}_{\lambda}(k;\tilde z)
\sum_{\lambda_2}
\int d\Phi(k_2)\,
\left|
\DC{1,\lambda_2}(k_2;\tilde z)
\right|^2
+\cdots.
\end{aligned}
\label{eq:inclusive-numerator-first-orders}
\]
The stripped denominator begins as
\[
\langle0|
\mathbb R_W^\dagger
\mathbb R_W
|0\rangle
=
1+
\sum_{\lambda_2}
\int d\Phi(k_2)\,
\left|
\DC{1,\lambda_2}(k_2;\tilde z)
\right|^2
+\cdots.
\label{eq:inclusive-denominator-first-orders}
\]
The factorised term in the numerator cancels against the corresponding contribution from the normalisation denominator. The inclusive waveshape is consequently
\[
\begin{aligned}
\alpha_{I,\lambda}(k,\tilde z)
={}&
\DC{1}_{\lambda}(k;\tilde z)
+ 
\sum_{\lambda_2}
\int d\Phi(k_2)\,
\DC{2,\lambda\lambda_2}(k,k_2;\tilde z)
\DC{1,\lambda_2}^*(k_2;\tilde z)
+\cdots.
\end{aligned}
\label{eq:first-inclusive-kernel-cut}
\]
The second term is the first nontrivial inclusive cut. The mode $(k,\lambda)$ is measured, while the second leg of the two-radiation kernel is summed over and contracted with the conjugate one-radiation kernel.

The same result can be recovered from the imaginary part of the eikonal. Once every matter line in eq.~\eqref{eq:inclusive-amplitude-adjacent-multiplicity} has been sewn together using Bopp shifts, we can endpoint reduce it and replace the eikonal with a centre-generating scalar exponential. The cut phase-space integrals may remain unperformed, but the amplitude symbols and their matter sewings must already be complete. Writing the exclusive radiation  operator as $e^{F[a^\dagger]}|0\rangle$, unitarity demands that
\[
e^{-\frac{2}{\hbar}\Im\chi_{\rm cut}}
\bra{0}e^{F^\dagger[a]}e^{F[a^\dagger]}\ket{0}
=
1,
\qquad
\frac{2}{\hbar}\Im\chi_{\rm cut}
=
\log\bra{0}e^{F^\dagger[a]}e^{F[a^\dagger]}\ket{0}.
\label{eq:exclusive-cut-eikonal-unitarity}
\]
Here $\chi_{\rm cut}$ is the zero-emission eikonal rather than the conservative real phase $\chi_{\rm C}$ (i.e. it still contains the absorptive part, as it must). The ordinary logarithm in eq.~\eqref{eq:exclusive-cut-eikonal-unitarity} is therefore only applicable after the endpoint limit has been taken and all star products have been evaluated.

We now take a functional derivative, treating $C_{1,\lambda}(k)$ and $C_{1,\lambda}^*(k)$ as independent variables and holding the higher exclusive kernels fixed. Since $F^\dagger[a]$ arose from a normal ordering procedure, it only contains annihilation operators, which commute, and we can write
\[
\frac{\delta F^\dagger[a]}
{\delta C_{1,\lambda}^*(k)}
=
a_\lambda(k),
\]
and hence
\[
\frac{2}{\hbar}
\frac{\delta\Im\chi_{\rm cut}}
{\delta C_{1,\lambda}^*(k)}
=
\frac{
\bra{0}e^{F^\dagger[a]}a_\lambda(k)e^{F[a^\dagger]}\ket{0}
}{
\bra{0}e^{F^\dagger[a]}e^{F[a^\dagger]}\ket{0}
}
=
\alpha_{I,\lambda}(k).
\label{eq:inclusive-waveshape-cut-eikonal}
\]
This formulation only makes sense as a functional derivative of the connected cut representation and not a derivative of the final numerical value of the eikonal: it must be kept in cut form, hence $\chi_{\rm cut}$. The higher $C_r$ are held fixed under the derivative, but still enter through the remaining factors in the cuts contained in $\chi_{\rm cut}$. Expanding the exponentials produces the adjacent overlaps $\cl R_n^\dagger\cl R_{n+1}$, and therefore gives the endpoint reduction of eq.~\eqref{eq:inclusive-amplitude-adjacent-multiplicity} after the common elastic factor has cancelled. Equation~\eqref{eq:inclusive-waveshape-cut-eikonal} is an alternative derivation of the inclusive cut, rather than an assumption that the exclusive state is coherent.

We have therefore organised the same waveshape in three ways: as the in-in commutator in eq.~\eqref{eq:inclusive-in-in-waveshape}, as the adjacent-multiplicity amplitude overlap in eq.~\eqref{eq:inclusive-amplitude-adjacent-multiplicity}, and as the cut-eikonal derivative in eq.~\eqref{eq:inclusive-waveshape-cut-eikonal}.
\subsection{A Coherent Source for the Classical Field}
In ref. \cite{Cristofoli:2021jas}, we motivated the idea that negligible variance of the radiation field implies a large-occupation coherent final state, for which the field strength correlators factorise. Here, we motivate the same coherent description of the classical field, noting that it would be a mistake to impose coherence on the exact exclusive state, and we should instead construct an \textit{inclusive} coherent state that obeys the vanishing variance requirement and includes all the required cuts. The full one-point function can be constructed from the exclusive state as detailed above. After evaluating all Bopp shifts, $F[a^\dagger]$ contains only creation operators, and hence acting with the annihilation operator differentiates the radiation generator. We can therefore write the inclusive waveshape as
\[
\alpha_{I,\lambda}(k,\tilde z)
=
\left\langle
\frac{\delta F[a^\dagger]}
{\delta a_\lambda^\dagger(k)}
\right\rangle.
\label{eq:inclusive-waveshape-functional-derivative}
\]
In the deterministic classical limit the connected fluctuations are $\hbar$-suppressed, and we have factorisation of the expectation value
\[
\left\langle
\prod_{a=2}^{r}a_{\lambda_a}^\dagger(k_a)
\right\rangle
=
\prod_{a=2}^{r}\alpha_{I,\lambda_a}^*(k_a,\tilde z)
+
\text{subleading connected terms}.
\]
We see then that every radiation leg can be represented by the coherent c-number source $\alpha_I^*$. A connected $r$-radiation kernel has $r$ equivalent choices for its measured leg, so differentiating the factor $1/r!$ in the radiation generator gives $1/(r-1)!$. We find
\[
\alpha_{I,\lambda}(k,\tilde z)
={}&
\left.
\frac{\delta F[a^\dagger]}
{\delta a_\lambda^\dagger(k)}
\right|_{a^\dagger=\alpha_I^*}
\\
={}&
\sum_{r\geq1}
\frac1{(r-1)!}
\sum_{\lambda_2,\ldots,\lambda_r}
\int
\prod_{a=2}^{r}d\Phi(k_a)
\DC{r}_{\lambda\lambda_2\cdots\lambda_r}
(k,k_2,\ldots,k_r;\tilde z)
\prod_{a=2}^{r}\alpha_{I,\lambda_a}^*(k_a,\tilde z).
\label{eq:inclusive-coherent-source-relation}
\]
Writing the first terms explicitly,
\[
\alpha_{I,\lambda}(k,\tilde z)
={}&
\DC{1}_{\lambda}(k;\tilde z)
+
\sum_{\lambda_2}
\int d\Phi(k_2)\,
\DC{2,\lambda\lambda_2}(k,k_2;\tilde z)
\alpha_{I,\lambda_2}^*(k_2;\tilde z)
\\
&+
\frac12
\sum_{\lambda_2,\lambda_3}
\int d\Phi(k_2)d\Phi(k_3)\,
\DC{3}_{\lambda\lambda_2\lambda_3}
(k,k_2,k_3;\tilde z)
\alpha_{I,\lambda_2}^*(k_2;\tilde z)
\alpha_{I,\lambda_3}^*(k_3;\tilde z)
+\cdots.
\label{eq:inclusive-kernel-relation}
\]
Equation~\eqref{eq:inclusive-coherent-source-relation} is the leading-classical coherent-waveshape representation of the exact adjacent-multiplicity overlap in eq.~\eqref{eq:inclusive-amplitude-adjacent-multiplicity}. It does not discard the higher $C_r$: it leaves one leg measured and saturates every other leg with a waveshape $\alpha_I^*$. At the first perturbative iteration, the factors $\alpha_I^*$ on the right-hand side may be replaced by their leading values $\DC{1}^*$.

The scaling below shows that each of these kernels can contribute at leading classical order after all but one of its legs have been cut.

Using the KMOC scaling for the radiated modes gives \cite{Cristofoli:2021vyo}
\[
\alpha_{I,\lambda}(\hbar\bar k)
&\sim
\hbar^{-3/2}\alpha^{\rm cl}_{I,\lambda}(\bar k),
\qquad
\DC{1,\lambda}(\hbar\bar k)
\sim
\hbar^{-3/2}C^{\rm cl}_{1,\lambda}(\bar k) .
\label{eq:alpha-KMOC-scaling-new}
\]

With this convention, a connected exclusive correlator with $r$ outgoing radiation modes has the coefficient scaling
\[
\DC{r}(\hbar\bar k_1,\ldots,\hbar\bar k_r)
\sim
\hbar^{-1-r/2}C^{\rm cl}_{r}(\bar k_1,\ldots,\bar k_r) .
\label{eq:Cr-KMOC-scaling-new}
\]
Thus a directly measured connected $r$-point correlator scales as
\[
\left(\hbar^{3/2}\right)^r\DC{r}
\sim
\hbar^{r-1}.
\]
Any individual correlator with $r\geq2$ vanishes in the deterministic classical limit due to this scaling, however this suppression does not allow us to discard $\DC{2},\DC{3},\ldots$ until their contributions to the inclusive waveshape have been established, which, by scaling arguments, can enter classically as
\[
\int_{2\cdots r}\DC{r}(1,\ldots,r)
\prod_{a=2}^{r}\alpha_I^*(a)
&\sim
\hbar^{-1-r/2}
\left(\hbar^2\hbar^{-3/2}\right)^{r-1}
\nonumber\\
&\sim
\hbar^{-3/2}.
\]
A generic $r\geq 2$ connected correlator is therefore not an independent deterministic classical object when all of its legs are measured, but it can be a leading classical correction when all but one of its legs are cut against the radiation already present in the state.
In eq.~\eqref{eq:inclusive-coherent-source-relation}, this radiation is represented by the coherent source $\alpha_I^*$, so the higher exclusive kernels dress the classical field without becoming independent deterministic observables.

Once the cuts have determined the exact one-point function $\alpha_I$, we can construct a coherent state that reproduces every observable linear in $a$ and $a^\dagger$ exactly, although it need not reproduce higher correlators of the original exclusive state. In other words, we can write
\[
\ket{\varphi'} = e^{-N_I/2}e^{\alpha_Ia^\dagger}\ket{0},~~~~~\ket{\varphi} = \cl{N}e^{F[a^\dagger]}\ket{0},
\qquad N_I=|\alpha_I|^2,
\]
with
\[
\braket{\varphi'|\bb A_\mu|\varphi'}-\braket{\varphi|\bb A_\mu|\varphi} = 0
\]
for a field operator linear in $a$ and $a^\dagger$. If $\alpha_I$ itself is retained only to leading classical order, the corresponding mismatch is subleading at the same order.
Including the real elastic phase, we can write this coherent representative as

\[
\ket{\Psi_{\rm inc}} = \exp\!\left[\frac{i}{\hbar}\Re\chi\right]\exp\!\left[-\frac12N_I\right]\exp\!\left[\Sigma[\alpha_I,a^\dagger]\right]\ket{0},
\]

where we have defined
\[
\Sigma[\alpha_I,a^\dagger]
=
\sum_\lambda\int d\Phi(k)\,
\alpha_{I,\lambda}(k)a_\lambda^\dagger(k),
\qquad
N_I
=
\sum_\lambda\int d\Phi(k)\,
|\alpha_{I,\lambda}(k)|^2.
\]

The state need not reproduce the squeezed correlations, emission probabilities or radiation noise of the exclusive state. The relation to the exclusive normalisation is particularly transparent in the one-mode example below.
\subsection{Example: Squeezed Exclusive vs Coherent Inclusive}
Interactions which are nonlinear in the radiation field generate terms with arbitrary numbers of creation operators, so the exclusive state is generally not coherent. In scalar QED, for example, gauge invariance demands that the final state is squeezed due to the seagull interaction term, while for gravity the final state has arbitrarily many radiating gravitons. The coefficient of the single creation operator in that state is an \textit{exclusive} one-radiation kernel. The classical waveform is characterised by the complex mode function $\alpha_\lambda(k)=\langle a_\lambda(k)\rangle$ (a single complex number only in the one-mode example below), and it is equal to the exclusive one-radiation coefficient only when the relevant state is coherent or at leading order before cut corrections enter. If the state is not coherent, then in principle it can be related to all the higher order coefficients. Consider a squeezed state
\[\label{eq:squeezed-state-example-inclusive-vacuum}
\ket{\phi} = \cl{N}\exp\left[\alpha a^\dagger + \frac12\beta a^\dagger a^\dagger\right]\ket{0},~~~~~|\beta|<1.
\]
Then, 
\[
a\ket{\phi} = (\alpha+\beta a^\dagger)\ket{\phi}
\]
and therefore the expectation value of the annihilation operator is given by 
\[\label{eq:squeezed-state-waveshape}
\braket{a} = \alpha + \beta\braket{a^\dagger} = \frac{\alpha + \beta\alpha^*}{1-|\beta|^2}.
\]
If $\beta$ is of order $\hbar^0$, then the classical field depends on it, and what's more, we expect $\alpha$ to be proportional to a five-point amplitude and $\beta$ a six-point amplitude: the $\beta\alpha^*$ term is therefore a cut contribution, and the one-mode prototype of the $C_2C_1^*$ cut derived above. 
To see that the formulation above is indeed correct, consider the squeezed final state we began with in eq.~\eqref{eq:squeezed-state-example-inclusive-vacuum}, where the operator $F[a^\dagger]$ is given by
\[
F[a^\dagger] = \alpha a^\dagger + \frac12\beta a^\dagger a^\dagger.
\]
From eq. \eqref{eq:squeezed-state-waveshape}, the inclusive waveshape is given by
\[
\alpha_I = \frac{\alpha + \beta\alpha^*}{1-|\beta|^2},
\]
and a coherent state with parameter $\alpha_I$ reproduces this one-point field, although it does not reproduce the full exclusive squeezed state.

The imaginary part of the eikonal is given by
\[
\Im\chi = \frac\hbar 2\log\braket{0|e^{F^\dagger[a]}e^{F[a^\dagger]}|0} = \frac\hbar 2\log\braket{0|e^{\alpha^* a + \frac12\beta^* a a}e^{\alpha a^\dagger + \frac12\beta a^\dagger a^\dagger}|0},
\]
where the real part of the eikonal drops out of the expectation value. Using $[a,a^\dagger] = 1$, we can use the Baker-Campbell-Hausdorff formula to write
\[
\braket{0|e^{\alpha^* a + \frac12\beta^* a a}e^{\alpha a^\dagger + \frac12\beta a^\dagger a^\dagger}|0} = \frac{1}{\sqrt{1-|\beta|^2}}
\exp\!\left[
\frac{
|\alpha|^2+\frac12\beta\alpha^{*2}+\frac12\beta^*\alpha^2
}{
1-|\beta|^2
}\right].
\]

Taking the derivative, we find the expected result
\[
\alpha_I = \frac{2}{\hbar}\frac{\delta \Im \chi}{\delta \alpha^\star} = \frac{\alpha + \beta\alpha^*}{1-|\beta|^2}.
\]
Note that $\Im\chi$ normalises the exclusive squeezed state, while the coherent state with the same one-point function is instead normalised independently as $\ket{\alpha_I}=e^{-|\alpha_I|^2/2}e^{\alpha_Ia^\dagger}\ket{0}$, and does not inherit this exclusive imaginary eikonal normalisation.
\section{Gravity and Nonlinear Memory}
\label{sec:beyond-leading-order-applications}
\subsection{Gravitational Radiation from On-Shell Kernels}
We can now play the same game in gravity, albeit with more difficulty because of its nonlinear nature. As in the scalar QED case, the five-point amplitude combines both matter-line orderings but this time it comes with the two-graviton matter contact and the cubic-graviton contribution. Similarly, the full six-point amplitude contains the two-radiation contact terms and all factorisation channels before the connected subtractions are made.

We therefore use eqs.~\eqref{eq:partial-weyl-amplitude} and \eqref{eq:onshell-full-C1} directly. For one outgoing graviton, $S_{1,W,\lambda}$ is the partial matter-Weyl transform of the full five-point amplitude and
\[
\DC{1,\lambda}(k;\tilde z)
=
\left[
E_W^{-\star_m}
\star_m
S_{1,W,\lambda}(k)
\right](\tilde z).
\label{eq:gravity-C1-onshell-main}
\]
For two outgoing gravitons, $S_{2,W,\lambda\tau}$ is obtained in the same way from the full six-point amplitude, after which the elastic star inverse and the radiation star logarithm perform the necessary subtractions according to eq.~\eqref{eq:C2-radiative-subtraction}. In particular, the one- and two-radiation kernels required below are the connected partial Weyl transforms of the five- and six-point amplitudes, with all Bopp shifts already performed.

While it would certainly be interesting to construct the full, finite waveform at two loops, this is a considerable undertaking, and we will instead sketch the calculation of the nonlinear memory effect, for which we only really need the tree-level 5pt amplitude. Formally, the nonlinear part of the waveform at this order is given by
\[
\Delta\alpha_{I,\lambda}^{\rm NL}(q)
=
\sum_\tau\int d\Phi(k)\,
\Delta\DC{2,\lambda\tau}(q,k)\DC{1,\tau}^*(k),
\label{eq:memory-preview-cut-new1}
\]
where the other terms that contribute come from the linear memory, essentially the five-point two-loop amplitude in the soft limit, where radiation is emitted from the massive legs. The nonlinear-memory contribution considered here is the radiative soft-pole part of the six-point amplitude in eq.~\eqref{eq:memory-preview-cut-new1}. Its residue can therefore be reconstructed from the five-point amplitude and soft factor. On the radiative pole, the connected six-point kernel factorises into a product of the one-radiation kernel and a soft-factor
\[
\Delta\DC{2,\lambda\tau}(q,k)
\xrightarrow{q\to0}
-\frac{\kappa}{2}
\frac{
\epsilon_{\mu\nu}^{(\lambda)*}(q)k^\mu k^\nu
}{
k\cdot q+i\epsilon
}
\DC{1,\tau}^{(0)}(k)
+
O(q^0).
\label{eq:gravity-radiative-leg-soft-kernel-amplitude}
\]

\subsection{The Inclusive Waveshape and Nonlinear Gravitational Memory}
\label{subsec:nonlinear-memory}
The gravitational kernels above are precisely the place where nonlinear gravitational memory enters the inclusive waveshape. As recently noted in \cite{Georgoudis:2025vkk}, the relevant soft graviton contribution to the non-linear memory is not produced solely by an uncut five-point waveform, but requires a $2\rightarrow 4$ amplitude (the 6pt) to be included in a cut. In \cite{Georgoudis:2025vkk}, this term arises from a two-loop expansion of the waveform in terms of the $N$-operator, which naturally contains higher-order terms in $a^\dagger$ and thus corresponds to the \textit{exclusive} expansion we alluded to above. The crucial piece for the non-linear memory --- the cut term --- is automatically included in our formalism in eq.~\eqref{eq:memory-preview-cut-new1}.

\begin{figure}[H]
\centering
\begin{tikzpicture}[scale=0.84,transform shape]
\path[use as bounding box] (-4.9,-1.55) rectangle (4.9,3.15);
\begin{feynman}
\vertex (lai) at (-4.6,0.85);
\vertex (lbi) at (-4.6,-0.85);
\vertex[draw,ellipse,minimum width=1.45cm,minimum height=0.9cm,fill=white]
 (Jstar) at (-2.35,0) {$\cl J_{1}^*$};
\vertex (ca) at (0,0.7);
\vertex (cb) at (0,-0.7);
\vertex (ck) at (0,2.15);
\vertex (u) at (-1.25,1.75);
\vertex[dot] (v) at (1.25,1.75) {};
\vertex[draw,ellipse,minimum width=1.45cm,minimum height=0.9cm,fill=white]
 (J) at (2.35,0) {$\cl J_{1}$};
\vertex (rao) at (4.6,0.85);
\vertex (rbo) at (4.6,-0.85);
\vertex (q) at (3.25,2.9);
\diagram*{
(lai) -- [scalar, edge label=$a$] (Jstar),
(lbi) -- [scalar, edge label'=$b$] (Jstar),
(Jstar) -- [scalar] (ca) -- [scalar] (J),
(Jstar) -- [scalar] (cb) -- [scalar] (J),
(J) -- [scalar, edge label=$a$] (rao),
(J) -- [scalar, edge label'=$b$] (rbo),
(Jstar) -- [graviton, edge label={$k,\tau$}] (u),
(u) -- [graviton] (ck),
(J) -- [graviton, edge label'={$Q=q+k$}] (v),
(v) -- [graviton] (ck),
(v) -- [graviton, edge label={$q,\lambda$}] (q),
};
\draw[white,solid,line width=5.25pt] (0,-1.3) -- (0,2.65);
\draw[red,densely dashed,line width=0.8pt] (0,-1.3) -- (0,2.65);
\node at (-2.35,-1.3) {$\DC{1,\tau}^{(0)*}(k)$};
\node at (2.35,-1.3) {$\Delta\DC{2,\lambda\tau}(q,k)$};
\end{feynman}
\end{tikzpicture}
\caption{The $5{\rm pt}^*\times6{\rm pt}$ contribution to the nonlinear-memory cut. The dashed line places the common hard final state and the unobserved graviton $k$ on shell, while the soft graviton $q$ remains measured. The five-point blobs denote the complete connected current. The hard-matter sewing is implicit in the reduced phase-space kernels.}
\label{fig:gravity-five-six-cut}
\end{figure}
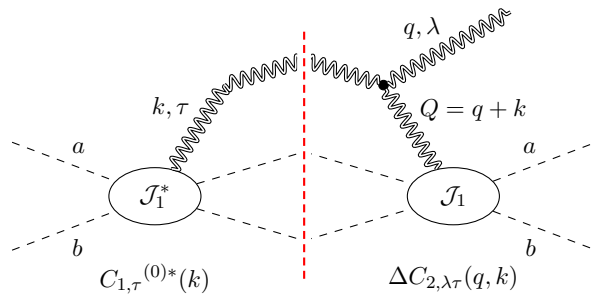

Specifically, the six-point contributes a measured soft graviton $q$ together with an unobserved hard graviton $k$ that is cut against the conjugate five-point amplitude.  These amplitudes are obtained from the adjacent-multiplicity cut formula: $\DC{1,\tau}^{(0)}(k)$ from $\cl A_5$ and $\Delta\DC{2,\lambda\tau}(q,k)$ from $\cl A_6$. This is the connected-kernel specialisation of the physical $5{\rm pt}^*\times6{\rm pt}$ cut shown in Fig.~\ref{fig:gravity-five-six-cut}, and gives the same $\DC{1}^*\DC{2}$ causal completion. Using the radiative-pole factorisation in eq.~\eqref{eq:gravity-radiative-leg-soft-kernel-amplitude}, we therefore find
\[
\Delta\alpha_{I,\lambda}^{\rm NL}(q)
=
-\frac{\kappa}{2}
\sum_\tau\int d\Phi(k)\,
\frac{\epsilon_{\mu\nu}^{(\lambda)*}(q)k^\mu k^\nu}{k\cdot q+i\epsilon}
\left|\DC{1,\tau}^{(0)}(k)\right|^2
+O(q^0).
\label{eq:memory-preview-result-new}
\]
Together with the inherited $i\epsilon$ prescription, eq. ~\eqref{eq:memory-preview-result-new} fixes the residue of the measured soft pole, rather than the complete finite-frequency cut. The residue fixes both the $\pi\delta(\omega)$ and $\operatorname{PV}(1/\omega)$ distributions (with sign determined by $i\epsilon$), but not the regular $O(\omega^0)$ six-point term, and therefore the complete finite-frequency waveform remains undetermined.
The corresponding cut is shown in Fig.~\ref{fig:gravity-five-six-cut}.
The result of ref.~\cite{Georgoudis:2025vkk} therefore provides an important confirmation that nonlinear memory is intrinsically an inclusive observable, and in particular that it cannot be obtained from the uncut
exclusive one-radiation kernel $\DC{1}$ alone. However, this does not require us to abandon the radiative eikonal or a coherent representation of the leading classical radiation field, rather it shows that the coherent displacement must be identified with the full inclusive waveshape $\alpha_I=\langle a\rangle$. It is important to point out, however, that the correct quantum description of the final state is the exclusive, noncoherent state, and that if we are interested in, for example, quantum observables such as the graviton noise as measured by a detector, then the exclusive state such as the one in \cite{Georgoudis:2025vkk} is the correct exact state to use. The coherent representative is relevant only for deterministic classical observables that can be derived from the one-point field.
\section{Quantum Geometry and Radiation Reaction}
\label{sec:coherent-state-geometry-radiation-reaction}

We have argued that the deterministic classical waveform can be represented by a coherent state in the radiation Hilbert space. However, we have also seen that this coherent state is not a fixed state in isolation, since it depends heavily on the dynamics of the matter sector, and so really we have identified a family of coherent states, whose waveshape depends on the hard momentum $p_i$, the impact parameter $x_\perp$, the masses/charges/spins etc. It is then natural to ask what the ordinary Hilbert-space overlap between neighbouring radiation states knows about the hard scattering problem. In other words, we are interested in understanding what happens to the state when we vary
\[
\ket{\psi(z)} \to \ket{\psi(z + \delta z)} 
\]
where $z$ are the phase-space parameters describing the matter scattering problem\footnote{This could be augmented to include charges, spin, mass etc --- we leave this idea to future work.} and $\ket{\psi(z)}$ is the coherent state with the eikonal part stripped off, i.e. it is $\ket{\alpha_I(z)}$. In other words, we are asking: given a scattering configuration $z^A$, a set of momentum and positions, say, which produces radiation with waveshape $\alpha_I(z)$, what happens to the radiation when we change the scattering configuration by a small amount $\delta z^A$?
We will again use $z^A$ to label the physical scattering phase-space, with derivatives taken while the constraints set out in appendix~\ref{app:off-shell-extensions} remain satisfied. 

Understanding how a state changes under a small change in parameters is a standard question in the geometry of quantum states \cite{Kibble:1979zz,Provost:1980nc,Bengtsson:2006rfv,Kolodrubetz:2017ofs,Hetenyi:2023hye}, and especially coherent states \cite{Glauber:1963tx,Sudarshan:1963ts}. Starting from a given coherent state, we can use the inner product to define a distance, a connection and a curvature on the space of parameters, usually defined in terms of the \textit{quantum geometric tensor} (QGT) \cite{Provost:1980nc}
\[
Q_{AB}
&\equiv
\braket{\partial_A\psi|
\left(1-\ket{\psi}\bra{\psi}\right)
|\partial_B\psi}
\\
&=
\braket{\partial_A\psi|\partial_B\psi}
-
\braket{\partial_A\psi|\psi}
\braket{\psi|\partial_B\psi}
\\
&=
g_{AB}
-
\frac{i}{2\hbar}\cl F_{AB}.
\]
The real, symmetric part of the tensor is known as the Fubini-Study metric \cite{Fubini:1904vd}, which defines a notion of distance on state space, while its imaginary antisymmetric part is the Berry curvature, which can be expressed in terms of the Berry connection \cite{Berry:1984jv,Simon:1983mh} $\cl{A}_A$ in the usual way
\[
\cl F_{AB} = \partial_A\cl{A}_B - \partial_B\cl{A}_A,
\qquad
\cl{A}_A = i\hbar\braket{\psi|\partial_A\psi}.
\label{eq:berry-curvature-connection}
\]
We note that the Berry connection is still a local function of the physical scattering data, and not itself a physical observable. When it contributes to the displacement $\Delta z_A$, we understand that it is projected against the sharply peaked Wigner function, as discussed in section \ref{subsec:phase-space-displacement}.

With $\cl{A}_A$ defined in this way, the Fubini-Study metric is given by
\[
g_{AB} = \gamma_{AB} - \frac{1}{\hbar^2}\cl{A}_A\cl{A}_B,
\] 
where $\gamma_{AB} = \Re\braket{\partial_A\psi|\partial_B\psi}$ is the \textit{naive} metric on state space. The Fubini-Study metric is invariant under a change of phase of the state (a gauge transformation), unlike the naive metric, while the Berry connection transforms as a gauge field. Notably, the transformation $\cl A_A \to \cl A_A + \partial_A\phi$ is equivalent to the phase transformation $\ket{\psi} \to e^{-i\phi/\hbar}\ket{\psi}$, which is equivalent to redefining the real part of the eikobal by $\phi$.

The Fubini-Study metric characterises the distinguishability or \textit{fidelity} of two nearby states, in our case two outgoing radiation states. The Berry connection measures the change in phase when comparing nearby states on phase-space. At a given point $z^A$ in phase space, consider two infinitesimal variations
\[\label{eq:infinitesimal-variations}
z^A \rightarrow z^A+u^A,
\qquad
z^A \rightarrow z^A+v^A.
\]
Positivity of the QGT implies \cite{Hetenyi:2023hye}
\[
\bigl(g_{AB}u^Au^B\bigr)\bigl(g_{CD}v^Cv^D\bigr)
-\bigl(g_{AB}u^Av^B\bigr)^2
\geq
\frac{1}{4\hbar^2}
\bigl(\cl{F}_{AB}u^Av^B\bigr)^2.
\label{eq:uncertainty-bound}
\]
This is the geometric form of the uncertainty relation. Its physical
meaning is that a non-zero Berry curvature requires the outgoing
radiation state to vary non-trivially in the corresponding two
phase-space directions. More precisely, the curvature is bounded by the
Fubini--Study area swept out by these variations.

The connection to radiation reaction arises because the Berry
connection contributes directly to the phase-space displacement,
\[
\Delta z_A=-\partial_A\Re\chi+\cl A_A.
\]
Taking the curl immediately gives
\[
\partial_A\Delta z_B-\partial_B\Delta z_A
=\cl F_{AB}.
\]
Hamiltonian evolution generates canonical transformations, or symplectomorphisms, which preserve the canonical antisymmetric pairing (the symplectic area) between any two infinitesimal variations of the phase-space data. This has been observed in the eikonal too \cite{Kim:2025ebl,Kim:2025sey}, and in the purely elastic case this means that $\cl F_{AB} = 0$, which follows directly from
\[
\Delta z_A=-\pd_A\Re\chi.
\]
Consider two infinitesimal variations $u^A$ and $v^A$ as in eq.~\eqref{eq:infinitesimal-variations}, which induce variations at the endpoints
\[
u_\pm^A
=
\frac{\pd z_{\pm}^A}{\pd z^B}u^B
=
u^A\pm\frac12u^B\pd_B\Delta z^A,
\]
and similarly for $v_\pm^A$. Plugging this into the canonical area form $\Omega_{AB}$ gives
\[
\begin{aligned}
\Omega_{AB}u_+^Av_+^B
-
\Omega_{AB}u_-^Av_-^B
&=
-u^Av^B
\left(
\pd_A\Delta z_B-
\pd_B\Delta z_A
\right)
\\
&=
-\cl F_{AB}u^Av^B.
\end{aligned}
\]
We see then that the massive particle scattering does not preserve the canonical area form on the massive-particle phase space when the Berry curvature is non-zero. We would like to interpret this as the symplectic area lost on the massive particle side shows up as an area gained by the radiative sector. To see this, we need to show that the right hand side above is indeed an area. We can see this by considering the fact that $\alpha_I$ is complex, and from its definition in eq. \eqref{eq:berry-curvature-connection}, we see that the curvature can be expressed as
\[
\cl{F}_{AB}u^Av^B &= i\hbar\sum_\lambda \int d\Phi(k)   
    \left[
        \pd_A\alpha^*_{I,\lambda}\pd_B\alpha_{I,\lambda}
        -
        \pd_B\alpha^*_{I,\lambda}\pd_A\alpha_{I,\lambda}
    \right]u^Av^B\\
    &= 2\hbar\sum_\lambda \int d\Phi(k)\left[\delta_v\Re\alpha_{I,\lambda}\delta_u\Im\alpha_{I,\lambda} - \delta_u\Re\alpha_{I,\lambda}\delta_v\Im\alpha_{I,\lambda}\right],
\label{eq:curvature-variation}
\]
where $\delta_u = u^A\pd_A$ and similarly for $\delta_v$.
This is indeed an area corresponding to the variations of the real and imaginary parts of the inclusive waveshape $\alpha_I$.

We can therefore interpret the result as a symplectic area balance law. The massive-particle sector considered alone does not preserve this area when radiation is present: the difference between the incoming and outgoing massive-particle phase-space areas is exactly balanced by the area spanned by the corresponding variations of the outgoing field. Intuitively, $u^A$ and $v^A$ define a small parallelogram of neighbouring massive-particle scattering configurations, and varying their directions produces a slightly different outgoing waveform. The two independent waveform variations span corresponding parallelograms in the complex planes of radiation modes, and the sum of their symplectic areas precisely accounts for the difference between the incoming and outgoing massive-particle areas. For gravity, the finite-frequency part of this balance is structurally analogous to the radiative phase space constructed in \cite{Ashtekar:1981bq} and can be described in terms of the symplectic flux through null infinity \cite{Wald:1999wa}. The regulated zero-frequency contribution indicates that the corresponding enlarged scattering phase space must also retain appropriate asymptotic or boundary-field data. It is also consistent with the radiation-eikonal construction of ref.~\cite{Kim:2025hpn}, where the radiative modes and their Poisson brackets are included explicitly in the scattering phase space. It would be interesting to enlarge our phase-space construction by treating the real and imaginary parts of the waveshape as explicit radiative phase-space coordinates with an associated Poisson bracket, although we will not pursue this here.

We can now revisit the uncertainty bound in eq.~\eqref{eq:uncertainty-bound}. Interestingly, we see that the magnitude of the canonical area transferred from the massive-particle sector to the outgoing field cannot exceed $2\hbar$ times the Fubini--Study area $A_{FS}$ spanned by the corresponding waveform variations. Defining $A_{FS}(u,v) = \sqrt{
\left(g_{AB}u^Au^B\right)
\left(g_{CD}v^Cv^D\right)
-
\left(g_{AB}u^Av^B\right)^2
}$, we have
\[
\left|
\Omega_{AB}u_+^Av_+^B
-
\Omega_{AB}u_-^Av_-^B
\right|
\leq
2\hbar A_{FS}(u,v)
~.
\]
or equivalently

\[
\left|
u^Av^B
\left(
\partial_A\Delta z_B-\partial_B\Delta z_A
\right)
\right|
\leq
2\hbar A_{FS}(u,v)
~.
\]
$A_{FS}$ is a measure of the distinguishability of the outgoing radiation states corresponding to the two infinitesimal variations $u^A$ and $v^A$. In essence, this inequality tells us that any particle dynamics that produces radiation that fails to preserve the canonical area of the massive particle phase space must produce two outgoing radiation states that have a minimum distinguishability. The more the massive-particle sector fails to preserve its canonical area, the more distinguishable the corresponding outgoing radiation states must be. This is a direct consequence of the uncertainty principle.

Let's now apply this construction to classical radiation. Let's consider the endpoint variables
\[
    z^A=(x_i,\tilde p_i),
\]
which play the role of our parameters, along with the inclusive coherent state expressed above
\[
    \ket{\Psi_{\rm inc}(z)}
    =
    \exp\!\left[
        \frac{i}{\hbar}\Re\chi(z)
    \right]
    \ket{\alpha_I(z)},
\]
where $\ket{\alpha_I(z)}$ denotes the separately normalised coherent state, including the factor $e^{-N_I(z)/2}$ defined above, rather than the exclusive no-emission factor $e^{-\Im\chi(z)/\hbar}$. We require that $N_I(z) < \infty$ to avoid IR divergences \cite{Prabhu:2022zcr}, and for the zero-frequency contribution this means we need to regulate the zero mode, as discussed in section~\ref{subsec:zero-frequency-dressings}. We take the following expressions with this zero-mode prescription, and in particular we perform the integrals at fixed causal $i\epsilon$, with $\varepsilon\rightarrow0^+$ taken only after the momentum integrals have been performed.

Here $\chi$ is the no-emission endpoint phase, using the radiative shorthand introduced in section~\ref{sec:radiation-inclusive-cuts}, and $\alpha_I$ is the inclusive one-point waveshape defined by eq.~\eqref{eq:inclusive-amplitude-adjacent-multiplicity}.
So: we have a classical waveform, represented by its inclusive coherent state, and so we should ask what the standard geometry of this family of states implies.

The overlap of two nearby radiation states can be found by Taylor expanding around $\delta z$, and is given by
\[
    \braket{\alpha_I(z)|\alpha_I(z+\delta z)}
    =
    \exp\!\left[
        -\frac{i}{\hbar}\cl A_A\,\delta z^A
        -\frac{i}{2\hbar}\pd_{(A}\cl A_{B)}\,\delta z^A\delta z^B
        -
        \frac12 g_{AB}\,\delta z^A\delta z^B
        +\cdots
    \right],
\]
where
\[
    \cl A_A
    =
    -\frac{\hbar}{2i}
    \sum_\lambda\int d\Phi(k)
    \left[
        \alpha^*_{I,\lambda}(k;z)\pd_A\alpha_{I,\lambda}(k;z)
        -
        \alpha_{I,\lambda}(k;z)\pd_A\alpha^*_{I,\lambda}(k;z)
    \right]
\label{eq:coherent-radiation-Berry-connection}
\]
and
\[
    g_{AB}
    =
    \Re
    \sum_\lambda\int d\Phi(k)\,
    \pd_A\alpha^*_{I,\lambda}(k;z)
    \pd_B\alpha_{I,\lambda}(k;z).
\]
For the zero-frequency contribution, we define the Berry connection by evaluating eq.~\eqref{eq:coherent-radiation-Berry-connection} with $\alpha_{{\rm ZF},\varepsilon}$ at fixed regulator and taking $\varepsilon\to0$ only after performing the momentum integral. The regulated observable has the finite limit computed below.

The first expression is the Berry connection of the coherent radiation, and the second expression is the distance on the space of outgoing radiation profiles. Roughly speaking, $g_{AB}$ measures how much the emitted waveform changes when we perturb the matter sector.

The curvature of the same connection is

\[
    \cl F_{AB}
    =
    \pd_A\cl A_B
    -
    \pd_B\cl A_A
    =
    i\hbar
    \sum_\lambda\int d\Phi(k)
    \left[
        \pd_A\alpha^*_{I,\lambda}\pd_B\alpha_{I,\lambda}
        -
        \pd_B\alpha^*_{I,\lambda}\pd_A\alpha_{I,\lambda}
    \right].
\]
For notational simplicity, we define the Berry connection as
\[
\cl A_{i,\mu} = \cl A_A\bigg|_{A = x_i^\mu},~~~~~\widetilde{\cl A}_{i}^{\mu} = \cl A_A\bigg|_{A = \tilde p_{i\mu}}.
\]
We first evaluate the impulse and displacement of the matter lines using the coherent state. Substituting $\ket{\Psi_{\rm inc}(z)}=e^{i\Re\chi(z)/\hbar}\ket{\alpha_I(z)}$ into the linear observable gives
\[
\Delta z_A^{\rm coh}
=
-\pd_A\Re\chi
+
\cl A_A.
\label{eq:coherent-lowered-displacement}
\]
We see then that the same Berry connection that compares neighbouring waveforms supplies the radiation-reaction contribution to the impulse and the displacement, at least with this phase convention. For $A=x_i^\mu$, the component relation below gives
\[
\Delta p_{i\mu}
=
\frac{\pd \Re\chi}{\pd x_i^\mu}
-
\cl A_{i,\mu},
\]
which is the impulse formula identified previously in ref.~\cite{Cristofoli:2021jas}. We can also define the asymptotic displacement of the matter lines as
\[
\Delta x_{i}^\mu
=
-\frac{\pd \Re\chi}{\pd \tilde p_{i\mu}}
+
\widetilde{\cl A}_{i}^\mu.
\]

\subsection{From Vanishing Variance to Radiation-State Geometry}
Vanishing variance suggests a coherent representative is appropriate for a deterministic radiation field, but it is clear that the coherent state must be non-trivially dependent on the matter phase-space scattering data. In ref.~\cite{Cristofoli:2021jas}, our coherent-state proposal contains a complex eikonal $\chi$ and a waveshape $\alpha$. Crucially, our minimal proposal assumes that the waveshape doesn't depend on the transfer momentum $q$, which we state explicitly, noting that this could be considered if desired, but was not required for the observables we considered. We keep that coherent-state description here, but with a slightly different state, whose waveshape is determined explicitly and can depend on $q$. 

The phase-space displacement is obtained by differentiating the coherent state, given in eq.~\eqref{eq:coherent-lowered-displacement}, repeated here for convenience:
\[
\Delta z_A
=
-\pd_A\Re\chi
+
\cl A_A.
\]
Although the split between $-\pd_A\Re\chi$ and $\cl A_A$ depends on the phase convention for $\ket{\alpha_I(z)}$, their sum is invariant. Crucially, though, the position shift contains a contribution from the Berry connection too, which is not present in our earlier proposal. This is because the earlier proposal assumed that the waveshape $\alpha$ was independent of the transfer momentum $q$, and therefore that there is no component of the Berry connection along the $q$-direction. In general, however, the inclusive waveshape depends on the complete endpoint data, and so the Berry connection contributes to the position shift.

This distinction is especially clear for the momentum-transfer derivative used in ref.~\cite{Cristofoli:2021jas}. With
\[
q^\mu=p_1'{}^\mu-p_1^\mu
=-\left(p_2'{}^\mu-p_2^\mu\right),
\qquad
\widetilde p_1=p_1'-\frac q2,
\qquad
\widetilde p_2=p_2'+\frac q2,
\]
the derivative at fixed outgoing momenta obeys
\[
\left.\frac{\pd}{\pd q_\mu}\right|_{p_1',p_2'}
=
-\frac12\frac{\pd}{\pd\widetilde p_{1\mu}}
+
\frac12\frac{\pd}{\pd\widetilde p_{2\mu}},
\]
and hence eq.~\eqref{eq:coherent-lowered-displacement} gives
\[
\frac12
\left(
\Delta x_2^\mu-\Delta x_1^\mu
\right)
=
-\frac{\pd\Re\chi}{\pd q_\mu}
+
\cl A_q^\mu,
\qquad
\cl A_q^\mu
=
\frac12
\left(
\widetilde{\cl A}_{2}^\mu
-
\widetilde{\cl A}_{1}^\mu
\right).
\]
The momentum $q$ transfers momentum from one worldline to the other, and so varying it moves the two midpoint momenta in opposite directions, each time by $\pm \frac12\delta q$ in midpoint variables. The coordinate conjugate to $q$ is therefore not an individual position shift, but half the change in the relative separation. This means that the displacement $x-b$ appearing in the stationary-phase analysis of ref.~\cite{Cristofoli:2021jas} is
\[
x^\mu-b^\mu=\frac12\left(\Delta x_2^\mu-\Delta x_1^\mu\right).
\label{eq:coherent-relative-displacement}
\]
The earlier stationary-phase relation is recovered in the conservative limit, or whenever the coherent waveshape has no dependence along the $q$-direction, so that $\cl A_q^\mu=0$.

We stress that $\cl A_A$ is the Berry connection of the inclusive coherent state, rather than that of the exact outgoing state, which is generally noncoherent and need not have the same quantum geometry. Within this coherent description, the curl of the phase-space displacement is
\[
\partial_A\Delta z_B
-
\partial_B\Delta z_A
=
\cl F^{}_{AB}.
\]
The curvature therefore measures the local obstruction to expressing the coherent-representative displacement as the gradient of a single phase, i.e. $\Delta z_A = \pd_A \varphi$. However, the converse is not necessarily true, since it is easy to imagine a nonzero waveform that gives rise to a vanishing curvature. For example, since the derivatives of $\cl A$ are taken with respect to $z^A$, we could imagine a waveform that depends only on a single combination of the scattering parameters, e.g. $\alpha_I(k;z)=\alpha_I(k;x_1\cdot p_2)$, which would give a vanishing curvature even though the waveform is nonzero.
\subsection{Momentum Balance}
Once we have radiation, momentum can be transferred from the hard particles to the radiation sector. Define the radiated momentum operator by
\[
\widehat P_{\rm rad}^\mu
\equiv
\sum_\lambda\int d\Phi(k)\,
k^\mu a_\lambda^\dagger(k)a_\lambda(k).
\]
Its expectation value in the exact normalised outgoing state, denoted by $\langle\cdot\rangle_F$, separates into mean-field and connected parts:
\[
P_{{\rm rad},F}^\mu
&\equiv
\sum_\lambda\int d\Phi(k)\,k^\mu
\left\langle
a_\lambda^\dagger(k)a_\lambda(k)
\right\rangle_F
\\
&=
\sum_\lambda\int d\Phi(k)\,k^\mu
\left[
|\alpha_{I,\lambda}(k;z)|^2
+
\rho_{{\rm conn},\lambda}(k;z)
\right],
\\
\rho_{{\rm conn},\lambda}(k;z)
&\equiv
\left\langle
\delta a_\lambda^\dagger(k)\delta a_\lambda(k)
\right\rangle_F,
\qquad
\delta a_\lambda(k)
\equiv
a_\lambda(k)-\alpha_{I,\lambda}(k;z).
\label{eq:exact-noncoherent-radiated-momentum}
\]
The second equality follows from $\alpha_{I,\lambda}=\langle a_\lambda\rangle_F$ and is exact: by definition $\langle\delta a_\lambda\rangle_F=0$, so the cross terms in $\langle(\alpha_I^*+\delta a^\dagger)(\alpha_I+\delta a)\rangle_F$ vanish. The connected two-point function $\rho_{\rm conn}$ need not vanish in the exact noncoherent state and is only suppressed in the deterministic classical limit.

For the coherent representative, $\rho_{\rm conn}=0$, and hence
\[
P_{\rm rad}^\mu[\alpha_I]
=
\sum_\lambda\int d\Phi(k)\,
k^\mu|\alpha_{I,\lambda}(k;z)|^2
=
\eta^{\mu\nu}\sum_i\cl A_{i,\nu}^{\rm rad}.
\]

Translation invariance of the full $S$-matrix implies that
\[
\sum_i\partial_{i}^\mu\!\left(S_W\ket0\right)
=
-\frac{i}{\hbar}\widehat P_{{\rm rad}}^\mu S_W\ket0,
\qquad
\sum_i\partial_{i}^\mu\!\left(\bra0 S_W^\dagger\right)
=
\frac{i}{\hbar}\bra0 S_W^\dagger\widehat P_{{\rm rad},}^\mu.
\label{eq:radiation-translation-covariance-operator}
\]
Substituting these identities into the exact displacement formula in eq.~\eqref{eq:exact-lowered-phase-space-displacement} gives
\[
\eta^{\mu\nu}\sum_i\Delta z_{x_i^\nu}
=
\frac{1}{\cl N}
\langle0|
S_W^\dagger\star_m
\widehat P_{\rm rad}^\mu
\star_m S_W
|0\rangle
=
P_{{\rm rad},F}^\mu.
\]
Using $\Delta p_i^\mu=-\eta^{\mu\nu}\Delta z_{x_i^\nu}$ therefore gives the exact balance law
\[
P_{{\rm rad},F}^\mu
=
\eta^{\mu\nu}\sum_i\Delta z_{x_i^\nu}
=
-\sum_i\Delta p_i^\mu,
\label{eq:exact-radiated-momentum-balance}
\]
without assuming coherence. In the deterministic classical limit where $\rho_{\rm conn}$ is suppressed, the exact radiation momentum reduces to the coherent expression in terms of $|\alpha_I|^2$. The radiation reaction is then the Berry connection written in local endpoint coordinates, while the classically radiated momentum is the momentum carried by the coherent state.

\section{Static Berry Connections and Position Shifts}
\label{sec:zero-frequency-berry}
\subsection[The Leading Scalar-QED Static Boundary Connection and Position Shift]{The Leading Scalar-QED Berry Connection and Position Shift}
\label{subsec:leading-radiation-berry-connection}
Let's now compute the Berry connection at leading order in scalar QED before using it to construct some observables. Recall that the inclusive waveshape is defined as
\[
\alpha_{I,\lambda}(k,\tilde z)
=
\frac{
\langle0|S_W^\dagger(\tilde z)\star_m a_\lambda(k)\star_m S_W(\tilde z)|0\rangle
}{
\langle0|S_W^\dagger(\tilde z)\star_m S_W(\tilde z)|0\rangle
} = E_W^{-\star_m}\star_m S_{1,W} + \cl{O}(e^5),
\]
As we saw in section \ref{subsec:zero-frequency-dressings}, the leading expansion of the inclusive waveshape is the zero-frequency term that arises from the straight-line dressing. That said, we shouldn't expect the static terms to contribute on their own to the Berry connection, so we will expand the waveshape to the next order in the coupling, which is the leading regular term.
The first terms in its expansion are
\[
E_W^{-\star_m}
&=
1-\cl E_0+\cl O(e^4),
&
S_{1,W,\lambda}
&=
S_{1,W,\lambda}^{\rm ZF}
+S_{1,W,\lambda}^{(0)}
+\cl O(e^5),
\]
and hence
\[
\alpha_{I,\lambda}^{\rm ZF}(k)
&=
S_{1,W,\lambda}^{\rm ZF}(k),
\\
\alpha_{I,\lambda,0}(k)
&=
S_{1,W,\lambda}^{(0)}(k)
-\cl E_0\star_m S_{1,W,\lambda}^{\rm ZF}(k).
\]
Here $S_{1,W,\lambda}^{\rm ZF}=\cl O(e)$ is the straight-line dressing, while $\cl E_0=\cl O(e^2)$ and $S_{1,W,\lambda}^{(0)}=\cl O(e^3)$ are the leading elastic and one-radiation coefficients found in appendix~\ref{sec:radiation-from-star-ordering}. The subtraction removes the four-point--three-point iteration term from the waveshape.

In section \ref{subsec:zero-frequency-dressings}, we discussed the fact that $\alpha_{I,\lambda}^{\rm ZF*}(k)$ was supported at zero frequency, and hence represents the static field. The first non-vanishing contribution to the Berry connection is the interference between the static field and the leading regular term $\alpha_{I,\lambda,0}(k)$
\[
\cl A_A
=
\frac{i\hbar}{2}
\sum_\lambda\int d\Phi(k)
\left[
\alpha_{I,\lambda}^{\rm ZF*}(k)
\overleftrightarrow{\pd}_A
\alpha_{I,\lambda,0}(k)
+
\alpha_{I,\lambda,0}^*(k)
\overleftrightarrow{\pd}_A
\alpha_{I,\lambda}^{\rm ZF}(k)
\right] + \cl O(e^6).
\]
To evaluate it, we write the waveshape in terms of its sources
\[
\alpha_{I,\lambda}^{\rm ZF}(k)
=
\frac{1}{\sqrt{\hbar}}\epsilon_{\lambda\rho}(k)F_{\rm ZF}^{\rho}(k),
\qquad
\alpha_{I,\lambda,0}(k)
=
\frac{1}{\sqrt{\hbar}}\epsilon_{\lambda\rho}(k)F_0^{\rho}(k).
\]
The zero-mode waveshape has to be handled with some care in both gauge theory and gravity. With the waveshape written as above, the Berry connection contains the full state sum over physical helicities $\lambda = \pm$, while the zero-mode profile is supported at $\omega=0$. The physical polarisation vectors (tensors in gravity) don't really make sense for $k^\mu = 0$, and so we need to be careful how we treat this. The strategy we will adopt is to use the completeness relation for the physical polarisations at non-zero $k$, and only \textit{then} introduce the causal $i\varepsilon$ prescription for the sources. This is consistent with the fact that without the $i\varepsilon$ prescription, the Ward identity holds and the sources are conserved, whereas the finite-$\varepsilon$ representatives are not exactly conserved. This holds in gravity as well, where the prescription is equivalent to picking out the de Donder numerator in the Berry connection, rather than using the full TT-projector. This is related to the distinction discussed in ref.~\cite{Heissenberg:2024umh}, where the TT-projected static expression excludes the Coulombic part of the field. With this in mind, we can write the zero-mode source as
\[
F_{{\rm ZF},\varepsilon}^{\rho}(k)
=
\sum_i e_i
\left[
\frac{\widetilde p_i^\rho}{\widetilde p_i\cdot k-i\varepsilon}
-
\frac{\widetilde p_i^\rho}{\widetilde p_i\cdot k+i\varepsilon}
\right].
\]
The source then obeys the Ward identity in the $\varepsilon\rightarrow0$ limit,
\[
k_\rho F_{{\rm ZF},\varepsilon}^{\rho}(k)
\longrightarrow 0,
\qquad
k_\rho\pd_A F_{{\rm ZF},\varepsilon}^{\rho}(k)
\longrightarrow 0,
\qquad
\varepsilon\rightarrow0.
\]
Here $A$ differentiates only the hard scattering data, and not the soft momentum $k$, and the Ward identities are satisfied due to charge conservation as usual.

Its antisymmetric momentum derivative vanishes,
\[
F_{{\rm ZF},\rho}^*
\overleftrightarrow{\pd}_{\widetilde p_i^\mu}
F_{\rm ZF}^{\rho}=0,
\]
so the straight-line source by itself gives no finite Berry connection as expected, and we can therefore write
\[
\cl A_A
=
-\frac{i}{2}
\int d\Phi(k)
\left[
F_{{\rm ZF},\rho}^*(k)
\overleftrightarrow{\pd}_A
F_0^\rho(k)
+
F_{0,\rho}^*(k)
\overleftrightarrow{\pd}_A
F_{\rm ZF}^\rho(k)
\right] + \cl O(e^6).
\]

The leading regular source follows from the pole term in eq.~\eqref{eq:onshell-sqed-full-C1}, with the elastic iteration removed by eq.~\eqref{eq:onshell-full-C1}. The action of the Bopp shifts in eq.~\eqref{eq:C1-bopp-momentum-routing-endpoint} then gives
\[
F_0^\rho(k)
={}&
-\frac12\sum_i e_i
\left[
\frac{\Delta p_i^\rho}
{\widetilde p_i\cdot k-i\varepsilon}
+
\frac{\Delta p_i^\rho}
{\widetilde p_i\cdot k+i\varepsilon}
\right.
\left.
-
\frac{\widetilde p_i^\rho(\Delta p_i\cdot k)}
{(\widetilde p_i\cdot k-i\varepsilon)^2}
-
\frac{\widetilde p_i^\rho(\Delta p_i\cdot k)}
{(\widetilde p_i\cdot k+i\varepsilon)^2}
\right].
\]
The first two terms come from the soft expansion of the numerator, while the double poles come from expanding the denominator. The scalar-QED seagull is regular as $k\rightarrow0$ and does not contribute to this leading soft term.

Using the endpoint momentum notation
\[
p_i^\pm=\widetilde p_i\pm\frac12\Delta p_i,
\]
we can combine both terms to first order in the impulse, to find
\[
F_\varepsilon^\rho(k)
\equiv
F_{{\rm ZF},\varepsilon}^\rho(k)+F_0^\rho(k)
=
\sum_i e_i
\left[
\frac{p_i^{-\rho}}{p_i^-\cdot k-i\varepsilon}
-
\frac{p_i^{+\rho}}{p_i^+\cdot k+i\varepsilon}
\right]
+\cl O(\Delta p^2).
\]
Since the $F_{\rm ZF}$ with itself vanishes and the $F_0$ with itself begins at higher order, the interference term can be written
\[
F_{{\rm ZF},\rho}^*
\overleftrightarrow{\pd}_A F_0^\rho
+
F_{0,\rho}^*
\overleftrightarrow{\pd}_A F_{\rm ZF}^\rho
=
F_{\varepsilon,\rho}^*
\overleftrightarrow{\pd}_A F_\varepsilon^\rho
 + \cl{O}(\delta p^2)
\]

The same-side incoming--incoming and outgoing--outgoing terms cancel in the antisymmetric derivative, leaving only the mixed terms involving both $p_a^+$ and $p_b^-$. By chain rule, we can separate the numerator and denominator derivative contributions, with the numerator contribution being
\[
\begin{aligned}
\left[
F_{\varepsilon,\rho}^*
\overleftrightarrow{\pd}_A
F_\varepsilon^\rho
\right]_{ab}^{\rm num}
={}&
e_ae_b
\left(
p_b^-\cdot\pd_Ap_a^+
-p_a^+\cdot\pd_Ap_b^-
\right)
\\
&\times
\left[
\frac{1}
{(p_a^+\cdot k-i\varepsilon)
 (p_b^-\cdot k-i\varepsilon)}
-
\frac{1}
{(p_a^+\cdot k+i\varepsilon)
 (p_b^-\cdot k+i\varepsilon)}
\right],
\end{aligned}
\]
while the pole derivatives give
\[
\begin{aligned}
\left[
F_{\varepsilon,\rho}^*
\overleftrightarrow{\pd}_A
F_\varepsilon^\rho
\right]&_{ab}^{\rm denom}
=
e_ae_b(p_a^+\cdot p_b^-)\\
&\times\bigg\{
(\pd_Ap_b^-)\cdot k
\left[
\frac{1}
{(p_a^+\cdot k-i\varepsilon)
 (p_b^-\cdot k-i\varepsilon)^2}
-
\frac{1}
{(p_a^+\cdot k+i\varepsilon)
 (p_b^-\cdot k+i\varepsilon)^2}
\right]
\\&-(\pd_Ap_a^+)\cdot k
\left[
\frac{1}
{(p_a^+\cdot k-i\varepsilon)^2
 (p_b^-\cdot k-i\varepsilon)}
-
\frac{1}
{(p_a^+\cdot k+i\varepsilon)^2
 (p_b^-\cdot k+i\varepsilon)}
\right]
\bigg\}.
\end{aligned}
\]
The full mixed term is the sum of these two contributions.

We now set $k=\omega n$ and use
\[
\int d\Phi(k)
=
\frac{1}{2(2\pi)^3}
\int_0^\Lambda d\omega\,\omega
\int d\Omega_n.
\]
It is useful to define
\[
\lambda_a=p_a^+\cdot n,
\qquad
\lambda_b=p_b^-\cdot n.
\]
Both radial integrations follow from the identity
\[
\cl I_{r,s}
\equiv
\lim_{\varepsilon\rightarrow0^+}
\int_0^\Lambda d\omega\,\omega^{r+s-1}
\left[
\frac{1}
{(\omega\lambda_a-i\varepsilon)^r
 (\omega\lambda_b-i\varepsilon)^s}
-
\frac{1}
{(\omega\lambda_a+i\varepsilon)^r
 (\omega\lambda_b+i\varepsilon)^s}
\right]
=
\frac{i\pi}{\lambda_a^r\lambda_b^s}.
\]
The numerator and denominator parts therefore become
\[
\begin{aligned}
\left.\cl A_{A,ab}^{\rm soft}\right|_{\rm num}
={}&
-\frac{i e_ae_b}{4(2\pi)^3}
\int d\Omega_n
\left(
p_b^-\cdot\pd_Ap_a^+
-p_a^+\cdot\pd_Ap_b^-
\right)
\cl I_{1,1},
\\
\left.\cl A_{A,ab}^{\rm soft}\right|_{\rm denom}
={}&
-\frac{i e_ae_b}{4(2\pi)^3}
\int d\Omega_n\,(p_a^+\cdot p_b^-)
\left[
(\pd_Ap_b^-)\cdot n\,\cl I_{1,2}
-(\pd_Ap_a^+)\cdot n\,\cl I_{2,1}
\right].
\end{aligned}
\]
After the radial integrations, their sum is
\[
\begin{aligned}
\cl A_{A,ab}^{\rm soft}
=
\frac{e_ae_b}{32\pi^2}
\int d\Omega_n\bigg[
&\frac{
p_b^-\cdot\pd_Ap_a^+
-p_a^+\cdot\pd_Ap_b^-}
{\lambda_a\lambda_b}
+(p_a^+\cdot p_b^-)
\left(
\frac{(\pd_Ap_b^-)\cdot n}
{\lambda_a\lambda_b^2}
-
\frac{(\pd_Ap_a^+)\cdot n}
{\lambda_a^2\lambda_b}
\right)
\bigg].
\end{aligned}
\]

The remaining angular integral is evaluated as
\[
\int d\Omega_n\,
\frac{1}{\lambda_a\lambda_b}
=
\frac{4\pi}{m_am_b}\Delta(\Gamma_{ab}),
\qquad
\Gamma_{ab}
\equiv
\frac{p_a^+\cdot p_b^-}{m_am_b},
\qquad
\Delta(\Gamma)
\equiv
\frac{\operatorname{arccosh}\Gamma}
{\sqrt{\Gamma^2-1}}.
\label{eq:angular-integral-Delta}
\]
Differentiating it with respect to the endpoint momenta gives
\[
\begin{aligned}
\int d\Omega_n\,
\frac{n^\rho}{\lambda_a^2\lambda_b}
&=
-\frac{4\pi}{m_a^2m_b^2}
\Delta'(\Gamma_{ab})p_b^{-\rho},
\\
\int d\Omega_n\,
\frac{n^\rho}{\lambda_a\lambda_b^2}
&=
-\frac{4\pi}{m_a^2m_b^2}
\Delta'(\Gamma_{ab})p_a^{+\rho}.
\end{aligned}
\]
The numerator and pole derivatives consequently combine into
\[
\cl A_A^{\rm soft}
=
\frac{1}{8\pi}
\sum_{a,b}\frac{e_ae_b}{m_am_b}
B(\Gamma_{ab})
\left(
p_b^-\cdot\pd_Ap_a^+
-p_a^+\cdot\pd_Ap_b^-
\right)
+\cl O(e^6),
\]
where we define
\[
B(\Gamma)
=
\frac{\Gamma}{\Gamma^2-1}
-
\frac{\operatorname{arccosh}\Gamma}{(\Gamma^2-1)^{3/2}},
\qquad
B(1)=\frac23.
\]
The part independent of the impulse is antisymmetric under $a\leftrightarrow b$ and cancels in the double sum. Defining
\[
\gamma_{ab}
\equiv
\frac{\widetilde p_a\cdot\widetilde p_b}{m_am_b},
\qquad
\delta\gamma_{ab}
\equiv
\frac{
\Delta p_a\cdot\widetilde p_b
-\widetilde p_a\cdot\Delta p_b}
{2m_am_b},
\]
and keeping the term linear in $\Delta p_i$, we finally obtain
\[
\begin{aligned}
\cl A_A
=
\frac{1}{8\pi}
\sum_{a,b}\frac{e_ae_b}{m_am_b}
\bigg[
&B(\gamma_{ab})
\left(
\widetilde p_b\cdot\pd_A\Delta p_a
-\Delta p_b\cdot\pd_A\widetilde p_a
\right)
\\
&+B'(\gamma_{ab})\delta\gamma_{ab}
\left(
\widetilde p_b\cdot\pd_A\widetilde p_a
-\widetilde p_a\cdot\pd_A\widetilde p_b
\right)
\bigg] + \cl O(e^6).
\end{aligned}
\label{eq:leading-radiation-berry-connection}
\]
The momentum part, choosing $A = x_i^\nu$, is given by 
\[
A_{i,\mathrm{rad}}^{\nu} &=
\frac{1}{8\pi}
\sum_{a,b}\frac{e_ae_b}{m_am_b}
\bigg[
B(\gamma_{ab})
\widetilde p_{b\rho}\frac{\pd \Delta p_a^\rho}{\pd\tilde x_{i\nu}}\bigg] + \cl O(e^6)\\
&=  \cl O(e^6),
\]

where we used the fact that the on-shell and soft conditions set $p_a\cdot\Delta p_b= \cl{O}(e^4)$, and hence $\widetilde p_{b\rho}\pd \Delta p_a^\rho/\pd\tilde x_{i\nu}=\cl{O}(e^4)$. The position part is chosen analogously by picking $A = p_i^\nu$, and is given by
\[
\widetilde A_{i,\mathrm{rad}}^{\nu} &=
\frac{1}{8\pi}
\sum_{a,b}\frac{e_ae_b}{m_am_b}
\bigg[
B(\gamma_{ab})
\left(
\widetilde p_{b\rho}\frac{\pd \Delta p_a^\rho}{\pd\tilde p_{i\nu}}
-\Delta p_b^\nu\delta_{ia}
\right)
\bigg] + \cl O(e^6)\\
&= -\frac{1}{4\pi}
\sum_{a,b}\frac{e_ae_b}{m_am_b}
B(\gamma_{ab})
\Delta p_b^\nu\delta_{ia} + \cl O(e^6)\\
&= -\frac{1}{4\pi}
\left[\frac23\frac{e_i^2}{m_i^2}\Delta p_i^\nu  + \frac{e_ie_j}{m_im_j}
B(\gamma_{ij})\Delta p_j^\nu 
\right]+ \cl O(e^6), ~~~~~j\neq i
\]
where we have used the fact that $\delta\gamma_{ab}=0$ for soft two-body scattering and again used the on-shell condition.

This result tells us two interesting things. First, it tells us that there is no radiation-reaction contribution to the impulse at $\cl{O}(e^4)$ and, second, that there \textit{is} a contribution to the position shift at this order:
\[
\Delta x_{i,\rm ZF}^{(4)\mu}
=
\frac{1}{4\pi}
\left[
-\frac{2e_i^2}{3m_i^2}
+\frac{e_ie_j}{m_im_j}B(\gamma)
\right]
\Delta p_i^\mu,
\qquad
\gamma=\frac{\widetilde p_1\cdot\widetilde p_2}{m_1m_2}.
\label{eq:zero-frequency-position-shift-rutherford}
\]
The self term is the endpoint displacement produced by the total-derivative part of the Lorentz--Dirac force, while the mixed term comes from the retarded interaction between the two worldlines. Both terms are contained in the second-order classical trajectories evaluated in ref.~\cite{Saketh:2021sri}.

This result has precisely the amplitude structure found by Higuchi and Martin in ref.~\cite{Higuchi:2006wu}. The position shift of a charged scalar wavepacket is written in their notation as
\[
\Delta x_i^\mu
=
-\frac{\pd\Re\cl F_i}{\pd p_{i\mu}}
-\frac{i}{2}
\int d\Phi(k)\,
\cl M_{i,\rho}^*(p_i,k)
\overleftrightarrow{\pd}_{p_{i\mu}}
\cl M_i^\rho(p_i,k),
\]
where $\cl F_a$ is the forward-scattering amplitude and $\cl M_a^\rho$ is the one-photon emission amplitude. In our notation, $\hbar\Re\cl F_i$ is the real zero-radiation phase, while the physical polarisation sum turns the second term into $\cl A_{\widetilde p_i^\mu}$. Their emission and forward-scattering contributions are therefore respectively the two terms in our position-shift formula. Higuchi and Martin further show that their sum agrees with the position shift obtained from the classical Lorentz--Dirac force. Equation~\eqref{eq:zero-frequency-position-shift-rutherford} is the covariant two-body version of the same statement.

\subsection{The Leading Gravitational Berry Connection and Position Shift}
\label{subsec:leading-gravitational-radiation-berry-connection}
We can now repeat the same calculation for gravity. We take $g_{\mu\nu}=\eta_{\mu\nu}+\kappa h_{\mu\nu}$ with $\kappa^2=32\pi G$, and use de Donder gauge. The expansion of the inclusive waveshape, the endpoint momenta, the regulated master integral and the angular identities are those collected in the previous section, so the only new ingredients are the tensor numerator and the spin-two state sum.  As discussed in the previous section, our use of the de Donder contraction does rather more than fix the gauge: it retains the Coulombic components of the gravitational field which are removed by the usual transverse--traceless projection. This reflects the broader distinction between radiative and Coulombic data discussed in ref.~\cite{1707.09914,Heissenberg:2024umh}. 

Soft gravitational radiation has been studied from several complementary points of view \cite{Weinberg:1965nx,Smarr:1977fy,Kovacs:1978eu,Gruzinov:2014nma,Addazi:2019mjh}. The zero-frequency limit was originally studied by Smarr \cite{Smarr:1977fy} and more modern treatments have since followed \cite{Ciafaloni:2018uwe,Gruzinov:2014nma,Addazi:2019mjh,Mougiakakos:2021ckm,DiVecchia:2022nna,DiVecchia:2022owy,DiVecchia:2022piu,Alessio:2022kwv,Heissenberg:2024umh}. This builds on puzzles surrounding radiation and its BMS interpretation \cite{Veneziano:2022zwh,Christodoulou:1991cr,Thorne:1992sdb,Favata:2010zu,Strominger:2014pwa}, and classical soft-theorem calculations \cite{Laddha:2018myi,Saha:2019tub,DiVecchia:2021ndb,DiVecchia:2022owy,Georgoudis:2023eke}.

The regulated straight-line source is obtained from the expansion of the inclusive gravitational waveshape
\[
F_{{\rm ZF},\varepsilon}^{\mu\nu}(k)
=
\frac{\kappa}{2}\sum_i\widetilde p_i^\mu\widetilde p_i^\nu
\left[
\frac{1}{\widetilde p_i\cdot k-i\varepsilon}
-
\frac{1}{\widetilde p_i\cdot k+i\varepsilon}
\right].
\]
The leading soft source, expanded to first order in the impulse, is
\[
F_0^{\mu\nu}(k)
={}&
-\frac{\kappa}{4}\sum_i
\bigg\{
\left(
\widetilde p_i^\mu\Delta p_i^\nu
+
\Delta p_i^\mu\widetilde p_i^\nu
\right)
\left[
\frac{1}{\widetilde p_i\cdot k-i\varepsilon}
+
\frac{1}{\widetilde p_i\cdot k+i\varepsilon}
\right]
\\
&\hspace{1.2cm}
-\widetilde p_i^\mu\widetilde p_i^\nu
(\Delta p_i\cdot k)
\left[
\frac{1}{(\widetilde p_i\cdot k-i\varepsilon)^2}
+
\frac{1}{(\widetilde p_i\cdot k+i\varepsilon)^2}
\right]
\bigg\}.
\]
The matter contact and cubic-graviton pieces of the current are regular as $k\rightarrow0$, so they do not change this leading pole, analogously to the seagull in scalar QED. Combining the two sources using the endpoint momenta gives
\[
F_\varepsilon^{\mu\nu}(k)
=
\frac{\kappa}{2}\sum_i
\left[
\frac{p_i^{-\mu}p_i^{-\nu}}
{p_i^-\cdot k-i\varepsilon}
-
\frac{p_i^{+\mu}p_i^{+\nu}}
{p_i^+\cdot k+i\varepsilon}
\right]
+\cl O(\Delta p^2).
\label{eq:gravitational-zero-frequency-endpoint-source}
\]
In four dimensions, the surviving projector is the usual one
\[
P_{\mu\nu,\rho\sigma}
=
\frac12\left(
\eta_{\mu\rho}\eta_{\nu\sigma}
+\eta_{\mu\sigma}\eta_{\nu\rho}
-\eta_{\mu\nu}\eta_{\rho\sigma}
\right).
\label{eq:gravitational-zero-frequency-function-projector}
\]
The leading gravitational Berry connection can therefore be written as
\[
\cl A_A
=
\frac{i}{2}
\int d\Phi(k)\,
\left[
F_{{\rm ZF},\mu\nu}^*(k)
P^{\mu\nu,\rho\sigma}
\overleftrightarrow{\pd}_A
F_{0,\rho\sigma}(k)
+
F_{0,\mu\nu}^*(k)
P^{\mu\nu,\rho\sigma}
\overleftrightarrow{\pd}_A
F_{{\rm ZF},\rho\sigma}(k)
\right] + \cl{O}(\kappa^6).
\]

As in the scalar-QED calculation, the same-side terms cancel and only the mixed $p_a^+$ and $p_b^-$ terms remain, and we get
\[
\left(p_a^+p_a^+\right)_{\mu\nu}
P^{\mu\nu,\rho\sigma}
\left(p_b^-p_b^-\right)_{\rho\sigma}
=
\left(p_a^+\cdot p_b^-\right)^2
-\frac12m_a^2m_b^2.
\]
The antisymmetric numerator derivative becomes
\[
2\left(p_a^+\cdot p_b^-\right)
\left(
p_b^-\cdot\pd_Ap_a^+
-p_a^+\cdot\pd_Ap_b^-
\right),
\]
while the pole derivative is weighted by the undifferentiated contraction above. We can now use the master radial integral and differentiated angular identities in section~\ref{subsec:leading-radiation-berry-connection}. Their sum gives
\[
\cl A_A
=
- G\sum_{a,b}
B_{\rm grav}(\Gamma_{ab})
\left(
p_b^-\cdot\pd_Ap_a^+
-p_a^+\cdot\pd_Ap_b^-
\right)
+\cl O(\kappa^6),
\]
where
\[
B_{\rm grav}(\Gamma)
\equiv
\left(\Gamma^2-\frac12\right)
\frac{1-\Gamma\Delta(\Gamma)}{\Gamma^2-1}
+2\Gamma\Delta(\Gamma),
\qquad
B_{\rm grav}(1)=\frac{11}{6}.
\label{eq:gravitational-zero-frequency-function}
\]
Expanding the endpoints to first order in the impulse gives the direct gravitational analogue of eq.~\eqref{eq:leading-radiation-berry-connection},
\[
\cl A_A
=
- G\sum_{a,b}\bigg[
&B_{\rm grav}(\gamma_{ab})
\left(
\widetilde p_b\cdot\pd_A\Delta p_a
-\Delta p_b\cdot\pd_A\widetilde p_a
\right)
\\
&+B_{\rm grav}'(\gamma_{ab})\delta\gamma_{ab}
\left(
\widetilde p_b\cdot\pd_A\widetilde p_a
-\widetilde p_a\cdot\pd_A\widetilde p_b
\right)
\bigg]
+\cl O(\kappa^6).
\label{eq:leading-gravitational-radiation-berry-connection}
\]

We now extract the two components that enter the displacement in eq.~\eqref{eq:coherent-lowered-displacement}, following the scalar-QED calculation in section~\ref{subsec:leading-radiation-berry-connection}. For two-body scattering, again we have
\[
\Delta p_1=-\Delta p_2,
\qquad
\widetilde p_a\cdot\Delta p_b=0,
\qquad
\delta\gamma_{ab}=0.
\]
These relations hold on the physical midpoint scattering phase space. Choosing $A=x_i^\nu$ in eq.~\eqref{eq:leading-gravitational-radiation-berry-connection} therefore gives
\[
\cl A_{i}^\nu
=
- G\sum_{a,b}
B_{\rm grav}(\gamma_{ab})
\widetilde p_{b\rho}
\frac{\pd\Delta p_a^\rho}{\pd\widetilde x_{i\nu}}
+\cl O(G^3)
=
\cl O(G^3),
\]
where we used $\widetilde p_a\cdot\Delta p_b=\cl{O}(G^2)$.

There is consequently no $\cl O(G^2)$ zero-frequency field contribution to the impulse, again as in scalar QED.

The position component, obtained by choosing $A=p_i^\nu$, is
\[
\widetilde{\cl A}_{i}^\nu
&=
- G\sum_{a,b}
B_{\rm grav}(\gamma_{ab})
\left(
\widetilde p_{b\rho}
\frac{\pd\Delta p_a^\rho}{\pd\widetilde p_{i\nu}}
-\Delta p_b^\nu\delta_{ia}
\right)
+\cl O(G^3)
\\
&=
 2 G\sum_b
B_{\rm grav}(\gamma_{ib})\Delta p_b^\nu
+\cl O(G^3),
\]
where the second line follows by differentiating $\tilde p_b\cdot\Delta p = 0$. The corresponding static field contribution is
\[
\Delta x_{i,{\rm ZF}}^\mu
=
2 G\sum_b
B_{\rm grav}(\gamma_{ib})\Delta p_b^\mu
=
-2 G\left[
B_{\rm grav}(\gamma)-B_{\rm grav}(1)
\right]\Delta p_i^\mu
+\cl O(G^3).
\label{eq:zero-frequency-gravitational-position-shift}
\]

In gravity, the static-field contribution to the position shift is not
invariant under BMS supertranslations, and eq.~\eqref{eq:zero-frequency-gravitational-position-shift} is evaluated
in the `intrinsic' BMS frame defined in ref.~\cite{Veneziano:2022zwh}. As discussed in section~\ref{subsec:zero-frequency-dressings}, the full supertranslation charge is $Q_s+Q_h$ (see e.g. \cite{Elkhidir:2024izo}).  The term in eq.~\eqref{eq:zero-frequency-gravitational-position-shift} is not preserved as a separate term under this supertranslation, since the transformation changes how the static field and the hard endpoint data are divided between the intrinsic and canonical BMS frames.  It should therefore be interpreted as an intrinsic-frame contribution, rather than as a separately BMS-invariant position shift, while the scattering operator commutes with the completed charge. The corresponding static-field contribution to the angular momentum is similarly frame dependent, as we show below.

Ref.~\cite{Elkhidir:2026ahd} gives another interesting interpretation of this result. There, changing the zero-frequency dressing induces a momentum-dependent phase whose derivative shifts the asymptotic impact parameter, much like our Berry connection if it were pure gauge (with vanishing $\cl F_{AB}$). After translating between their relative-impact-parameter convention and our midpoint variables, their leading gravitational shift agrees with the relative combination of our position shifts in eq. \eqref{eq:coherent-relative-displacement}. The same static contribution may therefore be viewed either as a Berry-connection correction to the displacement in a fixed BMS frame or as the accompanying redefinition of the hard position data under a change of asymptotic frame.

\section{Classical Angular Momentum}
\label{sec:momentum-angular-momentum}
\subsection{Angular-Momentum Balance}
For scalars, the orbital angular momentum is defined in terms of the phase-space coordinates
\[
L_i^{\mu\nu}
=
x_i^{[\mu}p_i^{\nu]},
\]
where antisymmetrisation has no factor of one half, and we work in the classical, deterministic limit throughout this section. Writing the endpoints in terms of the half-way variables as before
\[
z_{i\pm}^A
=
\widetilde z_i^A
\pm
\frac12\Delta z_i^A,
\label{eq:endpoint-centre-parameterisation}
\]
we find
\[
\Delta L_i^{\mu\nu}
&=
\left(\widetilde x_i+\frac12\Delta x_i\right)^{[\mu}
\left(\widetilde p_i+\frac12\Delta p_i\right)^{\nu]}
-
\left(\widetilde x_i-\frac12\Delta x_i\right)^{[\mu}
\left(\widetilde p_i-\frac12\Delta p_i\right)^{\nu]}
\\
&=
\widetilde x_i^{[\mu}\Delta p_i^{\nu]}
+
\Delta x_i^{[\mu}\widetilde p_i^{\nu]}.
\label{eq:exact-midpoint-angular-momentum}
\]
The term quadratic in $\Delta z$ that might otherwise appear vanishes in this formalism, since it occurs with the same sign at the two endpoints and cancels. The formula above is then classically exact and not a small-angle expansion. Expanding in terms of the coherent state parameters gives
\[
\Delta L_i^{\mu\nu} = \widetilde x_i^\mu
\frac{\partial\Re\chi}{\partial\widetilde x_{i\nu}}
-\widetilde x_i^\nu
\frac{\partial\Re\chi}{\partial\widetilde x_{i\mu}}
+
\widetilde p_i^\mu
\frac{\partial\Re\chi}{\partial\widetilde p_{i\nu}}
-\widetilde p_i^\nu
\frac{\partial\Re\chi}{\partial\widetilde p_{i\mu}}
-
\widetilde x_i^{[\mu}\cl A_{i}^{\nu]}
-
\widetilde p_i^{[\mu}
\widetilde{\cl A}_{i}^{\nu]}.
\]
This can be recast in terms of a Lorentz generator acting on the phase-space coordinates
\[
M^{\mu\nu}_i
=
\widetilde x_i^{\mu}\frac{\partial}{\partial\widetilde x_{i\nu}}
-
\widetilde x_i^{\nu}\frac{\partial}{\partial\widetilde x_{i\mu}}
+
\widetilde p_i^{\mu}\frac{\partial}{\partial\widetilde p_{i\nu}}
-
\widetilde p_i^{\nu}\frac{\partial}{\partial\widetilde p_{i\mu}}
\label{eq:hard-lorentz-vector}
\]
where now we can write
\[
\Delta L_i^{\mu\nu} = M^{\mu\nu}_i\Re\chi -
\widetilde x_i^{[\mu}\cl A_{i}^{\nu]}
-
\widetilde p_i^{[\mu}
\widetilde{\cl A}_{i}^{\nu]}.
\]
Using the definition of the Berry connection, we can in turn write the radiation part in terms of the Lorentz generator acting on the waveshape, finding
\[
\Delta L_i^{\mu\nu} = M^{\mu\nu}_i\Re\chi - \frac{i\hbar}{2}
\sum_\lambda\int d\Phi(k)\,
\alpha_{I,\lambda}^*(k)
\overleftrightarrow{M_i^{\mu\nu}}
\alpha_{I,\lambda}(k).
\]
Summing over all matter lines, we find
\[
\sum_i\Delta L_i^{\mu\nu} &= \sum_i M^{\mu\nu}_i\Re\chi - \frac{i\hbar}{2}
\sum_\lambda\int d\Phi(k)\,
\alpha_{I,\lambda}^*(k)
\overleftrightarrow{\bb M^{\mu\nu}}
\alpha_{I,\lambda}(k) \\&= -J_{\rm field}^{\mu\nu},~~~~~~~~~~~~~~\qquad~~~\bb M^{\mu\nu} \equiv \sum_iM_i^{\mu\nu}.
\]
The inclusive waveshape $\alpha_{I,\lambda}$ contains both strict zero-frequency and regular dynamical contributions, as seen in the expansion in eq. \eqref{eq:inclusive-waveshape-zero-frequency-expansion}. This means that $J_{\rm field}^{\mu\nu}$ contains both finite radiative and static contributions to the angular momentum, which we can separate by writing
\[
\alpha_{I,\lambda}(k)
=\alpha_{ZF,\lambda}(k)
+\alpha_{I,\lambda,{\rm reg}}(k),
\]
Using $\sum_i M^{\mu\nu}_i\Re\chi = 0$, we obtain the two definitions
\[
J_{\rm stat}^{\mu\nu} = \frac{i\hbar}{2}
\sum_\lambda\int d\Phi(k)\,\left(
\alpha_{ZF,\lambda}^*(k)
\overleftrightarrow{\bb M^{\mu\nu}}
\alpha_{I,\lambda,\rm{reg}}(k) + \alpha^*_{I,\lambda,\rm{reg}}(k)
\overleftrightarrow{\bb M^{\mu\nu}}
\alpha_{ZF,\lambda}(k)\right),
\]
and
\[
J_{\rm rad}^{\mu\nu} = \frac{i\hbar}{2}
\sum_\lambda\int d\Phi(k)\,
\alpha_{I,\lambda, {\rm reg}}^*(k)
\overleftrightarrow{\bb M^{\mu\nu}}
\alpha_{I,\lambda, {\rm reg}}(k),
\]
where we note that there is no contribution from two copies of the zero-frequency waveshape, i.e.
\[
\alpha_{ZF,\lambda}^*(k)
\overleftrightarrow{\bb M^{\mu\nu}}
\alpha_{ZF,\lambda}(k) = 0,
\]
and we take
\[
J_{\rm field}^{\mu\nu} = J_{\rm stat}^{\mu\nu} + J_{\rm rad}^{\mu\nu}.
\label{eq:angular-momentum-balance}
\]

Expressed in this way, eq. \eqref{eq:angular-momentum-balance} has the same radiative-plus-static structure as the mechanical angular-momentum balance law of \cite{Riva:2023xxm} and the decompositions in \cite{DiVecchia:2022owy,DiVecchia:2022piu}. In gravity, we shall see below that the mixed zero-frequency–regular term reproduces its leading $\cl O(G^2)$ static contribution, while the regular–regular radiative contribution begins at $\cl O(G^3)$.

To see that this definition makes sense, we can consider a spin-one waveshape, taking
\[
\alpha_{I,\lambda}(k) = \frac{1}{\sqrt\hbar}\epsilon_\rho^\lambda(k)F^\rho(\widetilde p_i,k),
\]
where $F^\rho$ is some source. Using the spin-1 completeness relation, we can therefore write
\[
J_{\rm field}^{\mu\nu} = -\frac{i}{2}
\int d\Phi(k)\,
F_\rho^*(\widetilde p_i,k)
\overleftrightarrow{\left(\sum_iM_i^{\mu\nu}\right)}
F^\rho(\widetilde p_i,k),
\]
where we have assumed that the source is conserved, $k_\rho F^\rho = 0$.
Using the identity
\[
\sum_iM_i^{\mu\nu} F^\rho
=
-D_k^{\mu\nu}F^\rho
+
\eta^{\rho\nu}F^\mu
-
\eta^{\rho\mu}F^\nu,
\]
where
\[
D_k^{\mu\nu} =
k^\mu\frac{\partial}{\partial k_\nu}
-
k^\nu\frac{\partial}{\partial k_\mu},
\]
we can write this as 
\[
J_{\rm field}^{\mu\nu} = \frac{i}{2}
\int d\Phi(k)\,
F_\rho^*(\widetilde p_i,k)
\left(\eta^{\rho\sigma}\overleftrightarrow{D}_k^{\mu\nu} + 2\eta^{\rho[\mu}\eta^{\nu]\sigma}\right)
F_\sigma(\widetilde p_i,k),
\]
which is the standard expression for the radiated angular momentum in terms of the source $F^\rho$ \cite{DiVecchia:2019myk}.

\subsection{Leading Static Contribution to Field Angular Momentum}
\label{subsec:rutherford-position-amplitudes}
\label{subsec:scalar-qed-rutherford-benchmark}
According to the formulas above, the field angular momentum is given by
\[
J^{\mu\nu}_{\rm field} &= -\sum_i\left(\widetilde x_i^{[\mu}\Delta p_i^{\nu]}
+
\Delta x_i^{[\mu}\widetilde p_i^{\nu]}\right) \\
&= \sum_i\left(
\widetilde x_i^{[\mu}\cl A_{i,\mathrm{rad}}^{\nu]}
+
\widetilde p_i^{[\mu}
\widetilde{\cl A}_{i,\mathrm{rad}}^{\nu]}
\right),
\]
so we can either compute it from the full impulse and position shift, or from the radiative Berry connection. In the second line we used the fact that the elastic phase is a Lorentz scalar, and hence $\sum_iM_i^{\mu\nu}\operatorname{Re}\chi=0$. Let's consider the leading-order radiated angular momentum in scalar QED to test this. We already saw that the radiative impulse vanishes at this order, but the radiative position shift does not, and is given by eq.~\eqref{eq:zero-frequency-position-shift-rutherford}.

Plugging this into the angular-momentum formula above gives the leading-order radiated angular momentum in terms of the impulses and midpoint momenta of the hard lines
\[
J_{\rm ZF}^{\mu\nu}
=\frac{1}{4\pi}\bigg[
&\left(
-\frac{2e_1^2}{3m_1^2}
+\frac{e_1e_2}{m_1m_2}B(\widetilde\gamma)
\right)
\widetilde p_1^{[\mu}\Delta p_1^{\nu]}
+ (1\leftrightarrow2)
\bigg]
+O(e^6).
\]
This agrees with the results in \cite{Saketh:2021sri,Biswas:2024ept}, and, in the nonrelativistic limit, with the older analysis of Aguiar and Barone~\cite{Aguiar:2009Rutherford}.

We can now play the same game in gravity, again using the results from the previous section.
Since the zero-frequency field gives no $\cl O(G^2)$ contribution to the impulse, its leading angular momentum comes entirely from the position shift term, analogously with scalar QED. Substituting eq.~\eqref{eq:zero-frequency-gravitational-position-shift} into the angular-momentum formula gives
\[
J_{{\rm ZF},{\rm grav}}^{\mu\nu}
&=
-\sum_{i=1}^{2}
\Delta x_{i,{\rm ZF}}^{[\mu}\widetilde p_i^{\nu]}
+\cl O(G^3)
\\
&=
-2 G
\left[
B_{\rm grav}(\gamma)-B_{\rm grav}(1)
\right]
\sum_{i=1}^{2}
\widetilde p_i^{[\mu}\Delta p_i^{\nu]}
+\cl O(G^3)
\label{eq:gravitational-zero-frequency-angular-momentum}
\]
This is the $\cl O(G^2)$ static-field contribution in the intrinsic BMS frame, matching results in \cite{Manohar:2022dea,DiVecchia:2022owy,Riva:2023xxm,Heissenberg:2024umh}.
\section{Discussion}
\label{sec:discussion}
In this paper, we have developed a formalism for computing classical observables scattering amplitudes represented on phase-space. More precisely, we have obtained an operator-valued phase-space symbol of the $S$-matrix and, from its inclusive one-point projection, a coherent representative of the deterministic classical field. The resulting symbol, $S_W(\tilde z)$, is the central object of interest, and we have used it to define certain classical observables. The star product of two symbols with adjacent numbers of legs $n$ and $n+1$ gives rise to the inclusive waveshape at some order, while the  star product of a derivative operator between two same-multiplicity symbols yields the displacement in phase-space, which is comprised of the impulse and asymptotic position shift. All objects are defined using midpoint variables, and in particular we presented a compact all-order formula for the angular momentum in terms of the midpoint displacement, which naturally contains both the radiative and static contributions, at least when properly regulated.

We have also made one conceptual point that is worth emphasising, which is the distinction between the inclusive and exclusive outgoing state in the classical limit. As pointed out in ref. \cite{Georgoudis:2025vkk}, the outgoing state is not generally coherent, and its exclusive kernels can not be discarded in the classical limit before the observable is constructed. That being said, the classical deterministic one-point function is completely characterised by $\alpha_I=\langle a\rangle$, which defines a coherent representative of the inclusive waveshape. As we have shown, the higher exclusive kernels are not discarded: when all but one of their legs are cut against radiation already present in the state, they can contribute at leading classical order to the inclusive waveshape, even though the same kernels are $\hbar$-suppressed when every leg is measured. The nonlinear-memory $5{\rm pt}^*\times6{\rm pt}$ cut is the first nontrivial example of this mechanism.

Another interesting point raised in ref. \cite{Georgoudis:2025vkk} is the idea that if the final state is not coherent and gives rise to classical physical observables, then there might exist classical correlations between the gravitational-wave signals at various detectors, arising from the nonlinear nature of GR, presumably manifesting in a non-zero connected correlator. In general, it is not the case that GR is necessarily deterministic, even if the initial conditions are known exactly \cite{and:2020pfm}. However, the nonlinear memory effect can be defined as a scattering observable for which there is a suitable Cauchy formulation and where spacetime is globally hyperbolic. Under these assumptions, the classical physics is fully deterministic and, given the equations of motion together with complete knowledge of the system (i.e. the initial conditions), classical observables can be predicted with certainty. There are several ways that these assumptions can be relaxed, such as stochastic gravity or ensembles of initial conditions, and classical correlations in detectors can then arise. Such correlations do not necessarily follow from the fact that the state is not coherent. Indeed, the scaling in eq.~\eqref{eq:Cr-KMOC-scaling-new} gives a directly measured connected $r$-point function of order $\hbar^{r-1}$. In particular,
\[
\braket{\delta \bar h(\bar q)\delta \bar h(\bar k)}_{\rm conn}
\sim
\hbar,
\]
while the same two-radiation kernel contributes at leading classical order when one leg is measured and the other is summed over in the inclusive cut. This does not rule out classical correlations between detectors for stochastic initial data or in an open system, but establishing them requires a separate calculation of an appropriately smeared connected two-point function and a demonstration that it contains a non-zero contribution of order $\hbar^0$. The state being either inclusive or exclusive does not by itself imply anything about the existence of such correlations, although the formalism we have developed here is well suited to probing this question further. In particular, we have mostly worked only with a sharply localised Wigner function, but the formalism is general: any initial Wigner function can be used to define the inclusive symbol, and hence we could easily introduce a source of classical uncertainty and compute the resulting connected correlators. Furthermore, it is clear how such correlators would arise from the symbol: they would come from a higher order projection in symbols with adjacency $\ell$, i.e. 
\[
&\langle0|
S_W^\dagger(\tilde z)
\star_m
a_{\lambda_1}(k)\ldots a_{\lambda_\ell}(k)
\star_m
S_W(\tilde z)
|0\rangle
\sim
\sum_n S_{n,W}^\dagger(\tilde z)
\star_m
S_{n+\ell,W}
(\tilde z),
\]
with the relevant integrals over momentum. This would certainly be interesting to pursue, and we leave this direction in future work.

We have also introduced a coherent-state geometry that aims to be physically useful. The family of states $\ket{\alpha_I(\tilde z)}$ induces a Berry connection, and this connection supplies radiative contributions to both the impulse and position shift. In particular, its zero-frequency sector is not merely a formal soft dressing: it reproduces the leading static-field position shifts and, through angular-momentum balance, the corresponding static contribution to radiated angular momentum in scalar QED and gravity. The coherent representative therefore knows about backreaction and asymptotic static fields even though it does not reproduce the exclusive emission probabilities, squeezing or radiation noise. It would be interesting to extend our understanding of this geometry, however, since the physical meaning of the accompanying Fubini--Study metric and its positivity bound deserves a more direct amplitude's interpretation. We could also easily ask how the geometry changes when we instead consider the exact noncoherent exclusive state, where it may retain information about noise and higher connected correlators. 

We have also introduced a symplectic area balance law, which gives a more direct interpretation of radiation reaction. On the hard phase space alone, the scattering map is not symplectic, but the missing area has not been destroyed, it simply flows into the radiation space. Including the outgoing waveform among the phase-space coordinates gives
\[
\Omega_{\rm hard}^{+}(u,v)
-
\Omega_{\rm hard}^{-}(u,v)
+
\Omega_{\rm rad}(\delta_u\alpha_I,\delta_v\alpha_I)
=
0,
\]
so the combined scattering map preserves the total hard-plus-radiative symplectic form. Locally, this means that radiation reaction remains Hamiltonian on the enlarged phase space: its apparently dissipative nature only comes about after the radiative degrees of freedom are projected out. In this sense, the Berry curvature measures the apparent departure from Hamiltonian evolution only when we ignore the radiation, rather than a failure of Hamiltonian evolution of the complete system.

We have also shown that the zero-frequency sector is naturally built into the formalism, although it points to a subtlety in the geometry of asymptotic states. At fixed regulator, the soft dressing can be treated as an ordinary coherent state, while its strict zero-frequency limit lies outside the standard Fock representation and should instead be regarded as belonging to a distinct infrared sector \cite{Prabhu:2022zcr}. Its interference with the regular waveform nevertheless gives finite contributions to the Berry connection, position shift and angular momentum, although in gravity the division between hard and soft contributions depends on the choice of BMS frame. We have explicitly picked a frame in our construction, but it would be much more satisfying to have a BMS-covariant description. One idea is to treat the hard scattering data and the accompanying IR boundary data together, and construct an enlarged state space which can be organised by BMS supertranslations and memory \cite{Strominger:2014pwa,Elkhidir:2024izo,Elkhidir:2026ahd}. Interestingly, preliminary calculations suggest that the memory contribution to the Berry curvature vanishes, at least in a fixed BMS frame with a particular phase convention. If this holds more generally, then the memory may depend on the asymptotic state as a whole rather than on its local geometry. This may help separate physical static effects from choices of regulator and BMS frame, and perhaps allow us to derive supertranslation-invariant elements of the angular momentum, matching \cite{Riva:2023xxm}, and position shift.

There are several other open questions which ought to be addressed. One immediate technical extension is to go beyond the endpoint approximation. The local Bopp shifts carry the $q/p$ expansion which the endpoint reduction discards, beginning with the so-called next-to-eikonal terms. It would be interesting to systematically retain these and see which quantities are sensitive to the discarded local history, making contact with the existing body of next-to-eikonal literature \cite{Laenen:2008gt,Laenen:2010uz,Akhoury:2013yua,Luna:2016idw,Fernandes:2024xqr}. Another obvious extension is to include spin, generalising the formalism developed in \cite{Luna:2023uwd} and deriving e.g. spinning radiation-reaction and spinning position shifts.

Although our focus in this paper has been radiation, the conservative endpoint sector raises a closely related open question. Ref.~\cite{Kim:2025magnusian} recently established an exact relation between Hamilton's principal function $S$ and the Magnusian $\chi$. Here we denote these objects by $\delta$ and $\Gamma$, respectively, reserving $\chi$ for the Marinovian, which appears in the phase-only ordinary exponential but depends on the midpoint coordinates\footnote{Note that with our sign convention, Marinov's original phase action is $\Phi_{\rm M}=-\chi$ \cite{Marinov1979PhaseAction}.}. An interesting next step is to clarify how all three descriptions fit into an amplitude-friendly framework.

Classically, $\delta$, $\chi$ and $\Gamma$ encode the same local canonical map, but they are distinct functions that represent it in different ways. With our conventions,
\[
\sd\delta
&=
 p_{{\rm out},a}\sd x_{\rm out}^a
-p_{{\rm in},a}\sd x_{\rm in}^a,
\\
\Delta z^A
&=
-\Omega^{AB}\pd_B\chi(\tilde z),
\qquad
z_{\rm out}^A
=
e^{\{\Gamma,\mathord\cdot\}_{\rm PB}}z_{\rm in}^A.
\]
Here $\delta(x_{\rm out},x_{\rm in})$ is the on-shell action with fixed endpoint positions, $\chi(\tilde z)$ determines the endpoint displacement implicitly from phase-space midpoint data (the impulse and position shift), and $\Gamma$ is the Magnusian that generates the map using nested Poisson brackets.

The endpoint and centre generating functions are related locally by the partial Legendre transform
\[
\chi(\tilde x,\tilde p)
=
\delta(x_{\rm out},x_{\rm in})
-\tilde p_a\Delta x^a,
\qquad
\left.\frac{\pd\delta}{\pd\Delta x^a}\right|_{\tilde x}
=
\tilde p_a.
\]
These three objects are therefore similar but distinct, and a systematic comparison of their classical descriptions is currently under development. 

The phase-space formulation may also provide a useful way to study the double copy beyond perturbation theory. Since several recent approaches formulate aspects of the classical double copy directly at the level of fields and equations of motion \cite{Lee:2018gxc,Monteiro:2020plf,Cheung:2021zvb,Moynihan:2021rwh,Adamo:2021dfg,Cheung:2022mix,Moynihan:2025vcs,Kim:2025eqd,Armstrong-Williams:2026wzi}, it is natural to ask whether the relation between gauge and gravitational radiation also extends to the geometry of their radiation states. In principle, the work of Ilderton and Lindved on the double copy of coherent-state gauge backgrounds into curved gravitational backgrounds could be useful here \cite{Ilderton:2025gug}. We look forward to exploring these questions in future work.
\paragraph{Acknowledgements}
I am grateful to Graham Brown, Andrea Cristofoli, Alessandro Georgoudis, Riccardo Gonzo, Carlo Heissenberg, Jung-Wook Kim, Donal O'Connell, Paolo Pichini, Trevor Scheopner and Chris White for helpful discussions, and in particular to Alexander Ochirov and Canxin Shi for coordinating their submission of \cite{Canxin_Amps26}. I have also benefited from discussions with intelligences implemented on less conventional substrates, and thank ChatGPT and Claude, especially for their help in writing Mathematica code using the Wolfbook VS Code extension (\href{https://github.com/vanbaalon/wolfbook}{github.com/vanbaalon/wolfbook}). This work was supported by the Science and Technology Facilities Council (STFC) Consolidated Grant ST/X00063X/1 “Amplitudes, Strings \& Duality”. No new data were generated or analysed during this study.
\appendix
\section{Off-Shell Extensions and the Projected Pairing}
\label{app:off-shell-extensions}
In the main text, we have used the notation $\pd_A$, or even $\pd_\mu$, for derivatives of on-shell functions. This is a fairly standard abuse of notation, but since the on-shell conditions reduce the number of independent variables it is worth exploring this in more detail.

Let's first define what we actually mean by derivatives of an on-shell function. Consider some independent coordinates $\xi^a$ for the physical scattering data, i.e. the reduced set of six phase-space variables after the on-shell and transverse conditions have been solved, and write $z^A=z^A(\xi)$. Formally, then, the derivative of a function on this restricted phase-space is defined by
\[
\frac{\pd}{\pd\xi^a}F(z(\xi))
=
\left.
\frac{\pd z^A}{\pd\xi^a}
\frac{\pd F(z)}{\pd z^A}
\right|_{z=z(\xi)}.
\]
In other words, we \textit{first} impose the physical conditions and then differentiate with respect to the independent variables. We nevertheless keep the standard $\pd_A$ or $\pd_\mu$ notation in the main text, with only these allowed variations understood. If two choices differ by
\[
F'(z)-F(z)=C_r(z)h^r(z),
\]
where $C_r(z)=0$ are the physical conditions, then
\[
\frac{\pd}{\pd\xi^a}
\left[(F'-F)(z(\xi))\right]
=
\frac{\pd}{\pd\xi^a}
\left[C_r(z(\xi))h^r(z(\xi))\right]
=0.
\]
This is the meaning of the ordinary derivatives of $\chi$ and of the radiation state used in the Berry connection and the quantum geometric tensor. The derivatives inside $\star_m$ are slightly different: they are taken \textit{before} the physical conditions are imposed, since they reproduce operator multiplication. They are not observables on their own, but thankfully any differences are projected away by the final pairing. The calculation below shows that this pairing does not depend on any particular off-shell continuation.

The on-shell matter symbol is not unique away from the final projected support, since we can augment it with any number of terms which are killed by the delta functions, and similarly for the Wigner function: only the projected pairing \eqref{eq:trace-phase-space} is physical. For example, two symbols related by an off-shell continuation will give rise to the same expectation value, but will differ as functions on phase-space
\[
\cl O'_W(\tilde p,q)
=
\cl O_W(\tilde p,q)
+
\left(
\tilde p^2+\frac{q^2}{4}-m^2
\right)h(\tilde p,q),
\]
but this is only true when paired with $\rho_W$. Any off-shell deformation of this type is allowed provided it can be smoothly projected to the positive-energy, on-shell domain mentioned above, and provided it has a sensible Fourier transform. The obvious place where this might break down is near the kinds of poles we get from propagators, and so it is understood that these are properly regulated with an $i\epsilon$ prescription which is kept until after wavepacket smearing and the final on-shell projection. More generally, two different allowed deformations may differ by the displayed mass-shell factor times $h$ plus $2\tilde p\cdot q\,k$, which obviously vanishes immediately under $\del(2\tilde p\cdot q)$, so it is enough to analyse the first case below.

More concerning, however, is the fact that the star product acts by taking derivatives with respect to $\tilde p$, which appears problematic if two off-shell deformations of an observable differ by a function that vanishes on shell but whose derivative does not, which is exactly the case for the example above. We have
\[
\frac{\pd}{\pd\tilde p^\mu}\cl O'_W
=
\frac{\pd}{\pd\tilde p^\mu}\cl O_W
+
2\tilde p_\mu h
+
\left(
\tilde p^2+\frac{q^2}{4}-m^2
\right)
\frac{\pd h}{\pd\tilde p^\mu},
\]
and so a generic product can distinguish the two symbols. However, the
included transverse delta function $\del(2\tilde p\cdot q)$ ensures that
this difference drops out after the final on-shell pairing. At leading order in the Moyal expansion, this is straightforward to see using eq. \eqref{eq:moyal-star-product}: for a product $\cl A_W\star (\cl O_W' - \cl O_W)$, the difference is proportional to $H\tilde p\cdot\partial_x A_W - H\leftrightarrow A$, which vanishes provided $A_W$ is also transversely supported, i.e.
\[
A_W(x,\tilde p)
=
\int\hat{\sd}^4\ell\,
e^{i\ell\cdot x/\hbar}
\del(2\tilde p\cdot\ell)\,
a(\tilde p,\ell),~~~~~H(x,\tilde p)
=
\int\hat{\sd}^4q\,
e^{iq\cdot x/\hbar}
\del(2\tilde p\cdot q)\,
h(\tilde p,q).
\]
The same is true for all higher-order terms in the expansion, but it is less obvious, so we will show this explicitly. Writing the on-shell condition as
\[
M(\tilde p,q)
\equiv
\tilde p^2+\frac{q^2}{4}-m^2,
\qquad
Q\equiv\ell+q,
\]
the difference between the two star products is therefore, at all orders using the positive-energy Bopp representation in eq. \eqref{eq:bopp-shift-identity},
\[
\begin{aligned}
A\star(\cl O'_W-\cl O_W)
=&
\int\hat{\sd}^4\ell\,\hat{\sd}^4q\,
e^{iQ\cdot x/\hbar}\del\left(2\left(\tilde p+\frac q2\right)\cdot\ell\right)\del\left(2\left(\tilde p-\frac\ell2\right)\cdot q\right)
\\
&\times
M\left(\tilde p-\frac\ell2,q\right)
a\left(\tilde p+\frac q2,\ell\right)h\left(\tilde p-\frac\ell2,q\right).
\end{aligned}
\]
From its definition, the shifted mass-shell factor satisfies
\[
M\left(\tilde p-\frac\ell2,q\right)
=
M(\tilde p,Q)
-
\left(\tilde p+\frac q2\right)\cdot\ell,
\]
where the second term vanishes on the support of the first transverse delta
function. Hence, under the integral we have
\[
\del\left(2\left(\tilde p+\frac q2\right)\cdot\ell\right)
M\left(\tilde p-\frac\ell2,q\right)
=
\del\left(2\left(\tilde p+\frac q2\right)\cdot\ell\right)
M(\tilde p,Q).
\]
It follows that the entire difference has the form
\[
A\star(\cl O'_W-\cl O_W)
=
\int\hat{\sd}^4Q\,
e^{iQ\cdot x/\hbar}\hat\delta(2\tilde p\cdot Q)
M(\tilde p,Q)\,
K(\tilde p,Q),
\]
for some kernel with transverse support $K$ determined by $a$ and $h$. 

Inside an expectation value, this expression is paired with $\rho_W$, the $x$ integral sets the
Fourier momentum of the state equal to $-Q$. The on-shell delta function in $\rho_W$ then enforces
\[
\del\left(
\tilde p^2+\frac{Q^2}{4}-m^2
\right)
=
\del\bigl(M(\tilde p,Q)\bigr),
\]
where we have used $M(\tilde p,-Q)=M(\tilde p,Q)$. Finally, we have
\[
\del\bigl(M(\tilde p,Q)\bigr)M(\tilde p,Q)=0,
\]
and therefore
\[
\int\hat{\sd}^4\tilde p\,\sd^4x\,
\rho_W\,
A\star\cl O'_W
=
\int\hat{\sd}^4\tilde p\,\sd^4x\,
\rho_W\,
A\star\cl O_W.
\]

The same conclusion holds when the modified observable appears on the
left of the star product. In that case the Bopp shifts produce
\[
M\left(\tilde p+\frac\ell2,q\right)
=
M(\tilde p,Q)
+
\left(\tilde p-\frac q2\right)\cdot\ell,
\]
and the second term is again removed by the shifted transverse delta
function. We see then that different, sufficiently regularised off-shell deformations can give different intermediate off-shell star products, but that they give identical physical expectation values after the final on-shell projection. This extends to any finite chain of star products, and therefore to functions which can be expressed as a finite series of star products, such as star exponentials and star logarithms, to some order in perturbation theory, which we will make use of below. The only requirement is that the off-shell continuation be smooth and well-defined in a neighbourhood of the on-shell support of the state, so that the final projection can be taken without encountering singularities. Our statements therefore only apply to poles with regulators for which wavepacket smearing and the on-shell projection are well defined at fixed $\epsilon>0$, and for which the subsequent $\epsilon\to0$ limit exists without a pinch. We always take the projection before the $\epsilon\to0$ limit, and do not assume that these two operations commute.
\section{Time Ordering, the Partial Weyl Transform and Wick's Theorem}
\label{app:time-order-weyl-brackets}\label{app:wicks-theorem}
It is useful to understand the distinctions between ordinary time-ordering of operators and the $\cl T_\star$-ordering of Weyl symbols, which is a key feature of the Weyl representation that we exploit in the main text. For simplicity, we will work with the second-order term in the Dyson expansion of the $S$-matrix, which is the first non-trivial order at which time-ordering and star products interact.
All matter star products in this appendix are assumed to be the positive-energy versions induced by composition on the positive-energy Hilbert space, and it is understood that the identities we derive are valid away from coincident times.

We define
\[
\bb H_a
\equiv
\bb J_a^\mu \bb A_{a\mu},
\qquad
\bb J_a^\mu\equiv \bb J^\mu(z_a),
\qquad
\bb A_{a\mu}\equiv \bb A_\mu(z_a),
\]
as well as the usual time-ordering step functions and their difference
\[
\theta_{12}\equiv \theta(z_1^0-z_2^0),
\qquad
\theta_{21}\equiv \theta(z_2^0-z_1^0),
\qquad
\varepsilon_{12}\equiv \theta_{12}-\theta_{21}.
\]
Up to second order in the interaction $J\cdot A$, we have
\[
\bb S_0
&=
1+\frac{i}{\hbar}\int d^4z_1\,\bb H_1
+\frac{1}{2}\left(\frac{i}{\hbar}\right)^2
\int d^4z_1\,d^4z_2\,
\cl T\!\left[\bb H_1\bb H_2\right]
+O\!\left((J\bb A)^3\right).
\]
The explicit second-order time ordering gives
\[
\cl T\!\left[\bb H_1\bb H_2\right]
&=
\theta_{12}\,
\bb J_1^\mu \bb A_{1\mu}\,
\bb J_2^\nu \bb A_{2\nu}
+
\theta_{21}\,
\bb J_2^\nu \bb A_{2\nu}\,
\bb J_1^\mu \bb A_{1\mu}.
\]
Since photon and matter operators act on different Hilbert spaces, they commute $[\bb A,\bb J]=0$, and we may therefore regroup within each ordering, finding
\[
\cl T\!\left[\bb H_1\bb H_2\right]
&=
\theta_{12}\,
\bb J_1^\mu\bb J_2^\nu\,
\bb A_{1\mu}\bb A_{2\nu}
+
\theta_{21}\,
\bb J_2^\nu\bb J_1^\mu\,
\bb A_{2\nu}\bb A_{1\mu}.
\label{eq:app-time-ordered-second}
\]

Now apply the partial Weyl transform $\cl W_m$, acting only on the matter Hilbert space. The photon operators are spectators in this operation, while the ordered products of matter operators become Moyal products of their symbols, i.e. 
\[
\cl W_m[\bb J_1^\mu\bb J_2^\nu]
&=
J_{1,W}^\mu\star_m J_{2,W}^\nu,
\qquad
\cl W_m[\bb J_2^\nu\bb J_1^\mu]
=
J_{2,W}^\nu\star_m J_{1,W}^\mu,
\]
where $J_{a,W}^\mu\equiv J_W^\mu(z_a)$. Therefore
\[
\cl W_m\!\left[\cl T(\bb H_1\bb H_2)\right]
&=
\theta_{12}\,
\left(J_{1,W}^\mu\star_m J_{2,W}^\nu\right)
\bb A_{1\mu}\bb A_{2\nu}
\\
&\quad+
\theta_{21}\,
\left(J_{2,W}^\nu\star_m J_{1,W}^\mu\right)
\bb A_{2\nu}\bb A_{1\mu}.
\label{eq:app-weyl-after-time-ordering}
\]

Now we can apply Wick's theorem to the photon operators, which allows us to exchange the time-ordered product of photon fields for a normal-ordered product, at the cost of introducing contractions between the photon fields, i.e. we use 
\[
\bb A_{1\mu}\bb A_{2\nu}
&=
\no{\bb A_{1\mu}\bb A_{2\nu}}
+
\braket{0|\bb A_{1\mu}\bb A_{2\nu}|0},~~~~~\text{and}~~~~~
\bb A_{2\nu}\bb A_{1\mu}
=
\no{\bb A_{1\mu}\bb A_{2\nu}}
+
\braket{0|\bb A_{2\nu}\bb A_{1\mu}|0},
\]
where we have used the fact that normal ordering is symmetric for bosonic operators. We can now expand the time-ordering in terms of this normal ordering and the contractions, giving
\[
\cl W_m[\cl T(\bb H_1^\mu\bb H_2^\nu)] &= \theta_{12}\,\left(J_{1,W}^\mu\star_m J_{2,W}^\nu\right)\,
\left[\no{\bb A_{1\mu}\bb A_{2\nu}}
+
\braket{0|\bb A_{1\mu}\bb A_{2\nu}|0}\right]
+
\theta_{21}\,\left(J_{2,W}^\nu\star_m J_{1,W}^\mu\right)\,
\left[\no{\bb A_{1\nu}\bb A_{2\mu}}
+
\braket{0|\bb A_{2\nu}\bb A_{1\mu}|0}\right]\\
&= \left(\theta_{12}\,\left(J_{1,W}^\mu\star_m J_{2,W}^\nu\right) + \theta_{21}\,\left(J_{2,W}^\nu\star_m J_{1,W}^\mu\right)\right)\,\left[\no{\bb A_{1\mu}\bb A_{2\nu}} + \theta_{12}\,\braket{0|\bb A_{1\mu}\bb A_{2\nu}|0} + \theta_{21}\,\braket{0|\bb A_{2\nu}\bb A_{1\mu}|0}\right].
\\
&= \left(\theta_{12}\,\left(J_{1,W}^\mu\star_m J_{2,W}^\nu\right) + \theta_{21}\,\left(J_{2,W}^\nu\star_m J_{1,W}^\mu\right)\right)\,\left[\no{\bb A_{1\mu}\bb A_{2\nu}} + \Delta^F_{\mu\nu}(z_1-z_2)\right]\\
&= \cl{T}_{\star_m}[J_{1,W}^\mu J_{2,W}^\nu]\,\cl{T}[\bb A_{1\mu}\bb A_{2\nu}],
\]
where, on the second line, we have used $\theta_{ij}^2 = \theta_{ij}$ and $\theta_{12}\theta_{21}=0$, and on the third line we have recognised the contraction as the Feynman propagator, and defined the $\cl T_{\star_m}$-ordering of the matter symbols as
\[
\cl{T}_{\star_m}[J_{1,W}^\mu J_{2,W}^\nu] \equiv \theta_{12}\,\left(J_{1,W}^\mu\star_m J_{2,W}^\nu\right) + \theta_{21}\,\left(J_{2,W}^\nu\star_m J_{1,W}^\mu\right).
\]
This definition of the $\cl T_{\star_m}$-ordering generalises to any number of symbols. Define the general time-ordering step function for $n$ operators as
\[
\Theta_\pi(z_1,\ldots,z_n)
\equiv
\theta(z^0_{\pi_1}-z^0_{\pi_2})
\theta(z^0_{\pi_2}-z^0_{\pi_3})
\cdots
\theta(z^0_{\pi_{n-1}}-z^0_{\pi_n}),
\]
where $\pi \in S_n$ is a permutation of the $n$ operators. The generalisation of the double theta function identities to $n$ operators is just 
\[
\Theta_{\pi}\Theta_{\sigma} = \delta_{\pi\sigma}\Theta_{\sigma},~~~~~\sum_{\pi \in S_n} \Theta_\pi = 1.
\]
Using these, the general $\cl T_{\star_m}$-ordering of $n$ symbols is given by
\[
\cl T_{\star_m}
\!\left[
J_1^{\mu_1}\cdots J_n^{\mu_n}
\right]
\equiv
\sum_{\pi\in S_n}
\Theta_\pi\,
J_{\pi_1}^{\mu_{\pi_1}}
\star_m\cdots\star_m
J_{\pi_n}^{\mu_{\pi_n}}, 
\]
which is simply the usual time-ordering of operators, but with the operator products replaced by star products of their symbols. Now, we can take the photon time-ordering and the matter $\cl T_{\star_m}$-ordering together, giving
\[
\cl W_m\!\left[\cl T(\bb H_1\cdots \bb H_n)\right]
&=
\cl T_{\star_m}
\!\left[J_1^{\mu_1}\cdots J_n^{\mu_n}
\right]
\cl T\!\left[
\bb A_{1\mu_1}\cdots \bb A_{n\mu_n}
\right].
\]
To see that this is the case, we can replace the time-orderings explicitly:
\[
\cl T_{\star_m}
\!\left[J_1^{\mu_1}\cdots J_n^{\mu_n}
\right]
\cl T\!\left[
\bb A_{1\mu_1}\cdots \bb A_{n\mu_n}
\right] &= \sum_{\pi\in S_n}\sum_{\sigma\in S_n}\Theta_\pi\Theta_\sigma\,J_{\pi_1}^{\mu_{\pi_1}}
\star_m\cdots\star_m
J_{\pi_n}^{\mu_{\pi_n}}
\bb A_{\sigma_1\mu_{\sigma_1}}\cdots \bb A_{\sigma_n\mu_{\sigma_n}}\\
&= \sum_{\pi\in S_n}\sum_{\sigma\in S_n}\delta_{\pi\sigma}\Theta_\sigma\,J_{\pi_1}^{\mu_{\pi_1}}
\star_m\cdots\star_m
J_{\pi_n}^{\mu_{\pi_n}}
\bb A_{\sigma_1\mu_{\sigma_1}}\cdots \bb A_{\sigma_n\mu_{\sigma_n}}\\
&= \cl{T}[\cl W_m[J_1^{\mu_1}\cdots J_n^{\mu_n}]\bb A_{1\mu_1}\cdots \bb A_{n\mu_n}],
\]
where we have used $\Theta_{\pi}\Theta_{\sigma} = \delta_{\pi\sigma}\Theta_{\sigma}$.

Crucially, this can then be used to straightforwardly generalise the textbook exponential formulation of Wick's theorem found in e.g. \cite{Itzykson:1980rh,Evans:1998yi} to star-products (and by extension, star-exponentials)
\[
\cl T_\gamma\cl T_{\star_m}
\exp_{\star_m}X
=
\cl T_{\star_m}\left\{
\exp_{\star_m}\!\left[{1\over2}\overbrace{XX}_{\star_m}\right]
\star_m
\normord{\exp_{\star_m}X}
\right\}.
\label{eq:star-wick-exp}
\]
The proof follows \cite{Evans:1998yi}, where $\cl T_\gamma$ denotes the ordinary photon time ordering. The contraction in \eqref{eq:star-wick-exp} is the two-vertex object
\[
{1\over2}\overbrace{XX}_{\star_m}
\equiv
- {1\over 2\hbar^2}
\int_{x,y}
\Delta^F_{\mu\nu}(x-y)\,
J_W^\mu(x)\star_m J_W^\nu(y),
\label{eq:star-contraction-block}
\]
where the Feynman propagator is
\[
\Delta^F_{\mu\nu}(x-y)
\equiv
\langle0|\cl T_\gamma\mathbb A_\mu(x)\mathbb A_\nu(y)|0\rangle .
\]

\section{The Global Endpoint Approximation}
\label{app:no-local-recoil-endpoint}

In this appendix we justify the steps taken in dropping certain terms when constructing the eikonal operator, in particular why we can trade the star exponential of a given function with a regular exponential of a corresponding reduced function evaluated on the midpoint variable $\tilde z$. For a given sequence of current insertions along the worldline, the star product records the total displacement between the endpoints of the trajectory, and also the local displacements that result from each individual insertion. For observables that are insensitive to the local recoil history, we will approximate all insertions at the \textit{global} midpoint only, keeping the total displacement between the endpoints but ignoring the local history. For a function undergoing this approximation, i.e.
\[
\exp_\star\left(\cl{F}\right)\longrightarrow \exp\left(\cl{F}_c(\tilde z)\right),
\]
the reduced function $\cl{F}_c(\tilde z)$ is associated with $\cl F$ and is known as the \textit{centre generating function} when it generates the corresponding endpoint canonical transformation\cite{deAlmeida:2013wal,deAlmeida:2020hgy}, and in general $\cl{F}_c(\tilde z)\neq\cl F(\tilde z)$.

The phase-space coordinates $z^A = (x^\mu, p_\mu)$ can be used to define the global midpoint $\tilde z$
\[
z_{\rm out}^A=z_{\rm in}^A+\Delta z^A,
\qquad
\tilde z^A
=
\frac12\left(z_{\rm in}^A+z_{\rm out}^A\right)
=
z_{\rm in}^A+\frac12\Delta z^A.
\]
Each local insertion at position $r$ picks up a phase-space coordinate $z^A_r = \tilde z^A + \delta z_r^A$, and in our approximation we drop the local displacements $\delta z_r^A$ and keep the global midpoint $\tilde z^A$. The dropped terms are often called next-to-eikonal corrections \cite{Laenen:2008gt}, not to be confused with next-to-leading order.

For a constant Poisson bivector $\Omega$, the positive-energy star product of $N$ matter symbols takes the ordinary Moyal form on any kinematic patch for which every intermediate on-shell momentum has positive energy,

\[
F_1\star_m\cdots\star_m F_N
=
\left.
\exp\!\left[
+\frac{i\hbar}{2}
\sum_{r<s}
\pd_A^{(r)}\Omega^{AB}\pd_B^{(s)}
\right]
\prod_{r=1}^N F_r(z_r)
\right|_{z_1=\cdots=z_N=z}.
\label{eq:app-general-N-moyal-product}
\]
Globally, the positive-energy factors $\theta(r_j^0)$ carried by the intermediate on-shell resolutions of the identity must also be retained.

The derivatives coupling distinct factors generate the Bopp shifts, and thereby define the local arguments $z_r=\tilde z+\delta z_r$. To see this, it is useful to express a given local function in terms of the local transfer momentum $q_{i,r}$ conjugate to $x_{i,r}$, i.e.

\[
F_{i,r}(x_i,p_i)
=
\int\hat{\sd}^4q_{i,r}\,
e^{+iq_{i,r}\cdot x_i/\hbar}
f_{i,r}(p_i,q_{i,r}).
\]
We write the factors in order, so the rightmost factor acts first, and the exact star product is then

\[
F_{i,1}\star_i\cdots\star_i F_{i,N_i}
&=
\int\prod_{r=1}^{N_i}\hat{\sd}^4q_{i,r}\,
\exp\!\left[+\frac{i}{\hbar}Q_i\cdot x_i\right]
\prod_{r=1}^{N_i}
f_{i,r}\!\left(\tilde p_{i,r},q_{i,r}\right),
\\
\tilde p_{i,r}
&=
p_i
-\frac12\sum_{a<r}q_{i,a}
+\frac12\sum_{b>r}q_{i,b},
\qquad
Q_i=\sum_{r=1}^{N_i}q_{i,r}.
\label{eq:app-N-factor-bopp-product}
\]

We can think of each factor as inserting the function $f$ at the local midpoint momentum $\tilde p_{i,r}$, with the local momentum $\tilde p_{i,r}$ shifted by $+\frac12$ of the total momentum transfer from all previous insertions ($b>r$), and by $-\frac12$ of the total momentum transfer from all subsequent insertions ($a<r$).

For two factors this gives

\[
F_{i,1}\star_i F_{i,2}
=
\int\hat{\sd}^4q_{i,1}\hat{\sd}^4q_{i,2}\,
e^{+i(q_{i,1}+q_{i,2})\cdot x_i/\hbar}
f_{i,1}\!\left(p_i+\frac{q_{i,2}}{2},q_{i,1}\right)
f_{i,2}\!\left(p_i-\frac{q_{i,1}}{2},q_{i,2}\right),
\]
in agreement with the Bopp-shift identity \eqref{eq:bopp-shift-identity}. We now note that each local term contains the base momentum $p_i$, which we replace everywhere with the global midpoint $\tilde p_i$, such that the local arguments in \eqref{eq:app-N-factor-bopp-product} are given by
\[
\tilde p_{i,r}
=
\tilde p_i
-\frac12\sum_{a<r}q_{i,a}
+\frac12\sum_{b>r}q_{i,b},
\qquad
p_{i,\rm in}=\tilde p_i-\frac{Q_i}{2},
\qquad
p_{i,\rm out}=\tilde p_i+\frac{Q_i}{2}.
\]
This ensures that the shifts of the individual factors only resolve the local recoil history, while $Q_i$ fixes the total endpoint recoil and $\tilde p_i$ fixes its midpoint. For several matter lines the full product factorises into one such expression for each line, since symbols on distinct matter phase spaces commute with respect to the star product.

The endpoint reduction is then defined by the replacement $z_r=\tilde z+\delta z_r\longrightarrow\tilde z$ for every local Bopp-shifted argument (i.e. we drop the $\frac12q_i$ terms), at fixed total displacement $\Delta z$, resulting in the full endpoint reduction
\[\label{eq:app-endpoint-reduction-map}
\!\left[F_1\star_m\cdots\star_m F_N\right]\bigg|_{\rm end}
\equiv
\prod_{r=1}^N F_r(\tilde z).
\]
We should note, however, that endpoint reduction is a prescription at a chosen stage of the perturbative construction: we first evaluate the star products needed to construct each complete amplitude or cut kernel, whose Bopp shifts can contribute at classical order, and only then drop the remaining local shifts. Importantly, this definition allows us to drop star-exponentials in favour of regular exponentials with their reduced functions. In the momentum-resolved example, it sends every $\tilde p_{i,r}$ in \eqref{eq:app-N-factor-bopp-product} to $\tilde p_i$, without setting $Q_i$ to zero.

Consider some matter symbols $\Phi_\alpha(z)$ that could be found in e.g. an interaction Lagrangian, perhaps in the form of a function
\[\label{eq:app-general-star-functional}
\cl F_{\star_m}[\Phi]
=
\sum_{N=0}^{\infty}
C^{\alpha_1\cdots\alpha_N}
\Phi_{\alpha_1}\star_m\cdots\star_m\Phi_{\alpha_N}.
\]
The coefficients $C^{\alpha_1\cdots\alpha_N}$ may contain spacetime integrals, propagators, or radiation creation/annihilation operators, all of which we will assume are spectators under the partial Weyl transform. Applying \eqref{eq:app-endpoint-reduction-map} term by term gives
\begin{equation}\label{eq:app-general-functional-reduction}
\begin{aligned}
\!\left[\cl F_{\star_m}[\Phi]\right]\bigg|_{\rm end}
=
\sum_{N=0}^{\infty}
C^{\alpha_1\cdots\alpha_N}
\prod_{r=1}^N\Phi_{\alpha_r}(\tilde z)
\equiv
\cl F[\Phi(\tilde z)].
\end{aligned}
\end{equation}
For multiple matter lines, $\tilde z$ in this formula represents the collection $(\tilde z_1,\ldots,\tilde z_n)$, with one global midpoint for each line. Any particular ordering carried by the spectator coefficients is left unchanged.

The same argument applies to an outer star-exponential. Writing its first few terms explicitly,
\[
\exp_{\star_m}\!\left(\cl F_{\star_m}\right)
=
1+\cl F_{\star_m}
+\frac{1}{2}\cl F_{\star_m}\star_m\cl F_{\star_m}
+\frac{1}{3!}\cl F_{\star_m}\star_m\cl F_{\star_m}\star_m\cl F_{\star_m}
+\cdots,
\]
we find
\[
\!\left[\exp_{\star_m}\!\left(\cl F_{\star_m}\right)\right]\bigg|_{\rm end}
=
1+\cl F[\Phi(\tilde z)]
+\frac{1}{2}\cl F[\Phi(\tilde z)]^2
+\frac{1}{3!}\cl F[\Phi(\tilde z)]^3
+\cdots
=
\exp\!\left(\cl F[\Phi(\tilde z)]\right).
\]
The same reduction removes the star time ordering of the matter currents. Writing $J_r\equiv J^{\mu_r}(x_r)$, we have
\[
\cl T_{\star_m}\!\left[J_1\star_m J_2\star_m\cdots\star_m J_n\right]
\bigg|_{\rm end}
=
\sum_{\pi\in S_n}\Theta_\pi\prod_{r=1}^n J_{\pi_r}
=
\left(\sum_{\pi\in S_n}\Theta_\pi\right)\prod_{r=1}^n J_r
=
J_1J_2\cdots J_n.
\]
Indeed, after endpoint reduction the currents are ordinary commuting functions, so each ordering gives the same product, while $\sum_{\pi\in S_n}\Theta_\pi=1$.

By contrast, photon time ordering and normal ordering remain intact even after the matter operators have been transformed and endpoint reduced, e.g.
\[
\!\left[
\cl T_\gamma\cl T_{\star_m}
\exp_{\star_m}X
\right]\bigg|_{\rm end}
=
\cl T_\gamma
\exp\!\left[X(\tilde z)\right],
\]
where any normal ordering on the two sides is understood to act only on the radiation operators.

Evaluation at the midpoint alone does not remove the Moyal derivatives: in general,

\[
\left.(F\star_m G)\right|_{z=\tilde z}
=
F(\tilde z)G(\tilde z)
+\frac{i\hbar}{2}
\pd_A F(\tilde z)\Omega^{AB}\pd_B G(\tilde z)
+\cl O(\hbar^2),
\]
and therefore $\left.\exp_{\star_m}F\right|_{z=\tilde z}$ need not equal $\exp F(\tilde z)$. The endpoint operation removes the local Bopp shifts before the full series is resummed.

The approximation is appropriate when the observable is insensitive to the local recoil history and depends only on asymptotic endpoint data, which covers a large class of observables that we are normally interested in. The impulse, displacement etc are explicitly kept through $z_{\rm out}=z_{\rm in}+\Delta z$, but we can not describe observables which resolve intermediate motion, the location of some particular emission along a worldline, or any other observable sensitive to the precise local recoil history.

\section{Bopp Representation of the Positive-Energy Matter Product}
\label{app:bopp-matter-poles}
We wish to consider the Weyl transform of products of operators, so let's consider $\bb C =\bb A\bb B$ and its Weyl symbol $C_W$
\[
C_W(x,\tilde p) &= \int\hat{\sd}^4 q\,
e^{i q\cdot x/\hbar}\,
\del(2\tilde p\cdot q)C_W(\tilde p,q)\,\\
&= 
\int\hat{\sd}^4 q\,\sd\Phi(r)
e^{i q\cdot x/\hbar}\,
\del(2\tilde p\cdot q)\,
\braket{\tilde p+\tfrac q2|\bb{A}|r}\braket{r|\bb{B}|\tilde p-\tfrac q2},
\]
We can evaluate this convolution by defining the momentum transfers through the two operators as
\[
\ell\equiv\tilde p-r+\frac q2,
\qquad
k\equiv r-\tilde p+\frac q2,
\qquad
Q\equiv\ell+k=q.
\]
The intermediate momentum and the midpoints of the two matrix elements are then
\[
r=\tilde p+\frac{k-\ell}{2},
\qquad
\frac{\tilde p+\frac Q2+r}{2}=\tilde p+\frac k2,
\qquad
\frac{r+\tilde p-\frac Q2}{2}=\tilde p-\frac\ell2.
\]
The change of variables from $(q,r)$ to $(\ell,k)$ has unit Jacobian, and therefore
\[
C_W(x,\tilde p)
&=
\int\hat{\sd}^4\ell\,\hat{\sd}^4k\,
e^{iQ\cdot x/\hbar}\,
\del(2\tilde p\cdot Q)\,
\del^+\!\left(
\left(\tilde p+\frac{k-\ell}{2}\right)^2-m^2
\right)
\\
&\qquad\times
A_W\!\left(\tilde p+\frac k2,\ell\right)
B_W\!\left(\tilde p-\frac\ell2,k\right).
\label{eq:app-direct-onshell-symbol-product}
\]
This motivates the exact positive-energy product of symbols, which we define as
\[
\left(A_W\star_m B_W\right)(x,\tilde p)
&\equiv
\int\hat{\sd}^4\ell\,\hat{\sd}^4k\,
e^{iQ\cdot x/\hbar}\,
\del(2\tilde p\cdot Q)\,
\del^+\!\left(
\left(\tilde p+\frac{k-\ell}{2}\right)^2-m^2
\right)
\\
&\qquad\times
A_W\!\left(\tilde p+\frac k2,\ell\right)
B_W\!\left(\tilde p-\frac\ell2,k\right).
\label{eq:app-positive-energy-star-product}
\]

Within the projected pairing, we can give the positive-energy product a convenient \textit{local} Bopp representative, where by local we mean in the $r^0>0$ patch. Substituting the two projected symbols into the local Moyal formula and retaining the intermediate positive-energy projector required by the definition of $\star_m$ gives
\[
A_W(x,\tilde p)\star_m B_W(x,\tilde p)
&=
\int\hat{\sd}^4\ell\,\hat{\sd}^4k\,
e^{iQ\cdot x/\hbar}
\\
&\qquad\times
\del\!\left(2\left(\tilde p+\frac k2\right)\cdot\ell\right)
\del\!\left(2\left(\tilde p-\frac\ell2\right)\cdot k\right)
\theta(r^0)
\\
&\qquad\times
A_W\!\left(\tilde p+\frac k2,\ell\right)
B_W\!\left(\tilde p-\frac\ell2,k\right).
\label{eq:app-star-onshell-symbol-product}
\]
The relation between the support of these expressions is particularly simple. Defining the on-shell and transverse combinations
\[
M\equiv\tilde p^2+\frac{Q^2}{4}-m^2,
\qquad
R\equiv
\left(\tilde p+\frac{k-\ell}{2}\right)^2-m^2,
\qquad
T\equiv2\tilde p\cdot Q,
\]
we find
\[
2\left(\tilde p+\frac k2\right)\cdot\ell
=M+\frac T2-R,
\qquad
2\left(\tilde p-\frac\ell2\right)\cdot k
=R-M+\frac T2,
\]
and hence, with no additional Jacobian,
\[
\del\!\left(2\left(\tilde p+\frac k2\right)\cdot\ell\right)
\del\!\left(2\left(\tilde p-\frac\ell2\right)\cdot k\right)
\theta(r^0)
=
\del(T)\del(R-M)\theta(r^0).
\]
The direct product in \eqref{eq:app-direct-onshell-symbol-product} contains $\del(T)\del^+(R)$. In the positive-energy product the factor $\theta(r^0)$ is part of the definition rather than a restriction imposed after multiplication. The state supplies $\del(M)$, and hence $\del(M)\del(T)\del(R-M)\theta(r^0)=\del(M)\del(T)\del^+(R)$. This proves the composition rule in eq.~\eqref{eq:rescaled-onshell-weyl-product} inside the projected on-shell pairing.

On a positive-energy patch, the matter product has the formal differential representation
\[
F\star_m G
=
F\exp\left[
\frac{i\hbar}{2}
\overleftarrow{\pd}_A\Omega^{AB}
\overrightarrow{\pd}_B
\right]G,
\]
so the Moyal product can be written exactly in terms of a Bopp
operator. Expanding the exponential gives
\[
\begin{aligned}
(F\star_m G)(\tilde z)
={}&
\sum_{n=0}^{\infty}
\frac{1}{n!}
\left(+\frac{i\hbar}{2}\right)^n
\left(\pd_{A_1}\cdots\pd_{A_n}F\right)(\tilde z)
\\
&\times
\Omega^{A_1B_1}\cdots\Omega^{A_nB_n}
\pd_{B_1}\cdots\pd_{B_n}G(\tilde z).
\end{aligned}
\]
This is precisely the Taylor expansion of $F$ about the
operator-valued displacement

\[
\delta\tilde z^A
=
\frac{i\hbar}{2}\Omega^{AB}\pd_B.
\]
Consequently,
\[
(F\star_m G)(\tilde z)
=
F\left(
\tilde z^A+\frac{i\hbar}{2}\Omega^{AB}\pd_B
\right)_{\rm Bopp}
G(\tilde z),
\]
where the derivatives in the shifted argument act only on factors to their right.  This identity is exact as a formal series (on the relevant patch), and for suitable types of symbol, defines the corresponding Bopp pseudodifferential operator \cite{deGosson:2025ljg}.

Suppose we are in the special case where $\chi$ depends only on a coordinate $b$, which has conjugate momentum $q$, then the coordinate Bopp
operators commute and the expression simplifies to the exact identity
\[
e^{\frac{i}{\hbar}\chi(b)}\star_m X_W(b,q)
=
\exp\left[
\frac{i}{\hbar}
\chi\left(
b+\frac{i\hbar}{2}\pd_q
\right)
\right]X_W(b,q).
\]
\subsection{Bopp Momentum Routing}
We will use the Bopp routing below only to extract the maximally matter-pole contribution. By $[\cdots]_{\rm pole}$ we mean that all $n-1$ internal matter denominators are retained. Equal-time terms, and terms in which a numerator cancels one or more of these denominators, are outside this projection.

Suppose we have a chain of $n$ current symbols, associated with worldline $a$, all multiplied via the positive-energy product $\star_a$. We retain the common $\hbar^{-3}$ factor from each current displayed in eq.~\eqref{eq:current-symbol}, although it plays no role in the momentum routing. The momentum-space current in an individual on-shell symbol is
\[
J_{aW}^\mu(q,\tilde p_a)
=
2e_a\tilde{p}^\mu_a
\del(2\tilde p_a\cdot q).
\]
The positive-energy Bopp shift then routes the local momentum through the chain, giving
\[
\begin{aligned}
&J^{\mu_1}_{aW}(x_1)
\star_a
\cdots
\star_a
J^{\mu_n}_{aW}(x_n)
\\
&\quad=
\frac{e_a^n}{\hbar^{3n}}
\int\!\prod_{l=1}^n\hat\sd^4q_l 
e^{\frac{i}{\hbar}\sum_l q_l\cdot(x_l-\tilde x_a)}
\prod_{j=1}^{n}
(r_{j-1}+r_j)^{\mu_j}\,
\del\!\left(r_{j-1}^2-r_j^2\right)
\prod_{j=1}^{n-1}\theta(r_j^0).
\end{aligned}
\label{eq:pole-lemma-A-raw}
\]
where, with $Q\equiv\sum_l q_l$ and
\[
\bar p
&\equiv
\tilde p_a-\frac{Q}{2},
&
\bar p'
&\equiv
\tilde p_a+\frac{Q}{2},
\label{eq:pole-endpoint-momenta}
\]
the midpoint momenta at each segment are
\[
r_j
\equiv
\bar p
+
\sum_{l>j}q_l,
\qquad
j=0,\dots,n,
\qquad
r_0=\bar p',
\quad
r_n=\bar p.
\label{eq:pole-segment-momenta}
\]
The delta functions and Heaviside functions in eq.~\eqref{eq:pole-lemma-A-raw} can be rewritten in a more recognisable form as
\[
\prod_{j=1}^{n}
\del\!\left(r_{j-1}^2-r_j^2\right)
\prod_{j=1}^{n-1}\theta(r_j^0)
=
\del\!\left(2\tilde p_a\cdot Q\right)
\prod_{j=1}^{n-1}
\del\!\left(r_j^2-\bar p^{\,2}\right)\theta(r_j^0),
\label{eq:pole-telescope}
\]
where we note the usual transverse delta function. The phase also reorganises itself as
\[
\sum_{l=1}^n q_l\cdot x_l
=
\bar p'\cdot x_1
-
\bar p\cdot x_n
+
\sum_{j=1}^{n-1}
r_j\cdot\xi_j,
\qquad
\xi_j\equiv x_{j+1}-x_j.
\label{eq:pole-phase}
\]
Finally, then, this means we can write the $n$-fold star product of currents as
\[
\begin{aligned}
&J^{\mu_1}_{aW}(x_1)
\star_a
\cdots
\star_a
J^{\mu_n}_{aW}(x_n)
\\
&\quad=
\frac{e_a^n}{\hbar^{3n}}
\int\!\hat\sd^4Q 
e^{\frac{i}{\hbar}
\left(\bar p'\cdot x_1-\bar p\cdot x_n-Q\cdot\tilde x_a\right)}
\del(2\tilde p_a\cdot Q)
\\
&\qquad\quad\times
\int\!\prod_{j=1}^{n-1}\hat\sd^4r_j 
e^{\frac{i}{\hbar}\sum_{j=1}^{n-1}r_j\cdot\xi_j}
\prod_{j=1}^{n-1}
\del(r_j^2-\bar p^{\,2})\theta(r_j^0)
\prod_{j=1}^{n}
(r_{j-1}+r_j)^{\mu_j}.
\end{aligned}
\label{eq:pole-lemma-A}
\]
\subsection{Time-Ordered Products}
With this pole projection understood, we define the time-ordered star product as we did earlier
\[
\cl T_\star\!\left[
J^{\mu_1}_{aW}(x_1)\cdots J^{\mu_n}_{aW}(x_n)
\right]
\equiv
\sum_{\sigma\in S_n}
\Theta_\sigma\,
J^{\mu_{\sigma(1)}}_{aW}(x_{\sigma(1)})
\star_a\cdots\star_a
J^{\mu_{\sigma(n)}}_{aW}(x_{\sigma(n)}),
\]
where
\[
\Theta_\sigma
\equiv
\prod_{j=1}^{n-1}
\theta\!\left(
x_{\sigma(j)}^0-x_{\sigma(j+1)}^0
\right).
\]
For each permutation $\sigma$, the segment midpoint momenta are
\[
r^\sigma_j
\equiv
\bar p
+
\sum_{l>j}q_{\sigma(l)},
\qquad
r^\sigma_0=\bar p',
\quad
r^\sigma_n=\bar p.
\label{eq:pole-permuted-segments}
\]
The external momenta are on shell, $\bar p^{\,2}=m_a^2$, and for a single internal segment we can define the kernel
\[
D_+(\xi)
\equiv
\int\hat\sd^4r\,
e^{\frac{i}{\hbar}r\cdot\xi}\,
\del(r^2-m_a^2)\theta(r^0).
\]
This is the positive-energy kernel inherited from the on-shell resolution of the identity.
The two opposite time orderings then convert the pair of positive-energy kernels to the Feynman pole through
\[
\theta(-\bar p\cdot\xi)D_+(\xi)
+
\theta(\bar p\cdot\xi)D_+(-\xi)
&=
\int\hat\sd^4r\,
e^{\frac{i}{\hbar}r\cdot\xi}\,
\del(r^2-m_a^2)
\Big[
\theta(-\bar p\cdot\xi)\theta(r^0)
+
\theta(\bar p\cdot\xi)\theta(-r^0)
\Big]
\\
&=
\int\hat\sd^4r\,
e^{\frac{i}{\hbar}r\cdot\xi}\,
\frac{i}{r^2-m_a^2+i0}.
\label{eq:pole-lemma-B}
\]
In the first equality we changed $r\to-r$ only in $D_+(-\xi)$, and therefore we have not introduced any negative energy states: the term with $\theta(-r^0)$ is the oppositely oriented Fourier variable of a positive-energy intermediate state. 

The explicit numerator factors of $r^\mu$ in eq.~\eqref{eq:pole-lemma-A} can be obtained in the usual way via
\[
r^\mu e^{\frac{i}{\hbar}r\cdot\xi}
=
-i\hbar\pd_{\xi_\mu}
e^{\frac{i}{\hbar}r\cdot\xi}.
\]
We note that repeated derivatives can also act on the ordering functions and generate equal-time terms, while contractions of the numerator can cancel a matter denominator, but we will not deal with those here: they must be restored, together with e.g. the scalar-QED seagull, when constructing the complete gauge-invariant result.

After the permutation sum has paired the two orientations of every internal segment, applying eq.~\eqref{eq:pole-lemma-B} gives
\[
\begin{aligned}
&\cl T_\star\!\left[
J^{\mu_1}_{aW}(x_1)\cdots J^{\mu_n}_{aW}(x_n)
\right]_{\rm pole}
\\
&\quad=
\frac{e_a^n}{\hbar^{3n}}
\int\!\prod_l\hat\sd^4q_l 
e^{\frac{i}{\hbar}\sum_l q_l\cdot(x_l-\tilde x_a)}
\del\!\left(2\tilde p_a\cdot Q\right)
\\
&\qquad\quad\times
\sum_{\sigma\in S_n}
\prod_{j=1}^{n}
\left(r^\sigma_{j-1}+r^\sigma_j\right)^{\mu_{\sigma(j)}}
\prod_{j=1}^{n-1}
\frac{i}
{\left(r^\sigma_j\right)^2-m_a^2+i0}.
\end{aligned}
\label{eq:pole-master}
\]
Equation~\eqref{eq:pole-master} is an identity for the matter-pole sector only.

\section{Resolved Star Ordering and Nonlinear Source Insertions}
\label{app:dyson-star-ordering}
\subsection{Nonlinear Insertions Beyond Leading Order}
In section \ref{sec:partial-weyl-dyson}, we have shown that the linear $S$-matrix symbol can be expressed as an exponential of the eikonal and a coherent state of radiation with linear waveshape $\alpha_\lambda(k)$. To go beyond leading order and include the non-linear interactions in the action, we will follow the techniques developed by Schwinger and Hori \cite{Schwinger:1951qf,Hori:1952ptp}, see also \cite{Matone:2015aib}. The idea is to replace the messenger  field operator $\bb A_\mu$ by a functional derivative with respect to the source, the current symbol, 
\[
\bb A_\mu \rightarrow -i\hbar\frac{\delta}{\delta J_W^\mu}.
\]

Doing this inside the Dyson series allows us to evaluate the photon-sector time ordering and express the full $S$-matrix in terms of the linear $S$-matrix $\bb S_0$,
\[
\cl W_m[\bb S] = \cl T_\star \left\{\exp_\star\left[\frac{i}{\hbar}V_W\left(\frac{\hbar}{i}\frac{\delta}{\delta J_W}\right)\right]\star\cl W_m[\bb S_0]\right\},
\]
where the photon time ordering is understood to have been evaluated via Wick's theorem as in section \ref{sec:partial-weyl-dyson}, while the matter factors retain their inherited $\star_m$-ordering. We use the functional derivative in the usual source-insertion sense: we shift $J_W\rightarrow J_W+\eta$, differentiate with respect to the auxiliary c-number source $\eta$, and then set $\eta=0$. The $\star$-exponential is defined as a power series in the matter star product \cite{Bayen:1977ha}. We note that the photon field $\bb{A}^\mu$ lives in the electromagnetic Fock space, while the matter current $\bb{J}^\mu$ and the scalar field $\hat{\phi}(x)$ both live in the matter Hilbert space. Since these are different sectors, and we are in the interaction picture, these objects commute: $[\bb{A}^\mu(x), \bb{J}^\nu(y)] = [\bb{A}^\mu(x), \hat{\phi}(y)] = 0$. We can therefore move all matter operators to the right so they act on the state first (all under the time-ordering), being careful not to commute the matter operators, since $[\bb{J}^\mu(x),\hat\phi(y)]\neq 0$.

It is perhaps more useful to write the final state
\[
\ket{\Psi_{0,\star}[J_W]}
\equiv
\cl W_m[\bb S_0[J_W]]\ket0
=
\cl T_{\star_m}\left\{
\exp_{\star_m}\!\left[\frac{i}{\hbar}\chi_{0,\star}[J_W]\right]
\star_m
\exp_{\star_m}\!\left[G_0[J_W,a^\dagger]\right]
\right\}\ket0,
\label{eq:functional-linear-endpoint-state}
\]
where
\[
\frac{i}{\hbar}\chi_{0,\star}[J_W]
&\equiv
-\frac{1}{2\hbar^2}
\int_{x,y}\Delta^F_{AB}(x-y)
J_W^A(x)\star_m J_W^B(y),\\
G_0[J_W,a^\dagger]
&\equiv
\sum_\lambda\int d\Phi(k)\,
C_{0,\lambda}[J_W](k)a_\lambda^\dagger(k),
\]
with
\[
C_{0,\lambda}[J_W](k)
=
\frac{i}{\hbar}\int_x
J_W^A(x)u^*_{\lambda A}(x;k).
\]
The coefficient $C_{0}[J_W]$ is the bare one-radiation kernel, which is linear in the source and effectively a 3pt interaction, while $\chi_{0,\star}[J_W]$ is the unresolved eikonal. We first consider the general case, where $A$ is an arbitrary collection of indices, and then specialise to the case of a single vector index $A=\mu$ for the photon in scalar QED.

To remain general, we will consider a generic bosonic messenger field $\bb Z$ and define an arbitrary nonlinear interaction of the form
\[
V[\bb Z]
=
\sum_{n\geq2}V_n[\bb Z],
\qquad
V_n[\bb Z]
=
\frac1{n!}
\int_{x_1,\ldots,x_n}
V_n^{A_1\cdots A_n}(x_1,\ldots,x_n)
\bb Z_{A_1}(x_1)\cdots\bb Z_{A_n}(x_n).
\]
We define
\[
D_A(x)
\equiv
\frac{\hbar}{i}
\frac{\overrightarrow{\delta}}{\delta J_W^A(x)},
\label{eq:endpoint-source-derivative}
\]
where the arrow means that the derivative acts only on the objects to its right, before $J_W$ is replaced by its explicit worldline form\footnote{If we wish to keep $J_W$ literally fixed, we can define the same operation by shifting $J_W\rightarrow J_W+\eta$, differentiating with respect to the auxiliary c-number source $\eta$, and then setting $\eta=0$. The auxiliary source carries no physics; it only makes clear which occurrence of the current is to be differentiated.}. Acting with the derivative inserts a messenger field coupled to the source, and we have the identity
\[
D_{A_1}(x_1)\cdots D_{A_n}(x_n)
\cl W_m[\bb S_0[J_W]]
=
\cl W_m\!\left[
\cl T\left\{
\bb Z_{A_1}(x_1)\cdots\bb Z_{A_n}(x_n)
\bb S_0[J_W]
\right\}
\right],
\label{eq:repeated-source-insertion}
\]
where the matter products on the right are understood to retain their original $\star_m$-ordering.

Using eq.~\eqref{eq:repeated-source-insertion} once for every field in every copy of $V$, we find
\[
\cl T_{\star_m}\left\{
\exp_{\star_m}\!\left[\frac{i}{\hbar}V_W[D]\right]
\star_m
\cl W_m[\bb S_0[J_W]]
\right\}
&=
\sum_{r=0}^{\infty}
\frac1{r!}\left(\frac{i}{\hbar}\right)^r
\cl W_m\!\left[
\cl T\left\{
V[\bb Z]^r\bb S_0[J_W]
\right\}
\right]
\\
&=
\cl W_m\!\left[
\cl T\left\{
\exp\!\left[\frac{i}{\hbar}V[\bb Z]\right]
\bb S_0[J_W]
\right\}
\right]\\
&=
\cl W_m[\bb S].
\label{eq:endpoint-functional-master-symbol}
\]
Each term of order $V^r$ contains exactly $r$ nonlinear vertices and the Dyson factor $1/r!$, so eq.~\eqref{eq:endpoint-functional-master-symbol} restores the nonlinear interaction to all orders.

The state including arbitrary insertions of the nonlinear interactions is therefore generated by
\[
\cl W_m[\bb S]\ket{ 0}
=
\cl T_{\star_m}\left\{
\exp_{\star_m}\!\left[\frac{i}{\hbar}V[D]\right]
\star_m
\ket{\Psi_{0,\star}[J_W]}
\right\}.
\label{eq:endpoint-functional-master-state}
\]

When a derivative acts on the eikonal or on $G_0$, it exposes a photon by removing one factor of $J_W$. The inserted photon can then contract with one of the currents, producing an $H_A$, or remain uncontracted as an outgoing on-shell mode, producing an $\bb R_A$, where we define
\[
H_A(x)
\equiv
\frac{\delta\chi_{0,\star}[J_W]}{\delta J_W^A(x)},
\qquad
\bb R_A(x)
\equiv
D_A(x)G_0[J_W,a^\dagger]
=
\frac{\hbar}{i}
\sum_\lambda\int d\Phi(k)\,
\frac{\delta C_{0,\lambda}[J_W](k)}
{\delta J_W^A(x)}
a_\lambda^\dagger(k).
\label{eq:endpoint-H-R-definitions}
\]
A term containing $r$ factors of $\bb R$ generates $r$ on-shell particles and an exclusive $r$-radiation coefficient after the star ordering has been evaluated.

While this state is an exponential in its unevaluated form, it isn't immediately obvious that it remains an exponential after the functional derivatives have acted, since a later derivative can act on a term produced by an earlier one. The simplest place to see the structure is scalar QED, where the only interaction nonlinear in the photon field is the seagull term,
\[
V_{\rm sg}[D]
=
\int_x K_W(x)D_\mu(x)D^\mu(x).
\]
Here $K_W(x)=\sum_aK_a(x)$ is the partial Weyl symbol of the scalar term multiplying the $e_a^2A_\mu A^\mu$ interaction, and $K_a$ is the contribution supported on worldline $a$.

The two derivative slots at one seagull vertex produce the ordered sectors
\[
\cl T_{\star_m}\left\{
H_\mu\star_m H^\mu
+H_\mu\star_m\bb R^\mu
+\bb R_\mu\star_m H^\mu
+\bb R_\mu\star_m\bb R^\mu
+ D_\mu H^\mu
+D_\mu\bb R^\mu
\right\},
\label{eq:endpoint-seagull-functional-expansion}
\]
which are the elastic, one-radiation and two-radiation sectors, together with the derivative terms $D_\mu H^\mu$ and $D_\mu\bb R^\mu$. Since $C_0$ is linear in $J_W$, we have $D_\mu\bb R^\mu=0$. The remaining term $D_\mu H^\mu=\Delta^F_\mu{}^\mu(0)$ is a coincident local tadpole. For the massless scaleless integral in dimensional regularisation it is set to zero, while in another regulator it is a local term fixed by the corresponding renormalisation prescription. We will drop these same-vertex tadpoles below.

The useful point is that we can perform the source differentiations before resolving the matter star products. At any fixed order in the Dyson series, we expand the star exponential as a power series, act with the source derivatives term by term, and retain the ordered list of matter symbols. 
The operator $D_A$ acts on the star products in the usual Leibniz way, so that for any two matter symbols $F$ and $G$ we have
\[
D_A(F\star_m G)
=
(D_AF)\star_m G
+F\star_m(D_AG).
\label{eq:endpoint-source-star-leibniz}
\]
We should stress that this does not mean that the naive exponential identity $D_A\exp_{\star_m}[F]=(D_AF)\star_m\exp_{\star_m}[F]$ holds -- it generally does not -- but expanding the star exponential as a power series and applying eq.~\eqref{eq:endpoint-source-star-leibniz} term by term does give the correct result, provided it's all done under $\cl T_\star$. The star ordering can therefore be evaluated afterwards, either directly or through the Bopp shifts, without differentiating the resummed star exponential as a single object\footnote{A simple example makes this point very clear. Consider
$F=e^{J^2x/2}$ and $G=e^{Jp}$, with $[x,p]_\star=i\hbar$. The Moyal product is then $F\star G = e^{\frac{J^2x}{2}+Jp+\frac{i\hbar}{4}J^3}$, which we can differentiate with respect to $J$ to find $\pd_J(F\star G) =\left(Jx+p+\frac{3i\hbar}{4}J^2\right)(F\star G)$. If instead we differentiate the factors separately, we find $\pd_JF\star G+F\star\pd_JG = \left(Jx+p+\frac{i\hbar}{4}J^2\right)(F\star G)$, which differs from the direct differentiation by $3i\hbar J^2/4$.

The same result follows directly from the Leibniz rule,
\[
\begin{aligned}
(\pd_JF)\star G+F\star(\pd_JG)
&=
(JxF)\star G+F\star(pG)\\
&=
\left(Jx+\frac{i\hbar}{2}J^2\right)(F\star G)
+
\left(p+\frac{i\hbar}{4}J^2\right)(F\star G),
\end{aligned}
\]
which reproduces the extra $3i\hbar J^2/4$ term. The apparent residual
therefore comes from the star products themselves. The same mechanism
works term by term in a star exponential, where repeated use of the
Leibniz rule generates the differentiated factor at every position in
the product.}.

To simplify things, let's introduce the notation
\[
Q_{i\mu}\equiv D_{i\mu}\left(\frac{i}{\hbar}\chi_{0,\star}+G_0\right)
=H_\mu(x_i)+\bb R_\mu(x_i),
\qquad
\Delta^F_{ij,\mu\nu}
\equiv
D_{i\mu}Q_{j\nu}
=D_{i\mu}H_\nu(x_j).
\]
Since $\chi_{0,\star}$ is quadratic in $J_W$ and $G_0$ is linear, we also have $D_i\Delta^F_{jk}=0$. Each source derivative inserts a photon at a nonlinear vertex. If it acts directly on the linear state, it produces a $Q_i$: the photon either ends on one of the linear source currents, giving $H_A$, or becomes an outgoing on-shell mode, giving $\bb R_\mu$. Since $H_\mu$ still contains one factor of $J_W$, a later derivative can remove that source endpoint, leaving the propagator $\Delta^F_{\mu\nu}$ that connects the new nonlinear vertex to the earlier one. Acting on $\bb R_\mu$ gives zero, and a further derivative of $\Delta^F_{\mu\nu}$ also vanishes, so these are the only possibilities.

Suppressing Lorentz indices for the moment, the first few cases are
\[
\begin{aligned}
D_1\ket{\Psi_{0,\star}}
&=
\cl T_{\star_m}\left\{
Q_1\star_m\ket{\Psi_{0,\star}}
\right\},
\\
D_2D_1\ket{\Psi_{0,\star}}
&=
\cl T_{\star_m}\left\{
\left[
Q_1\star_mQ_2+\Delta^F_{12}
\right]
\star_m\ket{\Psi_{0,\star}}
\right\},
\\
D_3D_2D_1\ket{\Psi_{0,\star}}
&=
\cl T_{\star_m}\left\{
\left[
Q_1\star_mQ_2\star_mQ_3
+\Delta^F_{12}Q_3
+\Delta^F_{13}Q_2
+\Delta^F_{23}Q_1
\right]
\star_m\ket{\Psi_{0,\star}}
\right\}.
\end{aligned}
\]
We see then that repeated differentiation simply generates the usual Wick contractions, with the star ordering left unresolved until the contraction pattern has been fixed.

The products between the $Q_i$ are matter star products, and the labels $1,2,3$ do not fix their chronological order: the single outer $\cl T_{\star_m}$ must then be evaluated to get the correct ordering, while the propagators $\Delta^F_{ij}$ are just spectators.

For the seagull interaction, it is useful to picture each nonlinear vertex as carrying two photon slots. Each slot either ends on the linear state, producing a $Q=H+\bb R$, or joins a slot on another vertex, producing a propagator $\Delta^F$. Same-vertex pairings give the local tadpoles removed above. At order $V_{\rm sg}^r$, the derivatives generate all possible diagrams built from $r$ such vertices, and we define $\cl B_r$ to be the sum of those diagrams in which all $r$ seagull vertices belong to one connected graph (in the graph-theoretic sense), with a lone vertex connected by convention.

To make this concrete, let's look at the connected contributions at one, two and three seagull vertices. Using the shorthand
\[
K_i\equiv K_W(x_i),
\qquad
\int_{1,\ldots,r}\equiv\int d^4x_1\cdots d^4x_r,
\]
we have
\[
\begin{aligned}
\cl B_1
&\equiv
\int_1
K_1\star_m Q_{1\mu}\star_m Q_1^{\mu},
\\[2mm]
\cl B_2
&\equiv
\int_{1,2}
K_1\star_m K_2\star_m
\left[
4\Delta_{12}^{F\,\mu\nu}
Q_{1\mu}\star_m Q_{2\nu}
+2\Delta^F_{12,\mu\nu}\Delta_{21}^{F\,\nu\mu}
\right],
\\[2mm]
\cl B_3
&\equiv
8\int_{1,2,3}
K_1\star_m K_2\star_m K_3\star_m
\left[
\sum_{\operatorname{cyc}(1,2,3)}
\Delta_{12}^{F\,\mu\nu}
\Delta^F_{23,\nu\rho}
Q_{1\mu}\star_m Q_3^{\rho}
+
\Delta^F_{12,\mu\nu}
\Delta_{23}^{F\,\nu\rho}
\Delta^F_{31,\rho}{}^{\mu}
\right].
\end{aligned}
\label{eq:endpoint-connected-seagull-kernels}
\]
The corresponding photon-slot contraction graphs are shown in Fig.~\ref{fig:endpoint-connected-seagull-kernels}.
\begin{figure}[H]
\centering
\begin{tikzpicture}[scale=0.90,transform shape]
\begin{feynman}

\node at (-6.65,1.65) {$\cl B_1:$};
\vertex (b1ql) at (-5.65,1.65) {$Q_1$};
\vertex[dot,label=below:{$K_1$}] (b1k1) at (-4.45,1.65) {};
\vertex (b1qr) at (-3.25,1.65) {$Q_1$};
\diagram*{
  (b1ql) -- [photon] (b1k1) -- [photon] (b1qr),
};

\node at (-1.95,1.65) {$\cl B_2:$};
\node at (-1.15,1.65) {$4$};
\vertex (b2ql) at (-0.55,1.65) {$Q_1$};
\vertex[dot,label=below:{$K_1$}] (b2k1) at (0.45,1.65) {};
\vertex[dot,label=below:{$K_2$}] (b2k2) at (1.95,1.65) {};
\vertex (b2qr) at (2.95,1.65) {$Q_2$};
\diagram*{
  (b2ql) -- [photon] (b2k1),
  (b2k1) -- [photon,edge label={$\Delta^F_{12}$}] (b2k2),
  (b2k2) -- [photon] (b2qr),
};
\node at (3.55,1.65) {$+$};
\node at (4.05,1.65) {$2$};
\vertex[dot,label=below:{$K_1$}] (b2dk1) at (4.70,1.65) {};
\vertex[dot,label=below:{$K_2$}] (b2dk2) at (6.20,1.65) {};
\diagram*{
  (b2dk1) -- [photon,half left,looseness=1.35] (b2dk2),
  (b2dk1) -- [photon,half right,looseness=1.35] (b2dk2),
};

\node at (-6.65,-1.25) {$\cl B_3:$};
\node at (-5.30,-1.25) {$8\!\displaystyle\sum_{\rm cyc}$};
\vertex (b3ql) at (-4.05,-1.25) {$Q_1$};
\vertex[dot,label=below:{$K_1$}] (b3k1) at (-3.05,-1.25) {};
\vertex[dot,label=below:{$K_2$}] (b3k2) at (-1.55,-1.25) {};
\vertex[dot,label=below:{$K_3$}] (b3k3) at (-0.05,-1.25) {};
\vertex (b3qr) at (0.95,-1.25) {$Q_3$};
\diagram*{
  (b3ql) -- [photon] (b3k1),
  (b3k1) -- [photon,edge label={$\Delta^F_{12}$}] (b3k2),
  (b3k2) -- [photon,edge label={$\Delta^F_{23}$}] (b3k3),
  (b3k3) -- [photon] (b3qr),
};
\node at (1.65,-1.25) {$+$};
\node at (2.15,-1.25) {$8$};
\vertex[dot,label=below left:{$K_1$}] (b3tk1) at (3.10,-1.75) {};
\vertex[dot,label=below right:{$K_2$}] (b3tk2) at (5.10,-1.75) {};
\vertex[dot,label=above:{$K_3$}] (b3tk3) at (4.10,0.00) {};
\diagram*{
  (b3tk1) -- [photon,edge label'={$\Delta^F_{12}$}] (b3tk2),
  (b3tk2) -- [photon,edge label'={$\Delta^F_{23}$}] (b3tk3),
  (b3tk3) -- [photon,edge label'={$\Delta^F_{31}$}] (b3tk1),
};
\end{feynman}
\end{tikzpicture}
\caption{Connected photon-slot contractions contributing to $\cl B_1$, $\cl B_2$ and $\cl B_3$. A filled dot is a seagull insertion $K_i$, a photon ending on $Q_i$ is an uncontracted slot acting on the linear state, while a photon between vertices is a Feynman contraction $\Delta^F_{ij}$.}
\label{fig:endpoint-connected-seagull-kernels}
\end{figure}
The two terms in $\cl B_2$ contain one and two propagators joining the two vertices, with factors $4$ and $2$ coming from the number of ways we can contract the derivatives. The two terms in $\cl B_3$ are the chain of seagulls terminated by $Q$'s and a seagull triangle, each with the overall factor $8$, visualised in figure \ref{fig:endpoint-connected-seagull-kernels}. With these definitions, the first three powers of the seagull operator give
\[
\begin{aligned}
&\cl T_{\star_m}\left\{
V_{\rm sg}[D]\star_m\ket{\Psi_{0,\star}}
\right\}
=
\cl T_{\star_m}\left\{
\cl B_1\star_m\ket{\Psi_{0,\star}}
\right\},
\\
&\cl T_{\star_m}\left\{
V_{\rm sg}[D]\star_m V_{\rm sg}[D]\star_m\ket{\Psi_{0,\star}}
\right\}
=
\cl T_{\star_m}\left\{
\left(
\cl B_1\star_m\cl B_1+\cl B_2
\right)
\star_m\ket{\Psi_{0,\star}}
\right\},
\\
&\cl T_{\star_m}\left\{
V_{\rm sg}[D]\star_m V_{\rm sg}[D]\star_m V_{\rm sg}[D]
\star_m\ket{\Psi_{0,\star}}
\right\}
=
\cl T_{\star_m}\left\{
\left(
\cl B_1\star_m\cl B_1\star_m\cl B_1
+3\cl B_1\star_m\cl B_2
+\cl B_3
\right)
\star_m\ket{\Psi_{0,\star}}
\right\}.
\end{aligned}
\label{eq:endpoint-seagull-first-orders-star}
\]

We can now write the corresponding terms in the exponential itself. Defining $\lambda\equiv i/\hbar$, the Dyson expansion through three nonlinear vertices is
\[
\begin{aligned}
&\cl T_{\star_m}\left\{
\exp_{\star_m}\!\left[\lambda V_{\rm sg}[D]\right]
\star_m\ket{\Psi_{0,\star}}
\right\}
\\
&\quad=
\cl T_{\star_m}\Biggl\{
\left[
1
+\lambda\cl B_1
+\frac{\lambda^2}{2!}
\left(
\cl B_1\star_m\cl B_1
+\cl B_2
\right)
\right.
\\[-1mm]
&\qquad\left.
+\frac{\lambda^3}{3!}
\left(
\cl B_1\star_m\cl B_1\star_m\cl B_1
+3\cl B_1\star_m\cl B_2
+\cl B_3
\right)
+\cl O(V_{\rm sg}^4)
\right]
\star_m\ket{\Psi_{0,\star}}
\Biggr\}.
\end{aligned}
\label{eq:endpoint-seagull-dyson-explicit}
\]
To compare this with an exponential of connected terms, define the connected generator at cubic order as
\[
\cl B_{\rm conn}^{[3]}
\equiv
\lambda\cl B_1
+\frac{\lambda^2}{2!}\cl B_2
+\frac{\lambda^3}{3!}\cl B_3.
\label{eq:endpoint-seagull-connected-generator-cubic}
\]
We can star exponentiate this and expand to get
\[
\begin{aligned}
\exp_{\star_m}\!\left[\cl B_{\rm conn}^{[3]}\right]
={}&
1
+\lambda\cl B_1
+\frac{\lambda^2}{2!}
\left(
\cl B_1\star_m\cl B_1
+\cl B_2
\right)
\\
&+
\frac{\lambda^3}{3!}
\left[
\cl B_1\star_m\cl B_1\star_m\cl B_1
+\frac32
\left(
\cl B_1\star_m\cl B_2
+\cl B_2\star_m\cl B_1
\right)
+\cl B_3
\right]
+\cl O(V_{\rm sg}^4).
\end{aligned}
\label{eq:endpoint-seagull-connected-exponential-cubic}
\]
Inside the single global $\cl T_{\star_m}$, the two mixed products $\cl B_1\star_m\cl B_2$ and $\cl B_2\star_m\cl B_1$ give the same integrated contribution after the dummy vertex variables are relabelled. This is not to say we identify $\cl B_1\star_m\cl B_2=\cl B_2\star_m\cl B_1$ in general,  the equality holds only after the time ordering. Their sum therefore gives the factor $3\cl B_1\star_m\cl B_2$ in eq.~\eqref{eq:endpoint-seagull-dyson-explicit}. We see then that the exponential of the connected terms reproduces the Dyson expansion through cubic order.

The same pattern continues at higher orders. The derivatives generate all possible ways of joining the nonlinear vertices by propagators. Some diagrams are fully connected, while others split into several separate connected groups, and are therefore disconnected. The Dyson series then sums these groups into the exponential of the connected contributions only, reproducing the disconnected terms in the expansion above. The star product changes the momentum routing associated with a diagram through the Bopp shifts, but it does not change which vertices belong to which connected piece\footnote{In some sense, this is the partial Weyl transform of the linked-cluster expansion: the product law is altered, but the contraction graph or its connected components are not.}. As a formal perturbative identity, we therefore find
\[
\cl T_{\star_m}\left\{
\exp_{\star_m}\!\left[\frac{i}{\hbar}V_{\rm sg}[D]\right]
\star_m\ket{\Psi_{0,\star}[J_W]}
\right\}
=
\cl T_{\star_m}\left\{
\exp_{\star_m}\!\left[
\sum_{r\geq1}\frac1{r!}
\left(\frac{i}{\hbar}\right)^r
\cl B_r
\right]
\star_m\ket{\Psi_{0,\star}[J_W]}
\right\}.
\label{eq:endpoint-connected-seagull-star-exponentiation}
\]
The outer $\cl T_{\star_m}$ acts on all elementary matter insertions contained in the $\cl B_r$, rather than treating each connected block as an indivisible object. The $\cl B_r$ contain both radiative and non-radiative terms, so the exponentiation holds in both sectors. The final state is not generally coherent, however, since the connected kernels need not be linear in the creation operators.

We should finally note that the star ordering cannot simply be dropped in the classical limit. The Fourier phases in the scattering symbols cancel the explicit powers of $\hbar$ in the Moyal product and generate the classical Bopp shifts which route recoil through each complete amplitude or cut kernel. Once these internal shifts have been evaluated, a further global endpoint reduction, if taken on the complete expression, replaces the remaining outer star products, star exponential and matter star ordering by their ordinary counterparts. In the radiative construction below, we will instead endpoint-reduce the completed conservative block while retaining its outer star product with the radiation kernel.

To see how the ordering works in practice, we will now consider the five-point amplitude, where the Bopp shifts route momentum from the exchange to the emission and the seagull term supplies the contact contribution required by gauge invariance.

\subsection{Radiation from Star Ordering}\label{sec:radiation-from-star-ordering}
Let's now derive the five-point amplitude --- the first non-trivial correction to the leading eikonal with radiation. Let's imagine that particle $a$ emits a photon of momentum $k$, and particle $b$ is a spectator, with a propagator of exchanged momentum $\ell$ from $a$ to $b$.

The endpoint approximation, when applied too early, doesn't work for radiation, and the reason is obvious: the naive term in the Dyson expansion gives a straight line three-point interaction, and straight lines don't radiate. As we have seen, however, the Bopp shift routes momentum from one current to another, and hence what we need to do is also obvious: we need to keep the star products so that they can route momentum from the currents in the eikonal, to the current sat in the radiation kernel, giving us a dynamical interaction that does produce physical radiation and recoil. 

Again starting from eq. \eqref{eq:three-current-product2}, but this time keeping the star products, we note that the spectator current commutes with the other two, and hence we can write
\[
\int_{x_1,x_2,x_3}J_{bW}^\rho(x_2)
\left[
\theta(x_3^0-x_1^0)
J_{aW}^\mu(x_3)\star_m J_{aW}^\nu(x_1)
+
\theta(x_1^0-x_3^0)
J_{aW}^\nu(x_1)\star_m J_{aW}^\mu(x_3)
\right].
\label{eq:partial-endpoint-radiative-orderings}
\]
The first term is exchange before emission, while the second is emission before exchange. The outgoing photon mode carries transfer $-k$ into the matter line, while the exchanged photon carries transfer momentum $\ell$. The Bopp shift formula in eq. \eqref{eq:bopp-shift-identity} means that, in momentum space, we can write the star products as 
\[
J_{aW}^\mu(-k)\star_m J_{aW}^\nu(\ell) &= j_a^\mu\left(\tilde p_a +\frac{\ell}{2},-k\right)j_a^\nu\left(\tilde p_a + \frac{k}{2},\ell\right),\\
J_{aW}^\nu(\ell)\star_m J_{aW}^\mu(-k) &= j_a^\nu\left(\tilde p_a -\frac{k}{2},\ell\right)j_a^\mu\left(\tilde p_a - \frac{\ell}{2},-k\right).
\]
We see then that the exchange current is evaluated at $\tilde p_a \pm k/2$, while the emission current is evaluated at $\tilde p_a \pm \ell/2$. In the second ordering these become $\tilde p_a-k/2$ and $\tilde p_a-\ell/2$, respectively. This is exactly what we need to route momentum from the exchange to the emission, and hence we can see that the star products are crucial for producing physical radiation. 
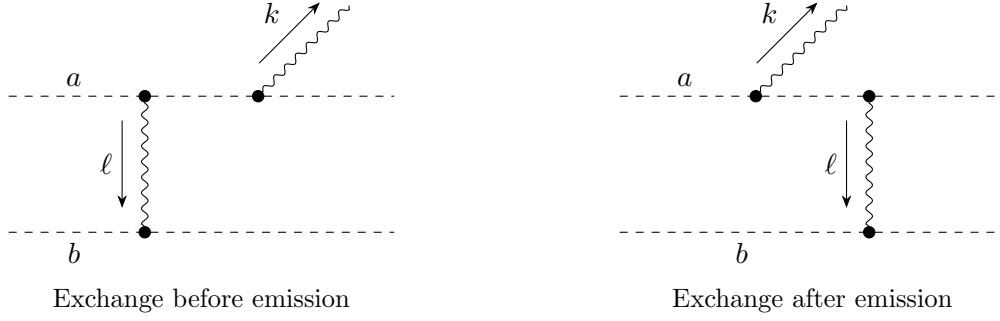
\begin{figure}[H]
\centering
\begin{minipage}{0.48\textwidth}
\centering
\begin{tikzpicture}
\begin{feynman}
\vertex (ai) at (-1.8,1);
\vertex[dot] (ax) at (0,1) {};
\vertex[dot] (ae) at (1.5,1) {};
\vertex (ao) at (3.3,1);
\vertex (bi) at (-1.8,-0.8);
\vertex[dot] (bx) at (0,-0.8) {};
\vertex (bo) at (3.3,-0.8);
\vertex (k) at (2.7,2.2);
\diagram*{
(ai) -- [scalar, edge label=$a$] (ax) -- [scalar] (ae) -- [scalar] (ao),
(bi) -- [scalar, edge label'=$b$] (bx) -- [scalar] (bo),
(ax) -- [photon, momentum'=$\ell$] (bx),
(ae) -- [photon, momentum=$k$] (k),
};
\end{feynman}
\end{tikzpicture}

\small Exchange before emission
\end{minipage}
\hfill
\begin{minipage}{0.48\textwidth}
\centering
\begin{tikzpicture}
\begin{feynman}
\vertex (ai) at (-1.8,1);
\vertex[dot] (ae) at (0,1) {};
\vertex[dot] (ax) at (1.5,1) {};
\vertex (ao) at (3.3,1);
\vertex (bi) at (-1.8,-0.8);
\vertex[dot] (bx) at (1.5,-0.8) {};
\vertex (bo) at (3.3,-0.8);
\vertex (k) at (1.2,2.2);
\diagram*{
(ai) -- [scalar, edge label=$a$] (ae) -- [scalar] (ax) -- [scalar] (ao),
(bi) -- [scalar, edge label'=$b$] (bx) -- [scalar] (bo),
(ax) -- [photon, momentum'=$\ell$] (bx),
(ae) -- [photon, momentum=$k$] (k),
};
\end{feynman}
\end{tikzpicture}

\small Exchange after emission
\end{minipage}
\end{figure}
We now have to evaluate the star-orderings, which turns these shifts into the matter poles as in eq.~\eqref{eq:pole-master}, following appendix~\ref{app:bopp-matter-poles}. Specifically, applying the $n=2$ case of eq.~\eqref{eq:pole-master} to the two orderings in eq.~\eqref{eq:partial-endpoint-radiative-orderings} gives
\[
\left(p_{a,{\rm out}}+k\right)^2-m_a^2+i\epsilon
=
2\tilde p_a\cdot k+\ell\cdot k+i\epsilon,
\]
and
\[
\left(p_{a,{\rm in}}-k\right)^2-m_a^2+i\epsilon
=
-2\tilde p_a\cdot k+\ell\cdot k+i\epsilon,
\]
for exchange before emission and emission before exchange, respectively.

Collecting the two terms gives the tree-level pole contribution in which particle $a$ radiates,
\[
\cl A_{5,a}^{(0),\rm{pole}}(k,\ell)
=
-\frac{2e_a^2e_b}{\ell^2+i\epsilon}
\tilde p_{b\nu}\epsilon_{\lambda\mu}^*(k)
\left[
\frac{
(2\tilde p_a+\ell)^\mu
(2\tilde p_a+k)^\nu
}{
2\tilde p_a\cdot k+\ell\cdot k+i\epsilon
}
-
\frac{
(2\tilde p_a-\ell)^\mu
(2\tilde p_a-k)^\nu
}{
2\tilde p_a\cdot k-\ell\cdot k-i\epsilon
}
\right],
\label{eq:partial-endpoint-five-point-amplitude}
\]
with the other ordering giving the same result with $a\leftrightarrow b$.

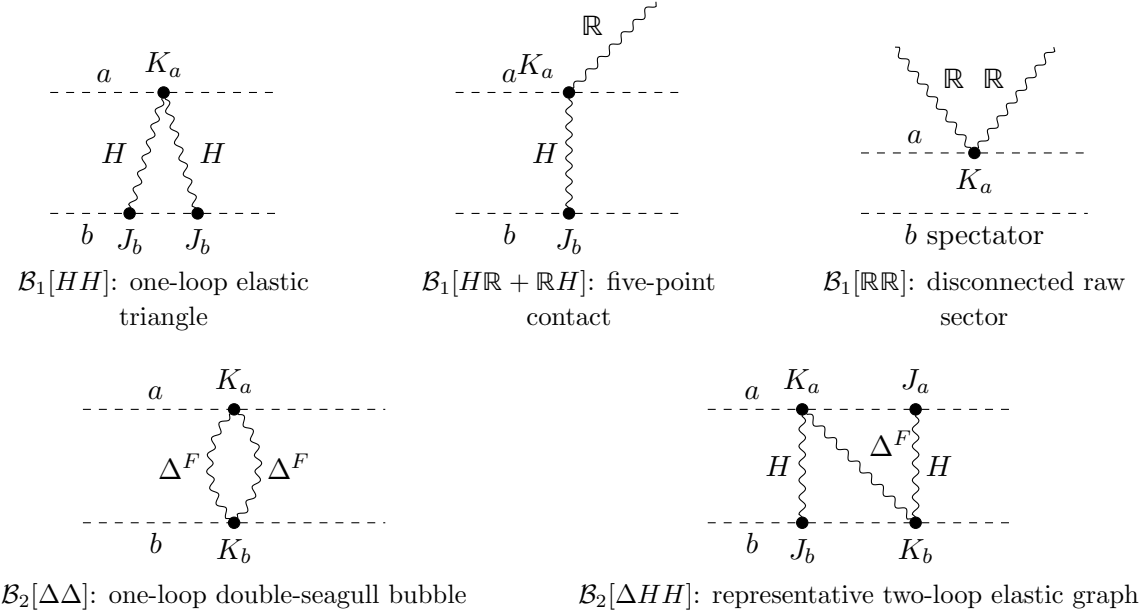
\begin{figure}[H]
\centering
\begin{minipage}[t]{0.31\textwidth}
\centering
\begin{tikzpicture}
\path[use as bounding box] (-1.7,-1.35) rectangle (1.7,2.1);
\begin{feynman}
\vertex (ai) at (-1.5,0.8);
\vertex[dot,label=above:{$K_a$}] (s) at (0,0.8) {};
\vertex (ao) at (1.5,0.8);
\vertex (bi) at (-1.5,-0.8);
\vertex[dot,label=below:{$J_b$}] (b1) at (-0.45,-0.8) {};
\vertex[dot,label=below:{$J_b$}] (b2) at (0.45,-0.8) {};
\vertex (bo) at (1.5,-0.8);
\diagram*{
(ai) -- [scalar, edge label=$a$] (s) -- [scalar] (ao),
(bi) -- [scalar, edge label'=$b$] (b1) -- [scalar] (b2) -- [scalar] (bo),
(s) -- [photon] (b1),
(s) -- [photon] (b2),
};
\end{feynman}
\node at (-0.65,0) {$H$};
\node at (0.65,0) {$H$};
\end{tikzpicture}

\small $\cl B_1[H H]$: one-loop elastic triangle
\end{minipage}
\hfill
\begin{minipage}[t]{0.31\textwidth}
\centering
\begin{tikzpicture}
\path[use as bounding box] (-1.7,-1.35) rectangle (1.7,2.1);
\begin{feynman}
\vertex (ai) at (-1.5,0.8);
\vertex[dot,label=above left:{$K_a$}] (s) at (0,0.8) {};
\vertex (ao) at (1.5,0.8);
\vertex (bi) at (-1.5,-0.8);
\vertex[dot,label=below:{$J_b$}] (b) at (0,-0.8) {};
\vertex (bo) at (1.5,-0.8);
\vertex (r) at (1.15,2.0);
\diagram*{
(ai) -- [scalar, edge label=$a$] (s) -- [scalar] (ao),
(bi) -- [scalar, edge label'=$b$] (b) -- [scalar] (bo),
(s) -- [photon, edge label'=$H$] (b),
(s) -- [photon, edge label=$\bb R$] (r),
};
\end{feynman}
\end{tikzpicture}

\small $\cl B_1[H\bb R+\bb R H]$: five-point contact
\end{minipage}
\hfill
\begin{minipage}[t]{0.31\textwidth}
\centering
\begin{tikzpicture}
\path[use as bounding box] (-1.7,-1.35) rectangle (1.7,2.1);
\begin{feynman}
\vertex (ai) at (-1.5,0);
\vertex[dot,label=below:{$K_a$}] (s) at (0,0) {};
\vertex (ao) at (1.5,0);
\vertex (bi) at (-1.5,-0.8);
\vertex (bo) at (1.5,-0.8);
\vertex (r1) at (-1.05,1.4);
\vertex (r2) at (1.05,1.4);
\diagram*{
(ai) -- [scalar, edge label=$a$] (s) -- [scalar] (ao),
(bi) -- [scalar, edge label'={$b\ {\rm spectator}$}] (bo),
(s) -- [photon, edge label'=$\bb R$] (r1),
(s) -- [photon, edge label=$\bb R$] (r2),
};
\end{feynman}
\end{tikzpicture}

\small $\cl B_1[\bb R\bb R]$: disconnected raw sector
\end{minipage}
\par\medskip
\begin{minipage}[t]{0.43\textwidth}
\centering
\begin{tikzpicture}
\path[use as bounding box] (-2.2,-1.35) rectangle (2.2,1.55);
\begin{feynman}
\vertex (ai) at (-2.0,0.75);
\vertex[dot,label=above:{$K_a$}] (sa) at (0,0.75) {};
\vertex (ao) at (2.0,0.75);
\vertex (bi) at (-2.0,-0.75);
\vertex[dot,label=below:{$K_b$}] (sb) at (0,-0.75) {};
\vertex (bo) at (2.0,-0.75);
\diagram*{
(ai) -- [scalar, edge label=$a$] (sa) -- [scalar] (ao),
(bi) -- [scalar, edge label'=$b$] (sb) -- [scalar] (bo),
(sa) -- [photon,bend left=42] (sb),
(sa) -- [photon,bend right=42] (sb),
};
\end{feynman}
\node at (-0.72,0) {$\Delta^F$};
\node at (0.72,0) {$\Delta^F$};
\end{tikzpicture}

\small $\cl B_2[\Delta\Delta]$: one-loop double-seagull bubble
\end{minipage}
\hfill
\begin{minipage}[t]{0.49\textwidth}
\centering
\begin{tikzpicture}
\path[use as bounding box] (-2.25,-1.35) rectangle (2.25,1.55);
\begin{feynman}
\vertex (ai) at (-2.0,0.75);
\vertex[dot,label=above:{$K_a$}] (ka) at (-0.75,0.75) {};
\vertex[dot,label=above:{$J_a$}] (ja) at (0.75,0.75) {};
\vertex (ao) at (2.0,0.75);
\vertex (bi) at (-2.0,-0.75);
\vertex[dot,label=below:{$J_b$}] (jb) at (-0.75,-0.75) {};
\vertex[dot,label=below:{$K_b$}] (kb) at (0.75,-0.75) {};
\vertex (bo) at (2.0,-0.75);
\diagram*{
(ai) -- [scalar, edge label=$a$] (ka) -- [scalar] (ja) -- [scalar] (ao),
(bi) -- [scalar, edge label'=$b$] (jb) -- [scalar] (kb) -- [scalar] (bo),
(ka) -- [photon,edge label'=$H$] (jb),
(ja) -- [photon,edge label=$H$] (kb),
(ka) -- [photon,edge label=$\Delta^F$] (kb),
};
\end{feynman}
\end{tikzpicture}

\small $\cl B_2[\Delta H H]$: representative two-loop elastic graph
\end{minipage}
\caption{Amplitude embeddings of the scalar-QED seagull kernels. Vertices labelled $K_a$ and $K_b$ are four-point seagull insertions on the corresponding scalar lines, while $J_a$ and $J_b$ are ordinary cubic scalar-current vertices. The $\cl B_1[\bb R\bb R]$ panel is intentionally disconnected: the free $b$ line is a spectator, and this raw two-radiation coefficient contributes to connected scattering only after composition with an elastic block. The lower row shows representative elastic embeddings of the two terms in $\cl B_2$. The exchange $a\leftrightarrow b$, other allowed worldline assignments and the matter orderings are implicit.}
\label{fig:endpoint-seagull-terms}
\end{figure}

As is easily checked, the Ward identity is not satisfied for this amplitude: we need to include the contribution coming from the seagull part of the Dyson expansion. To restore nonlinear vertices, we will use the techniques of the last subsection. Here, the $H H$, $H\bb R+\bb R H$ and $\bb R\bb R$ source sectors generate raw elastic, one-radiation and two-radiation contact coefficients, respectively, once their matter star ordering has been evaluated. In the scalar QED five-point example above, the $H\bb R+\bb R H$ sectors generate the additional seagull contribution to the five-point amplitude as required by gauge invariance. By itself, the $\bb R\bb R$ graph has a disconnected spectator matter line and is therefore a raw two-radiation source coefficient, not a connected $2\to2+2\gamma$ amplitude. It contributes to connected scattering only after it is composed with the eikonal or another elastic source through the matter star product, similarly to the bare current $C_0$. These sectors, together with two elastic embeddings of $\cl B_2$, are illustrated in Fig.~\ref{fig:endpoint-seagull-terms}.

For the elastic $2\to2$ sector, every open slot is resolved as $Q\to H$. The first graph in the lower row comes from the $2\Delta^F\Delta^F$ term and is the one-loop double-seagull bubble. The second comes from the $4\Delta^F H H$ term and shows one representative two-loop four-point embedding. The factors $2$ and $4$ are the source-contraction multiplicities appearing in $\cl B_2$; in a conventional amplitude they combine with the Dyson factorial and the usual diagram symmetry factors.

We see then that iterating the action of the potential in eq.~\eqref{eq:endpoint-functional-master-state} generates arbitrary numbers of nonlinear insertions. In particular here, repeated seagull insertions generate the corresponding triangle-type corrections to the eikonal, which can be composed with the radiation kernel, while the $H\bb R$ sectors give rise to contact terms. Extending this schematic organisation to a general $n$-loop five-point amplitude also requires the remaining interaction vertices and connected subtractions at that order.

The term containing one creation operator  generates the seagull part of the five-point kernel from $C_0$. Defining
\[
r_i^\mu(x)
\equiv
\frac{\hbar}{i}
\frac{\delta C_0(i;J_W)}{\delta J_{W\mu}(x)},
\]
we obtain
\[
\Delta_{\rm sg}C_1^{(0)}(i)
=
\frac{i}{\hbar}
\cl T_{\star_m}\left\{
\int_x K_W(x)\star_m
\left[
H_\mu(x)\star_m r_i^\mu(x)
+r_{i\mu}(x)\star_m H^\mu(x)
\right]
\right\}.
\label{eq:endpoint-sqed-one-radiation-correction}
\]
For the linearised source,
\[
C_{0,\lambda}[J_W](k)
=
\frac{i}{\hbar}
\int_x J_W^\mu(x)u^*_{\lambda\mu}(x;k),
\qquad
r_{\lambda}^\mu(x;k)
=
u_{\lambda}^{*\mu}(x;k).
\label{eq:endpoint-sqed-linear-radiative-derivative}
\]
Equation~\eqref{eq:endpoint-sqed-one-radiation-correction} then becomes
\[
C_{1,\lambda}^{(0),\rm sg}(k)
=
\frac{2i}{\hbar}
\sum_{a\neq b}
\int_x
K_a(x)H_{b\mu}(x)u_\lambda^{*\mu}(x;k).
\label{eq:endpoint-sqed-five-point-contact}
\]
In momentum space this is
\[
C_{1,\lambda}^{(0),\rm sg}(k;\tilde z)
&=
\frac{2i}{\hbar^{3/2}}
\sum_{a\neq b}
\int\dd^4\ell\,
K_a(\ell-k)
\frac{J_b(-\ell)\cdot\epsilon_\lambda^*(k)}
{\ell^2+i\epsilon}
\\
&=
\frac{i}{\hbar^{3/2}}
\sum_{a\neq b}
\int\dd^4\ell\,
e^{+\frac{i}{\hbar}[(\ell-k)\cdot b_a-\ell\cdot b_b]}
\del(2\tilde p_a\cdot(\ell-k))
\del(2\tilde p_b\cdot\ell)
\cl A_{5,a}^{\rm sg}(k,\ell).
\label{eq:endpoint-sqed-five-point-contact-momentum}
\]
where the stripped amplitude is
\[
\cl A_{5,a}^{\rm sg}(k,\ell)
=
\frac{4e_a^2e_b}{\ell^2+i\epsilon}
\tilde p_b\cdot\epsilon_\lambda^*(k).
\]
Therefore, eq.~\eqref{eq:endpoint-sqed-five-point-contact-momentum} is the tree-level five-point term coming from the seagull term. It is the contact contribution missing from eq.~\eqref{eq:partial-endpoint-five-point-amplitude}, which now becomes
\[
\cl A_{5,a}^{(0),\rm{full}}(k,\ell)
=
-\frac{2e_a^2e_b}{\ell^2+i\epsilon}
\tilde p_{b\nu}\epsilon_{\lambda\mu}^*(k)
\left[
\frac{
(2\tilde p_a+\ell)^\mu
(2\tilde p_a+k)^\nu
}{
2\tilde p_a\cdot k+\ell\cdot k+i\epsilon
}
-
\frac{
(2\tilde p_a-\ell)^\mu
(2\tilde p_a-k)^\nu
}{
2\tilde p_a\cdot k-\ell\cdot k-i\epsilon
}
-2\eta^{\mu\nu}
\right],
\label{eq:partial-endpoint-full-five-point-amplitude}
\] 
which correctly satisfies the Ward identity and vanishes when $\epsilon^*_{\lambda\mu}(k)\to k_\mu$.
Indeed, replacing the radiative polarisation by $k_\mu$ converts each numerator into the difference of its adjacent scalar inverse propagators, and the two resulting endpoint terms are cancelled by the $-2\eta^{\mu\nu}$ seagull contribution.

After acting with the nonlinear operator, removing the lower elastic and radiative iterations, and only then performing the endpoint approximation on the completed conservative sector, the most general outgoing radiation state has the form
\[
S_W| 0\rangle
=
\cl{T}_\star\left\{\exp\!\left[\frac{i}{\hbar}\chi(\tilde z)\right]\star_m
\exp_{\star_m} F_{\tilde z}[a^\dagger]\right\}| 0\rangle,
\label{eq:radiation-exclusive-state-new}
\]
where
\[
F_{\tilde z}[a^\dagger]
=
\sum_{r\geq1}\frac1{r!}
\int_{1\cdots r}
\DC{r}(1,\ldots,r;\tilde z)
a_1^\dagger\cdots a_r^\dagger ,~~~~\rm{and}~~~~\int_i\equiv \sum_{\lambda_i}\int d\Phi(k_i).
\]
The coefficients $\DC{r}$ are the exclusive radiation kernels, and we note that the matter $\star_m$-ordering applies to the matter dependence of the $\DC{r}$ terms, which will generally depend on a phase-space coordinate $z$.
The full connected exclusive one-radiation kernel is the term linear in the creation operator in $F_{\tilde z}[a^\dagger]$,
\[
C_{1,\lambda}(k;\tilde z)
\equiv
\left.
\frac{\delta F_{\tilde z}[a^\dagger]}
{\delta a_\lambda^\dagger(k)}
\right|_{a^\dagger=0}.
\label{eq:full-C1-generator-definition}
\]
This defines $C_{1,\lambda}$ to all perturbative orders after all iterations have been subtracted. At the leading scalar-QED order considered here, it is given by
\[
C_{1,\lambda}^{(0)}(k;\tilde z)
&=
C_{1,\lambda}^{(0),\rm pole}(k;\tilde z)
+
C_{1,\lambda}^{(0),\rm sg}(k;\tilde z)
\\
&=
\frac{i}{\hbar^{3/2}}
\sum_{a\neq b}
\int\dd^4\ell\,
e^{+\frac{i}{\hbar}[(\ell-k)\cdot b_a-\ell\cdot b_b]}
\del\!\left(2\tilde p_a\cdot(\ell-k)\right)
\del\!\left(2\tilde p_b\cdot\ell\right)
\cl A_{5,a}^{(0),\rm full}(k,\ell).
\label{eq:star-ordering-full-C1}
\]
This term is the leading contribution to the waveshape, however when nonlinear radiation kernels are present it does not determine the full waveshape by itself, since the higher $C_r$ contribute through cuts, as shown in section~\ref{sec:radiation-inclusive-cuts}.
\bibliographystyle{JHEP}
\bibliography{mainbib}
\end{document}